\pdfoutput=1 

\documentclass[11pt]{article}
\usepackage{amssymb}

\usepackage{amsthm} %for \theoremstyle{definition}

\usepackage{tikz}
\usetikzlibrary{trees}
\usepackage{qtree}

\usetikzlibrary{decorations.pathmorphing}
\usetikzlibrary{decorations.markings}
\usepackage{verbatim}

\usepackage{amsmath}
\usepackage{amsthm} %for \theoremstyle{definition}
\usepackage{amssymb,amsfonts,textcomp}
\usepackage{array}
\usepackage{hhline}
\usepackage{graphicx}

\usepackage[T3,T1]{fontenc} %for \textquotedbl
\usepackage{txfonts} %/pxfonts for \circleright

\usepackage{latexsym} %for \Box
\usepackage{mathrsfs} %for \mathscr

\usepackage{MnSymbol} %for more curved arrow symbol and many others are possible \lcurvearrowright

\usepackage{tikz}
\usetikzlibrary{arrows,shapes,chains,matrix,positioning,scopes}% shapes for circle split
\usetikzlibrary{decorations.pathmorphing}
\usetikzlibrary{decorations.markings}
\usepgflibrary{plotmarks}%for squares and other shapes on graphs lines , 
\usetikzlibrary{patterns}
\usepackage{pdfpages}

\usetikzlibrary{mindmap}%for circle connection bar

\usepackage{graphicx}

\usepackage{multirow}
\usepackage{multicol}

\usepackage{subfig} 

\usepackage{float}
\usepackage{wrapfig}
\restylefloat{figure}
\restylefloat{table}
\usepackage{color, colortbl} %to color rows
\usepackage{float}
\restylefloat{figure}

\theoremstyle{definition}

\newcommand{\STRUT}{\rule{0mm}{4ex}}

\newcommand{\pot}{o->}

\newcommand{\poti}{o-} %the start for line pots
\newcommand{\pote}{->} %the end for line pots

\newcommand{\jump}{right hook->}
\definecolor{currentcolor}{rgb}{0.8 0.4 0.2}%orange currentcolor is to be set to path color and then made lighter

\tikzstyle{stochasticjumpstyle}=[diamond,draw,fill=white,>=latex,>->,dashed]
\tikzstyle{stochasticPathstyle}=[>=latex,>->,dashed]
\tikzstyle{stochasticNodestyle}=[ellipse,inner sep=1pt,text=.,fill=.!20]%[fill=white,inner sep=1pt]%[ellipse,inner sep=1pt,draw,fill=white]
\tikzstyle{blankstyle}=[fill=white,inner sep=1pt]

\def\SnakeSegLen{0.6em}%defines snake segment length for signal jumps  in graphs
\def\SnakeAmp{0.11em}%defines snake amplitude for signal jumps  in graphs
\def\PrePostLen{5mm}

\tikzstyle{sendstyle}=[dashed,line width=1.1pt]%[dotted,ultra thick]

\tikzstyle{splitstyle}=[circle,draw]%not used

\newcommand{\sigjump}{*-to}
\newcommand{\sendsig}{*-latex}%{o->}%{triangle 60-*}%{triangle 60 reversed-*}%{*-*}%o-*

\tikzstyle{receivestyle}=[>->,line width=1.1pt,decorate, decoration={zigzag,segment length=\SnakeSegLen, amplitude=\SnakeAmp, pre length=\PrePostLen, post=curveto, post length=\PrePostLen},text=black]

\tikzstyle{receivesigstyle}=[draw,inner sep=2pt,fill=pink!20]
\tikzstyle{receivesigstyle3}=[draw,inner sep=2pt, fill=white]
\tikzstyle{receivesigstyle2}=[ellipse,shade, draw,double,fill=red!10]

\tikzstyle{sendsigstyle}=[diamond,draw,inner sep=1pt, text=black, fill=yellow!80]
\tikzstyle{sendsigstyle3}=[circle,draw, ball color=white]
\tikzstyle{sendsigstyle2}=[diamond,draw,double, inner sep=1pt, fill=white]

\tikzstyle{snakesendstyle}=[*->, decorate, decoration={snake, segment length=\SnakeSegLen, amplitude=\SnakeAmp,  pre length=\PrePostLen, post=curveto, post length=\PrePostLen}]

\tikzstyle{snakesendstyle1}=[line width=1.1pt, decorate, decoration={snake,segment length=\SnakeSegLen, amplitude=\SnakeAmp}]

\tikzstyle{snakesendstyle3}=[decorate, decoration={markings, mark=at position .75 with {\arrow[red,line width=5mm]{>}}, snake, segment length=\SnakeSegLen, amplitude=\SnakeAmp,  pre length=\PrePostLen, post=curveto, post length=\PrePostLen}]

\tikzstyle{snakesendstyle2}=[decorate, decoration={ zigzag,segment length=\SnakeSegLen, amplitude=\SnakeAmp, line around/.style={decoration={pre length=\PrePostLen,post length=\PrePostLen}}}]
\newcounter{foo}
\newcommand{\Cells}{\mbox{Cells}}

\usepackage{hyperref}

\makeatletter
\AtBeginDocument{\let\nameref\HyPsd@nameref}

\makeatother

\colorlet{anglecolor}{green!50!black}
\definecolor{darkgreen}{rgb}{0 0.6  0}

\definecolor{turquoise}{rgb}{0 0.41 0.41}
\definecolor{rouge}{rgb}{0.79 0.0 0.1}
\definecolor{vert}{rgb}{0.15 0.4 0.1}
\definecolor{mauve}{rgb}{0.6 0.4 0.8}
\definecolor{violet}{rgb}{0.58 0. 0.41}
\definecolor{orange}{rgb}{0.8 0.4 0.2}
\definecolor{bleu}{rgb}{0.39, 0.58, 0.93}

\definecolor{darkross}{rgb}{0.008,0.412,0.471}
\definecolor{middleross}{rgb}{0.012,0.580,0.663}
\definecolor{lightross}{rgb}{0.016,0.749,0.855}
\definecolor{darkblue}{rgb}{0.067,0.008,0.471}
\definecolor{middleblue}{rgb}{0.094,0.012,0.663}
\definecolor{lightblue}{rgb}{0.122,0.016,0.855}
\definecolor{darkpurple}{rgb}{0.471,0.008,0.412}
\definecolor{middlepurple}{rgb}{0.663,0.012,0.580}
\definecolor{lightpurple}{rgb}{0.855,0.016,0.749}
\definecolor{darkbrown}{rgb}{0.471,0.067,0.008}
\definecolor{middlebrown}{rgb}{0.663,0.094,0.012}
\definecolor{lightbrown}{rgb}{0.855,0.122,0.016}
\definecolor{darkolive}{rgb}{0.412,0.471,0.008}
\definecolor{middleolive}{rgb}{0.580,0.663,0.012}
\definecolor{lightolive}{rgb}{0.749,0.855,0.016}
\definecolor{darkgreen}{rgb}{0.008,0.417,0.067}
\definecolor{middlegreen}{rgb}{0.012,0.663,0.094}
\definecolor{lightgreen}{rgb}{0.016,0.855,0.122}
\definecolor{darkocre}{rgb}{0.471,0.298,0.008}
\definecolor{middleocre}{rgb}{0.663,0.420,0.012}
\definecolor{lightocre}{rgb}{0.855,0.541,0.016}

    \definecolor{lightblue}{rgb}{0,0,.7}
    \definecolor{orange}{rgb}{1,.7,0}
    \definecolor{darkorange}{rgb}{1,.4,0}
    \definecolor{darkgreen}{rgb}{0,.5,0}
    \definecolor{darkblue}{rgb}{0,0,.4}
    \definecolor{darkred}{rgb}{.4,0,0}
    \definecolor{gray}{rgb}{.2,.2,.2}
    \definecolor{darkgray}{rgb}{.2,.2,.2}
    \definecolor{shadecolor}{gray}{0.925}
    
\definecolor{darkred}{rgb}{0.65,0,0}
\definecolor{darkblue}{rgb}{0,0,.65}
\definecolor{darkgreen}{rgb}{0,0.5,0}
\definecolor{orange}{rgb}{1,.75,.25}
\definecolor{aqua}{rgb}{0,.25,.75}
\definecolor{grey}{rgb}{.5,.5,.5}
\definecolor{brown}{rgb}{.51,.35,.18}

\definecolor{lightblue}{rgb}{.3,.5,1}
\definecolor{orange}{rgb}{1,.7,0}
\definecolor{darkorange}{rgb}{1,.4,0}
\definecolor{darkgreen}{rgb}{0,.4,0}
\definecolor{darkblue}{rgb}{0,0,.4}
\definecolor{darkred}{rgb}{.56,0,0}
\definecolor{gray}{rgb}{.3,.3,.3}
\definecolor{darkgray}{rgb}{.2,.2,.2}

\definecolor{blue}{rgb}{0,0,1}
\definecolor{red}{rgb}{1,0,0}
\definecolor{pink}{rgb}{.933,0,.933}
\definecolor{green}{rgb}{0.133,0.545,0.133}
\definecolor{shadecolor}{gray}{0.925}

\definecolor{DarkBlue}{rgb}{0.000,0.000,0.545}
\definecolor{DarkChocolate}{rgb}{0.400,0.200,0.000}
\definecolor{DarkCyan}{rgb}{0.000,0.545,0.545}
\definecolor{DarkGoldenrod}{rgb}{0.720,0.525,0.044}
\definecolor{DarkGray}{rgb}{0.664,0.664,0.664}
\definecolor{DarkGreen}{rgb}{0.000,0.392,0.000}
\definecolor{DarkGrey}{rgb}{0.664,0.664,0.664}
\definecolor{DarkKhaki}{rgb}{0.740,0.716,0.420}
\definecolor{DarkLavender}{rgb}{0.400,0.200,0.600}
\definecolor{DarkMagenta}{rgb}{0.545,0.000,0.545}
\definecolor{DarkOliveGreen}{rgb}{0.332,0.420,0.185}
\definecolor{DarkOrange}{rgb}{1.000,0.550,0.000}
\definecolor{DarkOrchid}{rgb}{0.600,0.196,0.800}
\definecolor{DarkPeriwinkle}{rgb}{0.400,0.400,1.000}
\definecolor{DarkPurpleBlue}{rgb}{0.400,0.000,0.800}
\definecolor{DarkRed}{rgb}{0.545,0.000,0.000}
\definecolor{DarkRoyalBlue}{rgb}{0.000,0.200,0.800}
\definecolor{DarkSalmon}{rgb}{0.912,0.590,0.480}
\definecolor{DarkSeaGreen}{rgb}{0.560,0.736,0.560}
\definecolor{DarkSlateBlue}{rgb}{0.284,0.240,0.545}
\definecolor{DarkSlateGray}{rgb}{0.185,0.310,0.310}
\definecolor{DarkSlateGrey}{rgb}{0.185,0.310,0.310}
\definecolor{DarkSmoke}{rgb}{0.920,0.920,0.920}
\definecolor{DarkTurquoise}{rgb}{0.000,0.808,0.820}
\definecolor{DarkViolet}{rgb}{0.580,0.000,0.828}
\definecolor{DeepPink}{rgb}{1.000,0.080,0.576}
\definecolor{DeepSkyBlue}{rgb}{0.000,0.750,1.000}

\tikzstyle{mystyle}=[scale= \PicSize,  %[****Crit. PicSize is not defined*****]
baseline=(current bounding box.center),
nodedescr/.style={fill=white,inner sep=2.5pt},
nodedescr2/.style={draw,circle,fill=white},
description/.style={fill=white,inner sep=2pt},
cancer loop/.style={preaction={draw=white, -,line width=6pt}, line width=1.1pt},
pot1 loop/.style={preaction={draw=white, -,line width=6pt}, red, line width=1.1pt},
pot2 loop/.style={preaction={draw=white, -,line width=6pt}, green, line width=1.1pt},
pot1/.style={preaction={draw=white, -,line width=6pt}, gray, solid, line width=1pt}, 
pot2/.style={preaction={draw=white, -,line width=6pt}, gray, solid, line width=1pt},
inPot/.style={ out=45,in=90,distance=3cm, min distance=2cm, line width=1.1pt, black!75},
inPot1/.style={ out=45,in=120,distance=3cm, min distance=2cm, line width=1.1pt, black!75},
inPot2/.style={ out=-45,in=-120,distance=3cm, min distance=2cm, line width=1.1pt, black!75},
in100Pot1/.style={ out=80,in=90,distance=3cm, min distance=2cm, line width=1.1pt, darkgreen},
in100Pot2/.style={ out=-80,in=-90,distance=3cm, min distance=2cm, line width=1.1pt, brown},
in2Pot1/.style={ out=45,in=90,distance=3cm, min distance=2cm, line width=1.1pt, black!75},
in2Pot2/.style={ out=-65,in=-90,distance=3cm, min distance=2cm, line width=1.1pt, black!75},
selfloop1/.style={ out=45,in=120,distance=6cm, min distance=4cm, line width=1.1pt, red},
selfloop2/.style={ out=-45,in=-120,distance=6cm, min distance=4cm, line width=1.1pt, blue},
loop1red/.style={ out=45,in=120,distance=6cm, min distance=4cm, line width=1.1pt, red},
loop2red/.style={ out=-45,in=-80,distance=6cm, min distance=4cm, line width=1.1pt, red},
cross line/.style={preaction={draw=white, -,	line width=6pt}},
jump1/.style={ out=45,in=135,distance=4cm, min distance=2cm, line width=1.1pt, red, dashed},
jump2/.style={ out=-45,in=-135,distance=4cm, min distance=2cm, line width=1.1pt, blue, dashed},
jumpover/.style={ out=80,in=90,distance=4cm, min distance=3cm, line width=1.1pt, black, dashed},
stochastic jump/.style={>=latex,>->,dashed, line width=1.1pt, green}
]

\def\PicSize{ 0.5} % 0.5 defines constant PicSize for uniform scale of TikZ pictures

\def\ColSepNarrow{2.7em}

\def\ColSepTight{1.7em} %1.9 works for all but one

\def\ColSepExtraTight{1.5em} %1.9 works for all but one

\def\CellWallThickness{ 0.50mm}

\newcommand{\matrixsep}{0.75cm} %treat row and column separation same size

\numberwithin{equation}{section}
\usepackage[margin=1.4in]{geometry} %***set page dimensions easily. See geometryPageSize.pdf***

\usepackage[parfill]{parskip}    % Activate to begin paragraphs with an empty line rather than an indent

\begin{document}

\title{Stem Cell Networks}

\author{Eric Werner \thanks{Balliol Graduate Centre, Oxford Advanced Research Foundation (http://oarf.org), Cellnomica, Inc. (http://cellnomica.com). We gratefully acknowledge the use of Cellnomica's Software Suite to construct the cancer and stem cell networks used to model and simulate all the multicellular processes that generated the {\em in silico} cancers described and illustrated in this paper.  \copyright Werner 2016.  All rightsreserved. }\\
University of Oxford\\
Department of Physiology, Anatomy and Genetics, \\
and Department of Computer Science, \\
Le Gros Clark Building, 
South Parks Road, 
Oxford OX1 3QX  \\
email:  eric.werner@dpag.ox.ac.uk\\
}

\date{ } %This is to suppress the printing out of the date.

\maketitle

\thispagestyle{empty}

\begin{center}
\textbf{Abstract}

\begin{quote}
\it
We present a general computational theory of stem cell networks and their developmental dynamics.  Stem cell networks are special cases of developmental control networks. Our theory generates a natural classification of all possible stem cell networks based on their network architecture. Each stem cell network has a unique topology and semantics and developmental dynamics that result in distinct  phenotypes.   We show that the ideal growth dynamics of multicellular systems generated by stem cell networks have mathematical properties related to the coefficients of Pascal's Triangle.  The relationship to cancer stem cells and their control networks is indicated. The theory lays the foundation for a new research paradigm for understanding and  investigating stem cells.  The theory of stem cell networks implies that new methods for generating and controlling stem cells will become possible.  
\end{quote}
\end{center}

\pagebreak

\pagenumbering{roman}
\setcounter{page}{1}
\tableofcontents
%\listoffigures
%\listoftables
\newpage
\pagenumbering{arabic}

\section{Introduction}

Stem cells are crucial in embryogenesis as well as maintaining our bodies. They are the source of tissue regeneration, generating skin, hair, and replacing damaged cells in wound healing. They are also the source of many cancers.  Understanding the nature of stem cells and what controls them is, therefore, of great importance.  The aim of this essay is to understand the essence of what makes stem cells run.  To achieve this we present a general theory of developmental networks that lie hidden in stem cells and control them.  The theory is abstract and powerful describing the universal principles and architectures that underly all stem cell networks.  This allows us to model and simulate all stem cell types.   And that enables us to do computational experiments by running different networks in virtual stem cells and observing their behavior in a virtual space-time context of other cells.  

\label{sec:G3OctalTetrahedron}
\begin{figure}[H]
\subfloat[{\bf G3 OctalTetrahedron.}  ]{
\includegraphics[scale=0.38]{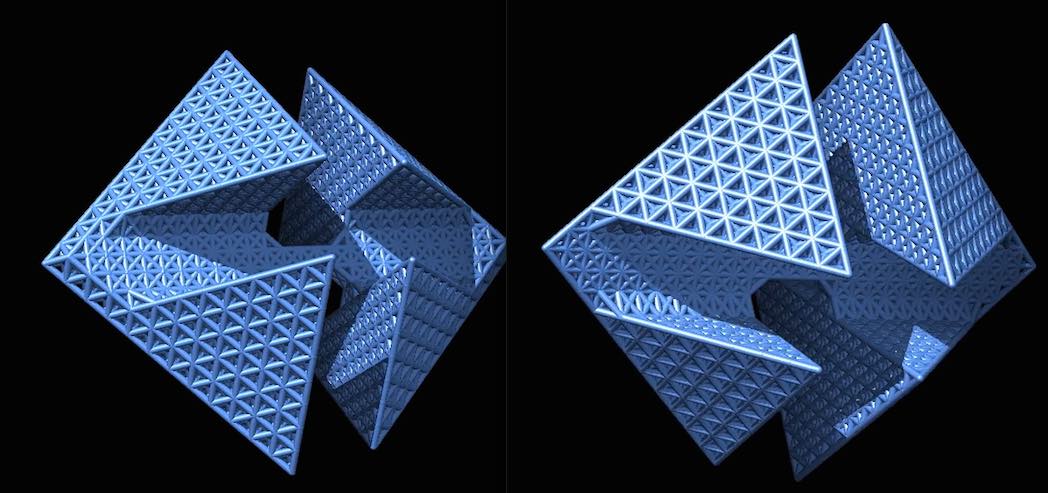}}
\caption{
{\bf Two views of an Octal Tetrahedron. } This Octal Tetrahedron was grown from a single cell. A cell first divides symmetrically several times to produce the resulting symmetries \cite{Werner2012a}. Then a 3rd order stem cell networks generates a tetrahedron. The result after several synchronous divisions is what you see. As shown in this article, these stem cell networks whether cancerous or normal have growth dynamics related to the classical Greek figurative numbers and Pascal's triangle. 
}
\label{fig:G3OctalTetrahedron}
\end{figure}

The result is a deep understanding of the properties of real stem cells and their controlling networks.  It also gives us a profound insight into the nature of the role of cancer stem cells in metastases.  The theory predicts that there exists a metastatic hierarchy that corresponds directly to the stem cell network hierarchy.  We will describe the stem cell network hierarchy and show that this network hierarchy has direct causal links to this previously unknown hierarchy of metastases. This has important implications for the classification, diagnosis and treatment of metastases in cancer.  The theory also implies that it is in principle possible to construct methodologies that can transform cancer stem cell networks into a harmless form giving us the potential to ultimately cure and even eliminate such cancers.   

\subsection{Developmental control networks}
We postulate that multicellular development is controlled by networks. We call these networks {\em developmental control networks} or {\em cenes} for control genes.  The global developmental control network is the entire network that controls the development of an organism.  This global cene is sometimes called the {\em cenome} \cite{Werner2011a}.  

{\em Stem cell networks} form a subclass of developmental networks.  They have a well defined network architectures with a range of topologies.  We will show that stem cell networks form a hierarchy based on their network topology.  The properties of stem cells are determined by the properties of their network architecture.  It is the network architecture that gives stem cells their primary defining property, namely, the capacity to self-renew while generating an unlimited set of progeny of various cell types.  The network locality determines the stem cell maturity and the capacity of its progeny to differentiate to various cell types. Thus, the potency of a stem cell's daughter cell is determined by the locality of that stem cell network in the global developmental network. 

\subsection{Definitions: Two kinds of stem cells}

We distinguish two independent properties of stem cells:  Self-renewal (iteropotency) and the capacity to transform into various cell types (transpotency).   These two properties are often confused in the scientific and popular science literature. 

\subsubsection{Iteropotent stem cells}
We define a cell to be {\em iteropotent} if it has the capacity of self-renewal.  When an {\em iteropotent stem cell} divides it generates two daughter cells, one that has the same control state as the parent and one that is in a new developmental control state.  The {\em developmental control state} of a cell is defined  by the set nodes in the developmental network that are active in that cell.  Examples of self-renewing, iteropotent stem cells are skin (epithelial) and hair stem cells, bone marrow stem cells, and cancer stem cells.  {\em Aniteropotent cells} are by definition non-self-renewing cells.  

\subsubsection{Transpotent stem cells} 
We define a cell to be {\em transpotent} if it has the capacity to differentiate into other cell types. The {\em degree of transpotency} is the variety and number of cell types into which a transpotent cell can differentiate. Transpotent stem cells are usually, roughly classified into embryonic, totipotent, pluripotent, and oligopotent stem cells.  Transpotency may be conditional and context dependent.  {\em Natural transpotency} is a natural, inherent property of the cell.  It is the natural property of a cell to transform itself based on its context and developmental network state. {\em Artificial transpotency} is an artificial, experimentally induced property of cells that would not naturally occur.  

\subsubsection{Relations between transpotency and iteropotency of stem cells}
Because self-renewal and transpotency are orthogonal properties of cells they can occur in any combination:  A transpotent stem cell need not have the capacity of self-renewal even though it may generate stem cells that do have that capacity.  So too, the two properties of stem cells can coexist in the same cell. For example, a self-renewing stem cell may also be transpotent, but if it differentiates it may or may not retain its property of self-renewal.   Finally, a self-renewing stem cell may may not be transpotent and yet generate transpotent cells.  All these different cases can be more precisely defined and modeled by different developmental stem cell networks.

\subsection{The graphical network formalism}
We will use an abstract graphical network language to describe developmental networks. Differential equations are often used to describe the quantitative output of cell proliferation.  While our graphical networks can in part be reductively formalized by differential equations, such reductions loose many of the architectural properties and advantages of graphical developmental networks.  A good deal of theoretical and practical insight is lost in reductions to differential equations.  Since many of the biological parameters and molecular implementation of stem cell networks is still unknown and perhaps never will be fully known, graphical representation languages of developmental networks help us to understand developmental processes and may even enable us to control developmental processes without necessarily knowing or understanding their full molecular implementation.  This is not to say that modeling attempts with differential equations is not worthwhile. Indeed, we use differential equations to model the physics of cell interactions.  
One goal is to link up lower level differential equation formalisms of molecular networks with our higher level abstract graphical representations of developmental control networks.  Any cooperations to this end are welcomed by the author. 

\subsection{Plan}
The main focus of this essay will be on self-renewing, iteropotent stem cell networks. We first look at deterministic stem cell networks. We describe meta-stem cell networks and their mathematical properties. We discover a stem cell hierarchy of networks and show that their proliferative properties are related to the geometric numbers and the coefficients of Pascal's Triangle.  Next,  we describe the close relationship between normal stem cell networks and cancer stem cell networks.  We apply the theory of the hierarchy of stem cell networks to metastases. We show that there is a direct correspondence between the stem cell network hierarchy and a metastatic hierarchy. The theory of a hierarchy of metastases makes testable predictions and has significant implications for cancer diagnosis and treatment.  We then add stochasticity to stem cell networks and investigate the range of stochastic stem cell networks. It turns out that adding stochasticity makes stem cell networks extremely flexible.  Next, we add cell communication to stem cell networks.  Finally, we combine deterministic, stochastic and communication stem cell networks.  

\section{A hierarchy of stem cell networks} %
\label{sec:NG}
Traditionally, stem cells have been viewed as cells that endlessly self-regenerate while producing terminal daughter cells of some type regulated by some terminal developmental network.  However, there may exist stem cells produce other stem cells.  We refer to stem cells that produce stem cells as {\em meta-stem cells} or as {\em higher order stem cells}.  The order of a stem cell depends on the number of linear loops in its developmental control network.  A {\em 3rd-order stem cell network} contains three loops and produces 2nd-order and 1st-order stem cells. In turn, a {\em 2nd-order stem cell network} contains two loops and produces only 1st-order stem cells.  {\em 1st-order stem cells} are controlled by single loop and produce terminal cells or cells with limited proliferative potential. Cells with limited proliferative potential that can engage in a small number of further divisions are called  {\em progenitor cells}.  We will see that stem cell networks have proliferation or growth potential related to the geometric numbers and the coefficients of Pascal's Triangle.  Without preconditions in an idealized setting without physical constraints, a {\em $k-$th order stem cell network} would have the proliferations properties of an $k$-th order geometric network (see \ref{sec:Gk}).  We will thus sometimes refer to stem cell networks as {\em geometric networks} when emphasizing their geometric growth properties and to distinguish them from exponential networks (Werner~\cite{Werner2011b}).

\subsection{First order linear stem cells}
\label{sec:NSC1}
A first order stem cell produces no additional stem cells.  Instead a first order stem cell $A$ produces cells $B$  that is in a different control state than its parent cell $A$. The stem cell thus produces one daughter cell that inherits the control state of its parent which is $A$ itself and one daughter cell $B$ that is in a new control state.  Thus stem cells have a self reflexive control system.  Note, that the cell $B$ may still be multi-potent  in that it may be controlled by a network that generates a whole multicellular system. 

First order stem cells have conditional control networks. This means that their activation depends on not just being linked into another network that activates them but that they can only be active if the conditions $\Phi$ are satisfied.  

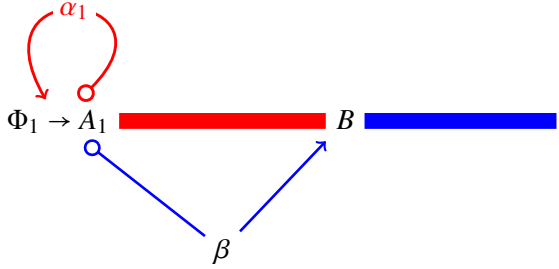
\begin{figure}[H]
\begin{tikzpicture}[style=mystyle]
\matrix (m) [matrix of math nodes, row sep=3em,
column sep=3em, text height=1.5ex, text depth=0.25ex]
{ \vphantom{a} & \vphantom{b}  &  \vphantom{c}  & \vphantom{c}  & \\
\Phi_{1} \rightarrow A_{1}  &&  B  &&  \vphantom{C}  && \vphantom{D} \\
  \vphantom{a} &  \beta & \vphantom{b} &  \vphantom{d}  & \\ };
 \path[\poti]
(m-2-1) edge [pot2, blue](m-3-2);
 \path[\pote]
(m-3-2) edge [pot2, blue] (m-2-3); 
\path[\pot]
(m-2-1) edge [selfloop1, cross line] node[nodedescr] {$\alpha_{1}$} (m-2-1);

\path[solid, red, line width=6pt]
(m-2-1) edge (m-2-3);
\path[solid, blue, line width=6pt]
(m-2-3) edge (m-2-5);
%(m-2-5) edge (m-2-7);
\end{tikzpicture} 

  \caption{
    {\bf Network G1: 1st Order (Linear) Stem  Cell Network} The cell of type $A_{1}$ is  a regular stem cell that divides conditionally.  If condition $\Phi_{1}$ hold, then  $A_{1}$ produces cells of type $B$.  One daughter of  $A_{1}$ self loops giving a daughter cell of the same cell type $A_{1}$ as the parent.  The other daughter cell differentiates to type $B$.  The a stem cell controlled by this network exhibits slow, linear growth.  The proliferation potential is given by \autoref{eq:G1}. 
  }
\label{fig:G1}
\end{figure}

In terms of network topology, first order stem cells are similar to first order geometric cancer networks in that both contain one loop Werner~\cite{Werner2011b}  They contain one loop and generate cells linearly, one at a time. The newly generated cells $B$ are terminal cells that do not proliferate or they may be progenitor cells that are controlled by a terminal developmental network (see Werner~\cite{Werner2011b, Werner2011a} for more on developmental control networks, stem cell and cancer networks.  The cells $B$ may also be transpotent cells that can be induced naturally or artificially to differentiate into various cell types. 

\subsection{Second order geometric meta-stem cells}
\label{sec:DSC2}
Basic first order stem cells can only produce terminal cells. A first order stem cell cannot generate any further stem cells and a regular developmental network that links to a single stem cell network will only generate one stem cell.  Thus, to produce multiple stem cells we need another type of developmental network.  One way multiple stem cells can be produced is by networks of the type NI$_k$ (\autoref{fig:NIkG1}) that could produce $2^k$ identical first order stem cells.  But this is still limited by number of loops $k$ in those networks. Another method is to have meta-stem cell networks that can endlessly produce 1st order stem cells.  It is possible that organisms use both strategies. 

Thus, there are at least two methods of producing stem cells.  A developmental network may link to a stem cell network to generate one or more stem cells.  Or there are meta-stem cells that can produce an unlimited supply of stem cells.   A third method uses stochastic dedifferentiation of terminal or progenitor cells to their 1st order stem cell parent network. (\cite{Werner2011b}). 

Meta-stem cells, like first order stem cells, can also have conditional activation in that being linked into another network is not sufficient for their potential activation. Their preconditions $\Phi_{2}$ must be satisfied at each pass of the loop to execute the next loop.  In the case of cancer stem cells the preconditions may be permanently switched on or they may be null. 

 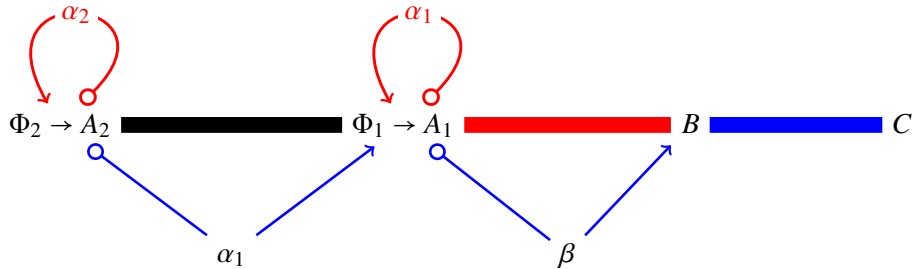
\begin{figure}[H]
\begin{tikzpicture}[style=mystyle]
\matrix (m) [matrix of math nodes, row sep=3em,
column sep=3em, text height=1.5ex, text depth=0.25ex]
{ \vphantom{a} & \vphantom{b}  &  \vphantom{c}  & \vphantom{c}  & \\
 \Phi_{2} \rightarrow A_{2} &&  \Phi_{1} \rightarrow A_{1} &&  B  && C \\
  \vphantom{a} &  \alpha_{1} & \vphantom{b} &  \beta  & \\ };
 \path[\pot]
(m-2-1) edge [selfloop1, cross line] node[nodedescr] {$ \alpha_{2} $} (m-2-1);
 \path[\poti]
(m-2-1) edge [pot2,blue](m-3-2);
 \path[\pote]
(m-3-2) edge [pot2,blue] (m-2-3); 
\path[\pot]
(m-2-3) edge [selfloop1, cross line] node[nodedescr] {$ \alpha_{1} $} (m-2-3);
 \path[\poti]
(m-2-3) edge [pot2,blue](m-3-4);
 \path[\pote]
(m-3-4) edge [pot2,blue] (m-2-5);

\path[solid,  line width=6pt]
(m-2-1) edge (m-2-3);
\path[solid, red, line width=6pt]
(m-2-3) edge (m-2-5);
\path[solid, blue, line width=6pt]
(m-2-5) edge (m-2-7);
\end{tikzpicture}
\caption{
    {\bf Second order, meta-stem cell network $NG_{2}$:} The cell $A_{2}$ is a {\em meta-stem cell} that given condition $\Phi_{2}$ holds,  produces stem cells of type $A_{1}$ which in turn  conditionally produces cells of type $B$. Thus, the cell of type $A_{1}$ is  a regular stem cell.  If condition $\Phi_{1}$ hold, then  $A_{1}$ produces cells of type $B$. When their preconditions are satisfied, both $A_{2}$ and $A_{1}$ proliferate linearly such that the joint proliferation potential is given by \autoref{eq:G2}.  
}
    \label{fig:G2}
\end{figure} 

\subsection{3rd-Order stem cell networks NG$_3$ with $3$ loops}
\label{sec:G3}
Adding another loop to a meta-stem cell network results in a meta-meta-stem cell network or third order stem cell network.  

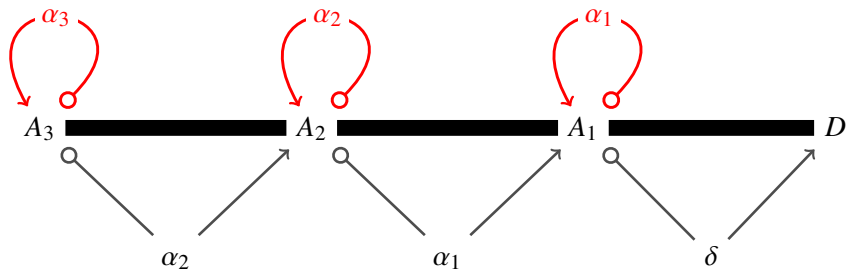
\begin{figure}[H]
\begin{tikzpicture}[style=mystyle]
\matrix (m) [matrix of math nodes, row sep=3em,
column sep=3em, text height=1.5ex, text depth=0.25ex]
{ \vphantom{a} & \vphantom{b}  &  \vphantom{c}  & \vphantom{c}  & \\
 A_3  &&  A_2  &&  A_1  && D \\
  \vphantom{a} &  \mbox{$\alpha_2$} & \vphantom{b} &  \mbox{$\alpha_1$}  &  \vphantom{a3}  & \delta \\ };
 \path[\pot]
(m-2-1) edge [selfloop1, cross line] node[nodedescr] {$ \alpha_3 $} (m-2-1);
\path[\poti]
(m-2-1) edge [pot2](m-3-2);
\path[\pote]
(m-3-2) edge [pot2] (m-2-3);
\path[\pot]
(m-2-3) edge [selfloop1, cross line] node[nodedescr] {$ \alpha_2 $} (m-2-3);
\path[\poti]
(m-2-3) edge [pot2](m-3-4);
\path[\pote]
(m-3-4) edge [pot2] (m-2-5);
\path[\pot]
(m-2-5) edge [selfloop1, cross line] node[nodedescr] {$ \alpha_1 $} (m-2-5);
\path[\poti]
(m-2-5) edge [pot2](m-3-6);
\path[\pote]
(m-3-6) edge [pot2] (m-2-7);

\path[solid, line width=6pt]
(m-2-1) edge (m-2-3);
\path[solid, line width=6pt]
(m-2-3) edge (m-2-5);s
\path[solid, line width=6pt]
(m-2-5) edge (m-2-7);
\end{tikzpicture}
\caption{
  {\bf Network NG$_{3}$ A 3rd-Order Stem Cell Network:} A stem cell network $3$ loops: The proliferation potential is give by \autoref{eq:G3}. Each cell of type $A_i$, $i < 3$,  is a higher order meta-stem cell; it produces stem cells of type $A_{i-1}$.  Thus, cell type $A_{3}$ is a meta-meta-stem cell that produces meta-stem cells of type $A_2$, etc. The cell type $A_2$ is a stem cell that produces a stem cell $A_{1}$ which in turn produces a non-stem cells of some type D.  }
  \label{fig:G3}
\end{figure} 

\subsection{Meta-stem cells}

Geometric networks  of the above type suggest a new category of stem cells, namely, meta-stem cells (see \autoref{sec:DSC2}).  A {\em linear}  or {\em first order stem cell} is a cell that produces other cells (that are not stem cells) by means of a linear developmental network.  {\em Meta-stem cells} are stem cells that produce stem cells.  Thus a 2nd-order geometric network with two loops contains a meta-stem cell $A_{2}$ that produces linear stem cells of type $A_{1}$.  A 3rd-order three loop geometric network contains a {\em meta-meta-stem} cell $A_{3}$ that produces meta-stem cells $A_{2}$ that produce stem cells $A_{1}$ that produce cells of type D. And, so on.  

Whether a geometric meta-stem cell network is a cancerours network will depend on the properties of the cells and how they function in the organism.  Any stem cell or meta-stem cell network can by mutations become an exponential cancer (see below).  A transformation of linear stem cell network into a meta-stem cell network will make the stem cell proliferate according to \autoref{eq:TriangularNo}  below and be potentially harmful.  Indeed, even the metastatic behavior of cancer stem cells will be seen to be directly related to the properties of their geometric, meta-stem cell networks (\autoref{sec:GMetastases}). 

\subsection{Geometric stem cell networks NG$_k$ with $k$ loops}
\label{sec:Gk}

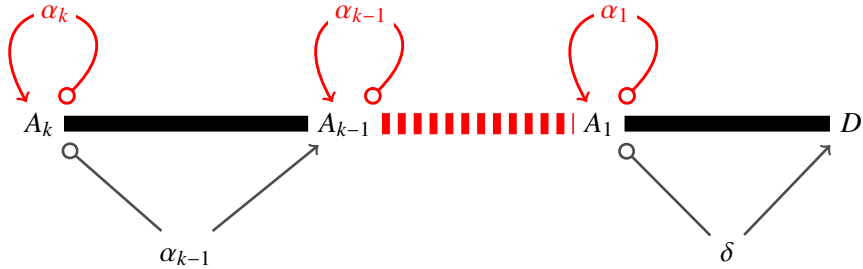
\begin{figure}[H]
\begin{tikzpicture}[style=mystyle]
\matrix (m) [matrix of math nodes, row sep=3em,
column sep=3em, text height=1.5ex, text depth=0.25ex]
{ \vphantom{a} & \vphantom{b}  &  \vphantom{c}  & \vphantom{c}  & \\
 A_k  &&  A_{k-1}  &&  A_1  && D \\
  \vphantom{a} &  \mbox{$\alpha_{k-1}$} & \vphantom{b} &  \vphantom{a3}  &  \vphantom{a3}  & \delta \\ };
 \path[\pot]
(m-2-1) edge [selfloop1, cross line] node[nodedescr] {$ \alpha_k $} (m-2-1);
\path[\poti]
(m-2-1) edge [pot2](m-3-2);
\path[\pote]
(m-3-2) edge [pot2] (m-2-3);
\path[\pot]
(m-2-3) edge [selfloop1, cross line] node[nodedescr] {$ \alpha_{k-1} $} (m-2-3);
\path[\poti]
(m-2-5) edge [pot2](m-3-6);
\path[\pote]
(m-3-6) edge [pot2] (m-2-7);

\path[\pot]
(m-2-5) edge [selfloop1, cross line] node[nodedescr] {$ \alpha_1 $} (m-2-5);

\path[solid, line width=6pt]
(m-2-1) edge (m-2-3);
\path[dashed, red, line width=8pt]
(m-2-3) edge (m-2-5);s
\path[solid, line width=6pt]
(m-2-5) edge (m-2-7);
\end{tikzpicture}
\caption{
  {\bf Network NG$_{k}$: An k-th order stem cell network:} A geometric cancer with $k$ loops: Each of the cell types $A_i$, $i = k \ldots 1$ are  cancer cells whose joint proliferation potential is give by \autoref{eq:Gnk}. Each cell of type $A_i$, $i < k$,  is a higher order meta-stem cell; it produces stem cells of type $A_{i-1}$.  Thus, cell type $A_{k}$ is a meta-stem cell that produces stem cells of type $A_{k-1}$, etc. The cell type $A_1$ is a first order stem cell that produces regular non-stem cells of type D.  The difference between stem cells and cancer stem cells has to do with their functionality in the system as a whole. 
  }
  \label{fig:Gk}
\end{figure} 

\subsection{Mathematical properties of geometric stem cell networks}
\label{sec:Pascal}
A stem cell network of the above type with one or more loops in a linear connected sequence leads has interesting mathematical properties. It turns out that such networks are directly related to Pascal's Triangle, binomial coefficients and geometric numbers. Under the appropriate interpretation, one loop simple produces a line. Two loops produce a triangle. Three loops produce an equilateral pyramid (tetrahedron), four loops a pentalope (a four dimensional pyramid), etc.  In general, an network of $k$ linearly connected loops will produce an $k$-dimensional pyramid. Or viewed numerically, after $n$ synchronous rounds of ideal division, a cancer network with $k$ linearly connected single loops produce the sum of the first $k$ binomial coefficients at level $n$ of Pascal's triangle. 

Recall that by definition:
\begin{equation}
 {n \choose k}   =  \frac{n!}{k!(n - k)!}
\end{equation}

Then the number of cells produced by a single cell that is controlled by an stem cell network with $k$ linearly connected loops, after undergoing $n$ rounds of synchronous cell division division is given by the following  formula (for  $n > 0$): 
\begin{equation}
{\mbox Cells}(n, k) = \sum_{i = 0}^k  {n \choose i} = \sum_{i = 0}^k \frac{n!}{i!(n - i)!}
\label{eq:Gnk}%was GGformula
\end{equation}

The above formula shows the direct relationship between such stem cell networks and the binomial coefficients of Pascal's triangle, i.e., the coefficients the binomial theorem. 

Given the following standard definitions: 
\begin{enumerate}

\item Linear number: \begin{equation}
 Lin(n) = 1 + n  = 1 + {n \choose 1} \mbox{ when } n \geq 1 
\end{equation}

\item Triangular number: 
\begin{equation}
 Tri(n) = \frac{n^2 + n}{2} = \frac{n(n + 1)}{2}  = {n + 1 \choose 2} %\left( \begin{array}{c} n \\ 2 \end{array} \right)
 \label{eq:TriangularNo}
\end{equation}

\item Tetrahedral number: 
\begin{equation}
Tet(n) =  \frac{n(n + 1)(n+2)}{6} = {n + 2 \choose 3} %\left( \begin{array}{c} n \\ 3 \end{array} \right)
 \label{eq:TetrahedralNo}
\end{equation}

\item Pentalope number:  
\begin{equation}
Pen(n) = \frac{n(n + 1)(n+2)(n + 3)}{24} = {n +3 \choose 4} %\left( \begin{array}{c} n \\ 4 \end{array} \right)
\end{equation}
\end{enumerate}

Then the number of cells, Cells$(n,k)$, that develop after $n$ rounds of synchronous division obey the sum of the coefficients of Pascal's Triangle where $n$ is the height of the triangle and $k$ is horizontal coordinate corresponding to the number of single connected loops in the regulatory network.  These sums correspond to the volumes of the corresponding $k$-dimensional geometric form in an ideal $k$-dimensional discrete space:  

\begin{enumerate}
\item For one loop we get  a $1$-dimensional structure where its length gives the number of cells: 
\begin{equation}
Cells(n, 1) = Lin(n) = 1 + n
\label{eq:G1}
\end{equation}

\item Two loops add the triangular number to give the area of a $2$-dimensional triangle: 
\begin{eqnarray} Cells(n, 2) & = & Lin(n) + \textcolor{red}{Tri(n-1)} \\ \STRUT
& = & 1 + n + \textcolor{red}{\frac{n(n - 1) }{2}} \\  %\\ \STRUT
& = & \sum_{i = 0}^2 {n \choose i} = {n \choose 0} + {n \choose 1} +  \textcolor{red}{{n \choose 2} }
\label{eq:G2}
\end{eqnarray}

\item For three loops add the tetrahedral number to give the volume of $3$-dimensional pyramid: 
\begin{eqnarray} Cells(n, 3) & = & Lin(n) + Tri(n-1) + \textcolor{red}{Tet(n-2)} \\ \STRUT
& = & 1 + n + \frac{n(n - 1)}{2} + \textcolor{red}{ \frac{n(n-1)(n-2)}{6}} \\ \STRUT
& = & \sum_{i = 0}^2 {n \choose i} = {n \choose 0} + {n \choose 1} + {n \choose 2} + \textcolor{red}{{n \choose 3} }
 \label{eq:G3}
\end{eqnarray}

\item For four loops add the pentalope number to give the volume of a $4$-dimensional pyramid:  
\begin{eqnarray} 
 Cells(n, 4) & = & Lin(n) + Tri(n-1)+ Tet(n-2) + \textcolor{red}{Pen(n-3)} \\ \STRUT
 & = & 1 + n + \frac{n(n - 1)}{2} + \frac{n(n-1)(n-2)}{6} + \textcolor{red}{\frac{n(n-1)(n-2)(n-3)}{24} }\\  \STRUT
 & = & {n \choose 0} + {n \choose 1} +   {n \choose 2} +  {n \choose 3} + \textcolor{red}{{n \choose 4} }
  \label{eq:G4}
 \end{eqnarray}
  
 \item For $k$ loops sum the sequence of numbers through ${\frac{n!}{k!(n - k)!}} $ to give the volume of a $k$-dimensional pyramid. Given $n > 0$ :  
\begin{eqnarray} 
 Cells(n, k) & = & Lin(n) + Tri(n-1)+ Tet(n-2) + \ldots + \textcolor{red}{{n \choose k}}\\ \STRUT
 & = & 1 + n + \frac{n(n - 1)}{2} + \frac{n(n-1)(n-2)}{6} + \dots 
 + \textcolor{red}{\frac{n!}{k!(n - k)!}} \\ \STRUT
 & =  &  {n \choose 0} + {n \choose 1} + {n \choose 2} +  \dots  + \textcolor{red}{{n \choose k} }\\ \STRUT
 & = & \sum_{i = 0}^k {n \choose i} =  \sum_{i = 0}^k \frac{n!}{i!(n - i)!} 
 \label{eq:longGnk}
 \end{eqnarray}
 
 \item So, in general we have: 
\begin{eqnarray}
 Cells(n, k) = \left\{ 
 \begin{array}{ll} 
 		1 & \mbox{if $n = 0$} \\ 
 		\sum_{i = 0}^k \frac{n!}{i!(n - i)!} 
 		= \sum_{i = 0}^k {n \choose i} %\left( \begin{array}{c} n \\ i \end{array} \right) 
 		& \mbox{otherwise}
 \end{array} 
 				\right.
\end{eqnarray}
\label{eq:generalGnk} %was generalGk

\end{enumerate}

This view shows the direct relationship between this type of cancer and stem cell network and the geometric numbers. 
What is fascinating is that these ideal networks have a generative competence with such interesting numerical properties.  Their mathematical history spans back from the binomial theorem, to Pascal's triangle, to the Greek's discovery of geometric numbers.  

\section{Stem cell production in development}
%was {Multi linear stem cell networks NLI$_k$ of identical cells}

Stem cells are created in a developmental context. This is done by linking in a stem cell network into the normal developmental network. The locality of origin of the link in the developmental network to the stem cell network determines the properties of the cells that a stem cell network generates.  For example, heart stem cells are different from liver stem cells even if one can coaxed into assuming the properties of the other. 

\subsection{Multi-identical linear stem cell network}
A linear $k$-identical cell stem cell network $NI_{k}L = NI_{k}G_{1}$ links an identical cell network NI$_k$ that generates $k$ identical cells with a downstream linear stem cell network $NL = G_{1}$. The simplest case is where the number of identical linear stem cells is $k=2$: 

\begin{figure}[H]
\begin{tikzpicture}[style=mystyle]
\matrix (m) [matrix of math nodes, row sep=3em,
column sep=3em, text height=1.5ex, text depth=0.25ex]
{ \vphantom{a} & \vphantom{a}  &  \vphantom{c}  & \vphantom{c}  & \\
 A  &&  B  &&  C  && D \\
  \vphantom{n} & \vphantom{a} & \vphantom{b} &  c  & \\ };
 \path[\pot]
(m-2-1) edge [inPot1, DarkGreen]node[nodedescr] {$ b $}( m-2-3);
 \path[\pot] 
(m-2-1) edge [inPot2, DarkBlue]node[nodedescr] {$ b $}( m-2-3); 
%to A
%(m-1-2) edge  [cross line, pot1] (m-2-3)
 \path[\poti]
(m-2-3) edge [pot2, blue](m-3-4);
 \path[\pote]
(m-3-4) edge [pot2, blue] (m-2-5);
%(m-1-2) edge [pot1 loop, bend left=-80] node[nodedescr] {$ \alpha_1 $} (m-1-1)
 \path[\pot]
(m-2-3) edge [selfloop1, cross line] node[nodedescr] {$ b $} (m-2-3);

\path[solid,black!20, line width=6pt]
(m-2-1) edge (m-2-3);
\path[solid,red!90, line width=10pt]
(m-2-3) edge (m-2-5);
\path[solid,black!20, line width=6pt]
(m-2-5) edge (m-2-7);
\end{tikzpicture}
\caption{
    {\bf Network $NI_{1}L = NI_{1}G_{1}$ : A binary-linear stem cell network}. An initial network NI$_{1}$ is linked to a linear stem cell network G$_{1}$. The network NI$_{1}$ generates $2$ identical stem cells B which are controlled by $NL = NG_{1}$. The parent cell $A$ divides into two identical daughter stem cells of type $B$.  $B$ is controlled by a linear self-renewing stem cell sub-network which generates terminal cells of type $C$. 
  }
    \label{fig:NI1G1}
\end{figure}
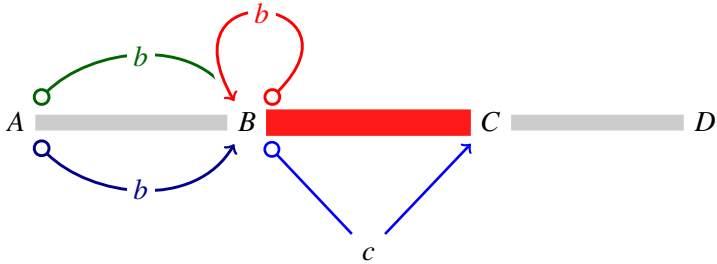 
%$NI_{1}L = NI_{1}G_{1}$  was NLI}$_2$
If $2$ is the number of identical cells controlled by the network and $n$ is the number of loops synchronously executed by those cells then in this case the number of cells after $n$ cell divisions is:
\begin{equation}
\Cells(n) = 2 + (2 \times n)
\end{equation}

The \textbf{phenotype} of such a network would have identical linearly growing stem cells in several places at once. They would have started simultaneously (up to some error range).  

More generally, let NI$_k$ be a network of $k$ identical divisions that generates $2^k$ cells.  Then we can generate a stem cell network NLI$_k$ if we link an upstream NI$_k$ to a simple linear stem cell network of type G1 (see \autoref{fig:G1}), for example as follows: 

\begin{figure}[H]
\begin{tikzpicture}[style=mystyle]
\matrix (m) [matrix of math nodes, row sep=3em,
column sep=2em, text height=1.5ex, text depth=0.25ex]
{ \vphantom{a} & \vphantom{b} &  \vphantom{c}  & \vphantom{c}  & & \vphantom{c} &\\
 A_0  &&  A_1  & \vphantom{b} &  A_2 && A_{k-1} && A_k = B && C \\
 \vphantom{a} &  \vphantom{b} & \vphantom{B} &  \vphantom{c}  & \vphantom{C} & \vphantom{d} & \vphantom{D}\\ };
 \path[\pot]

(m-2-1) edge [inPot1,DarkGreen, cross line] node[nodedescr] {$\alpha_1$} (m-2-3) %A to Nx
(m-2-1) edge [inPot2, DarkBlue, cross line] node[nodedescr] {$\alpha_1$} (m-2-3) %A to Ny
(m-2-3) edge [inPot1, DarkGreen , cross line] node[nodedescr] {$\alpha_2$} (m-2-5) %A to Nx
(m-2-3) edge [inPot2,  DarkBlue, cross line] node[nodedescr] {$\alpha_2$} (m-2-5) %A to Ny
(m-2-7) edge [in100Pot1, DarkGreen, cross line] node[nodedescr] {$\alpha_{k}$} (m-2-9) %A to Nx
(m-2-7) edge [in100Pot2, DarkBlue, cross line] node[nodedescr] {$\alpha_{k}$} (m-2-9)%A to Ny

%(m-2-9) edge [selfloop2,blue,  cross line] node[nodedescr] {$ a_1$} (m-2-1) %end of Ny = D to A
(m-2-9) edge [selfloop1, red, cross line] node[nodedescr] {$\alpha_{k}$} (m-2-9) % Nx = C to A;
(m-2-9) edge [inPot2, blue, cross line] node[nodedescr] {$c$} (m-2-11); %A to Ny
\path[solid,red!80, line width=6pt]
(m-2-1) edge (m-2-3);
\path[solid,red!80, line width=6pt]
(m-2-3) edge (m-2-5);
\path[dashed,red!90, line width=6pt]
(m-2-5) edge (m-2-7);
\path[solid,red!80, line width=6pt]
(m-2-7) edge (m-2-9);
\path[solid,blue!80, line width=6pt]
(m-2-9) edge (m-2-11);

\end{tikzpicture}
\caption{
    {\bf Network NI$_k$G$_{1}$ Multi-linear stem cell network starting from $2^k$ identical cells:} This network generates $k$ identical divisions to produce $2^k$ identical daughter stem cells of type $A_k = B$ from one founder cell A$_1$.  The $B$ stem cells are controlled by a linear stem cell sub-network that after $n$ rounds of division, jointly produce $2^k \times n$ terminal cells of type $C$ while retaining a constant $2^{k}$ of linear stem cells of type B.  Starting from one founder cell, the total number of cells after $n$ rounds of synchronous division is:  $\mbox{Cells}(n, k) = 2^k + 2^{k}\times(n - k + 1) = 2^k\times(n -k +2)$ if $n > k$ and $\mbox{Cells}(n, k) = 2^n$ otherwise.}
    \label{fig:NIkG1}
\end{figure}
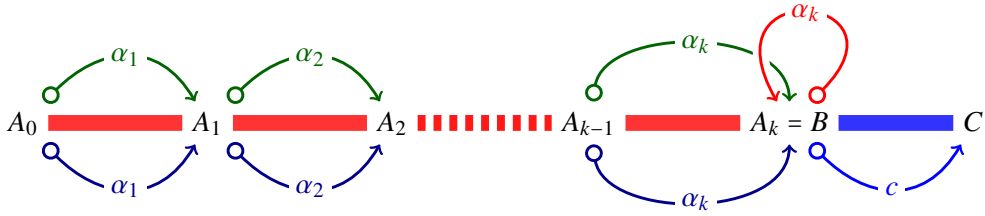 

The below is function describing the ideal rate of growth after $n$ synchronous divisions where $k$ is the number of identical daughter cell divisions nodes in the network in \autoref{fig:NIkG1}. 

\begin{eqnarray*}
 \mbox{Cells}(n,k) = \left\{
 \begin{array}{ll} 
 		 2^n  & \mbox{if $n <= k$} \\ 
 		 2^k + 2^{k}\times(n - k + 1) = 2^k\times(n -k +2) & \mbox{for $n > k$}
 						\end{array} 
 				\right.
\end{eqnarray*}

\subsection{Stem cells that generate identical cells}

\begin{figure}[H]
\begin{tikzpicture}[style=mystyle]
\matrix (m) [matrix of math nodes, row sep=3em,
column sep=2em, text height=1.5ex, text depth=0.25ex]
{ \vphantom{a} & \vphantom{b} &  \vphantom{c}  & \vphantom{c}  & & \vphantom{c} &\\
 A_{L}  &&  A_0  & \vphantom{b} &  A_1 && A_{2} && A_{k-1}  && A_{k}=B \\
 \vphantom{a} &  \vphantom{b} & \vphantom{B} &  \vphantom{c}  & \vphantom{C} & \vphantom{d} & \vphantom{D}\\ };
 \path[\pot]

(m-2-3) edge [inPot1, DarkGreen , cross line] node[nodedescr] {$\alpha_1$} (m-2-5) %A to Nx
(m-2-3) edge [inPot2,  DarkBlue, cross line] node[nodedescr] {$\alpha_1$} (m-2-5) %A to Ny
(m-2-5) edge [inPot1,DarkGreen, cross line] node[nodedescr] {$\alpha_2$} (m-2-7) %A to Nx
(m-2-5) edge [inPot2, DarkBlue, cross line] node[nodedescr] {$\alpha_2$} (m-2-7) %A to Ny
(m-2-9) edge [in100Pot1, DarkGreen, cross line] node[nodedescr] {$\alpha_{k}$} (m-2-11) %A to Nx
(m-2-9) edge [in100Pot2, DarkBlue, cross line] node[nodedescr] {$\alpha_{k}$} (m-2-11)%A to Ny

%(m-2-9) edge [selfloop2,blue,  cross line] node[nodedescr] {$ a_1$} (m-2-1) %end of Ny = D to A
(m-2-1) edge [selfloop1, red, cross line] node[nodedescr] {$\alpha_{L}$} (m-2-1) % Nx = C to A;
(m-2-1) edge [inPot2, blue, cross line] node[nodedescr] {$\alpha_{0}$} (m-2-3); %A to Ny
\path[solid,red!80, line width=6pt]
(m-2-1) edge (m-2-3);
\path[solid,red!80, line width=6pt]
(m-2-3) edge (m-2-5);
\path[solid,red!90, line width=6pt]
(m-2-5) edge (m-2-7);
\path[dashed,red!80, line width=6pt]
(m-2-7) edge (m-2-9);
\path[solid,blue!80, line width=6pt]
(m-2-9) edge (m-2-11);

\end{tikzpicture}
\caption{
    {\bf Network NG1I$_k$ Linear stem cell producing $2^k$ identical cells each loop:} This linear stem cell network generates $2^k$ identical daughter cells of type $A_k = B$ at each stem cell loop.  The $A_{L}$ stem cell is controlled by a linear stem cell sub-network that after $n$ rounds of division, that ultimately produces $n \times 2^k$ terminal cells of type $B$.  Starting from one founder cell, the total number of cells after $n$ rounds of synchronous division is:
     $\mbox{Cells}(n, k) = 2^k + 2^{k}\times(n - k + 1) = 2^k\times(n -k +2)$ if $n > k$ and $\mbox{Cells}(n, k) = 2^n$ otherwise.}
    \label{fig:NG1Ik}
\end{figure}
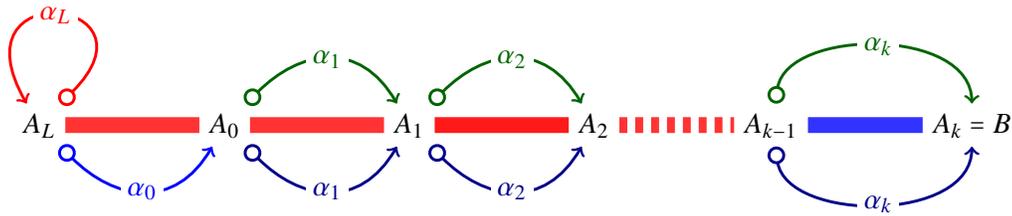 

\subsection{Stem cell networks commute with identical cell networks}
It turns out that in terms of ultimate cell production the network in \autoref{fig:NG1Ik} is equivalent to the network in  \autoref{fig:NIkG1}. This is because the simple network NG$_{1}$ commutes with NI networks.  More generally, given NI$_{k}$ then any element  NG$_{1}$ in NG$_{1}$ commute with the elements of NI$_{k}$ to generate the same ultimate cell proliferation even while the multicellular form may vary. 

\section{Stem cell networks and cancer stem cells}
\label{sec:NSC}
Normal networks that come closest to cancer networks are stem cell networks.  What distinguishes normal from cancer stem cell networks is not primarily their network architecture but rather when and where a stem cell network is linked and activated, and what further networks the stem cell network activates.  A cancer stem cell network is either formed by mutation in the wrong place in the global developmental network, it is activated inappropriately in the multicellular system, or it activates inappropriate, nonfunctional networks with abnormal and possibly pathological cellular or multicellular phenotypes relative to the overall multicellular system. 

From our theory it follows that there are at least two main kinds of stem cell networks, linear and geometric.  However, linear stem cell networks are just a special case of geometric stem cell networks, namely, 1st order geometric stem cell networks.  These are also fundamental cancer stem cell networks.  A normal linear stem cell produces a cell of a particular type that is either terminal itself or activates a terminal network. A second order geometric stem cell network is a meta-stem cell network that produces linear stem cells.  Thus, linear stem cells produce no further stem cells. Meta stem cells do produce stem cells.  

Since a linear stem cell cannot produce further stem cells there must exist meta-stem cells that produce more than one stem cell. These in turn are produced by yet further upstream embryonic networks.  The original embryonic fertilized egg is controlled by the global embryonic network that is mostly terminal with the exception of its stem cells.  It consists of many subnetworks that may be multiply employed. 

A linear cancer stem cell produces no additional cancer cells.  Interestingly, we will see that the order of a cancer meta-stem cell network controlling a cancer cell determines the possible metastases producible by that cancer cell (see \autoref{sec:GMetastases}). 

A further possibility is that there exist normal {\em exponential stem cell networks} that are activated by cell signaling, in effect a communication network linked with a cytogenic control network to produce cells quickly on demand. However, such networks are be dangerous, leading to a proliferative explosion if something goes wrong with the communication network. 

\subsection{Normal stem cells and cancer stem cells}

Stem cells have the function of continually producing new cells in tissue that is meant to regenerate. Stem cells have  functional developmental control networks with preexisting loops. This makes the progression to cancer a shorter route, because the linear, proliferating network already exists.  We just need to mutate in one more loop to makes it exponential and then with additional mutations it becomes invasive.  Any slow growing cancers that suddenly become very fast growing may be an instance of this process.  

If a stem cell proliferates conditionally, i.e., based on some contextual cellular or environmental condition, then a stem cell can transform into a cancer stem cell by a situation or process that constantly activates the antecedent condition necessary for the stem cell to divide. This can happen if the activating signal is constant or the receptor system for that signal is mutated to be constantly on.  It may be the case that a stem cell may already be potentially, linearly, geometrically or exponentially, proliferative, but its antecedent conditions may not be satisfied. Hence, if the conditional system remains functional the potential stem cell proliferation whether it be linear or exponential will not show itself.  However, when the antecedent conditions for proliferation are fulfilled or the antecedent testing system is mutated to a constant on-state then the stem cell cancer will flower.  An example of this would be a stem cell whose proliferation depends on some signal such as a hormone that, together with a signal transduction pathway, activates the stem cell proliferative loop.  Then if the signal is constantly turned on or the signal transduction mechanism is in a constant "signal received" state, then the stem cell will proliferate either linearly, geometrically or exponentially depending on the structure of its controlling network. 

A further danger with some types of  stem cells is that they may already be functionally, tissue invasive in order to move to the appropriate site, as, for example, in wound healing. This invasive property would then make a stem cell cancer even more dangerous. 

Stem cells by themselves are not cancerous, but they have the properties of conditional, possibly invasive, linear, geometric or exponential cancers.  The boundary between a cancer and normal cell may not always be clear since in the wrong context a normal stem cell may be cancerous. 

Distinguishing linear cancer cells from stem cells based purely on the network architecture is difficult if not impossible.  Stem cells generate normal cell that are functional in the context of the organism.  Any stem cell that generates such "normal" cells inappropriately is a cancer cell. Thus, the dividing line between linear cancers and stem cells may be indistinct.   Thus, a normal stem cell network becomes cancerous if it is activated inappropriately with respect to the overall functioning of a biological system in the organism, or if the cell differentiation networks the stem network activates generate abnormal cells (e.g., invasive cells, or having  abnormal phenotype) or cells that generate structures inappropriately in the overall system. 

\subsection{Cancer susceptibility of stem cells}
We see why stem cells are especially susceptible to develop into cancer networks.  First, all that has to happen is that the first order or second order conditional antecedents are mutated to be always true, then the stem cell networks, once activated produces cells indefinitely independent of outside signaling.  Second, any of the links from the other non-proliferative daughter cell differentiation site could loop back to a first or second order loop activation point to produce an exponential cancer. 

Having two or more proliferative loops does not necessarily result in an exponential cancer.  However, they can transform into exponential cancers.  For example, using the WHO-grading scheme, astrocytoma grad I, which is slow growing and benign, appears to be a linear cancer of type NG, astrocytoma grade II -> grade III.  Grade III is relatively fast growing compared to grade II, but can be present for several years and then suddenly change to grade IV Glioblastoma Multiforme (GBM). GBM are fast growing and spread quickly.   \nocite{Nakada2011}.

\section{Hierarchy of metastases from geometric cancer networks}

One of the most important consequences of the hierarchical nature of stem cell networks is that they imply the existence of a corresponding metastatic hierarchy.  Geometric cancer stem cell networks form a hierarchy of interlinked stem cell networks such that an k-th order stem cell network generates a cell controlled by an k-1 order stem cell network. Thus, 3rd order or meta-meta-stem cell networks produce cells controlled by 2nd order, meta-stem cells.  A 2nd order or meta-stem cell generates a cell controlled by a 1st order, linear stem cell network.  At the base of the hierarchy is a terminal network controlling a cell generated by a 1st order, linear stem cell. 

This hierarchy of control in higher order stem cell networks has direct consequences for the types of metastases that can, in principle, be  generated by such higher order geometric networks.  In other words. the metastatic potential of a cancer stem cell is determined by the order of its geometric control network.  As we will see, this the dynamic phenotypic properties of the metastases are a result of their place in the metastatic hierarchy.  This can be used for diagnosis and reverse inference as to the type of stem cell network controlling the metastatic tumor.  

Note, however, that there may also be functional uses of higher order stem cell networks for the production and distribution of stem cells in healthy multicellular systems.  Meta stem cells and their stem cell network hierarchy could be used by organisms to generate and deliver lower level stem cells to various parts of the body. 

\label{sec:GMetastases}
\begin{figure}[H]
\subfloat[{\bf Terminal metastases from a primary 1st order geometric cancer network.}  ]{
\includegraphics[scale=0.4]{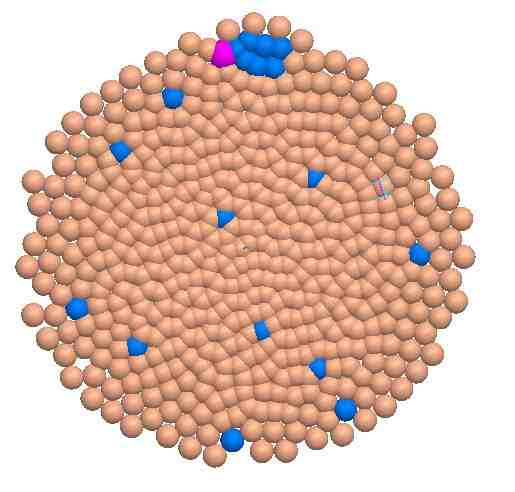}
\label{fig:G1Metsubfig1}
}
\hspace{1.0cm}
\subfloat[{\bf 1st order and terminal metastases from a 2nd order geometric cancer network.}  ]{
\includegraphics[scale=0.4]{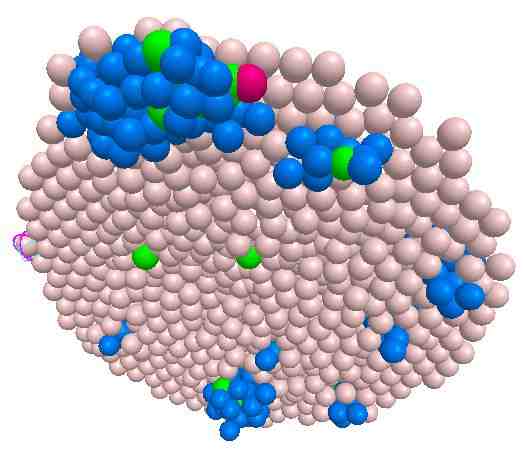}
\label{fig:G2metsubfig2}
}
\caption{
{\bf Two views of metastasis by 1st order and 2nd order geometric networks.}  In Figure \autoref{fig:G1Metsubfig1}:  One primary 1st order geometric, linear cancer cell (in red) generates terminal metastatic cells (in blue). But the blue terminal cells are terminal generating no further cells. 
The right hand Figure \autoref{fig:G2metsubfig2}: A 2nd order geometric cancer network generates 1st order geometric and terminal metastases.  Against a background of identical cells (in light pink) there is one primary 2nd order geometric cancer cell  (in red) that generates metastases consisting of 1st order geometric cancer cells (in green). These in turn, generate secondary metastatic passive terminal cells (in blue). The blue terminal cells may metastasize, moving to other regions, but they do not generate further cells.  
}
\label{fig:G1G2Metastasis}
\end{figure}

A 2nd order geometric cancer cell generates 1st order linear cancer cells (see \autoref{fig:G2}). If these 1st order linear cancer cells metastasize then we have a phenotype of multiple linear cancer growths in different regions of the body.  For example, one might have multiple slow growing cysts in different regions of the body. In this case, these cysts are metastases that were generated from a single meta-stem cell that generates progenitor cells that generate cysts.  Other examples, might include slowly progressing cancers that move in the lymphatic system.  Here again in may, but need not, be the case that a single meta-cancer stem cell is generating sub-cancers that have the overt phenotype of a uniform cancer progressing in some direction across a region or through the whole body. 

\subsection{Secondary and tertiary metastases generated by 3rd order cancer networks}

\begin{figure}[H]
\subfloat[{\bf 1st oder and terminal metastases from a primary 2nd order geometric cancer network.}  ]{
\includegraphics[scale=0.4]{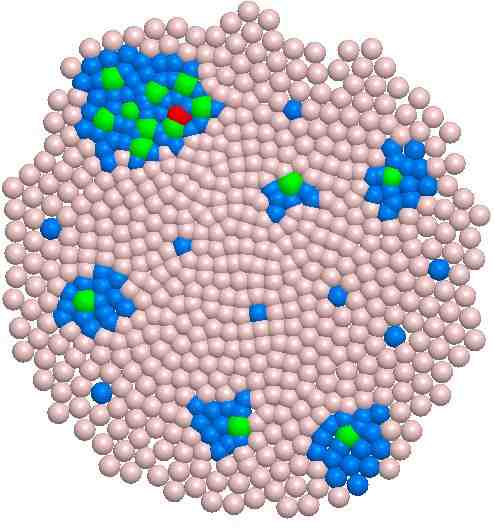}
\label{fig:G2Metsubfig1b}
}
\hspace{0.5cm}%{1.0cm}
\subfloat[{\bf 2nd, 1st order, and terminal metastases from a primary 3rd order geometric cancer network.}  ]{
\includegraphics[scale=0.4]{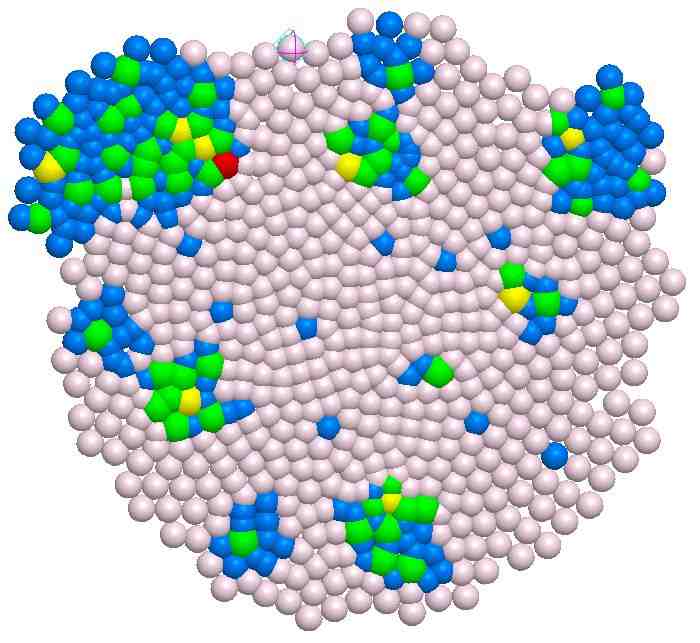}
\label{fig:G3metsubfig2}
}
\caption{
{\bf Comparing metastases by 2nd order and 3rd order geometric cancer networks.}  The {\bf left hand Figure} \autoref{fig:G2Metsubfig1b}: A 2nd order geometric cancer network (\autoref{fig:G2}) generates 1st order geometric and terminal metastases.  Against a background of identical cells (in light pink) there is one primary 2nd order geometric cancer cell  (in red) that generates  metastases consisting of 1st order geometric cancer cells (in green). These in turn, generate secondary metastatic passive terminal cells (in blue). The blue terminal cells may metastasize, moving to other regions, but they do not generate further cells.  
In the {\bf right hand Figure} \autoref{fig:G3metsubfig2}: A primary tumor controlled by a 3rd order geometric cancer network generates 2nd order and 1st order geometric and terminal metastases. The primary 3rd order geometric cancer network (G3) generates 2nd order (G2) metastases. These G2 networks in turn generate 1st order (G1) metastases. The 1st order geometric (G1) cancer cells in turn generate terminal (G0) cells that are also potentially metastatic (invasive) but do not generate further cells. 
} 
\label{fig:G2G3Metastasis}
\end{figure}

In general, k-th order geometric cancer networks $G_{k}$ (see \autoref{fig:Gk}) generated cells controlled by $k-1$ order geometric cancer networks $G_{k-1}$.  These $G_{k-1}$ networks, in turn, generate cells controlled by $k-2$ order cancer networks $G_{k-2}$, etc. Therefore, third-order geometric cancer networks $G3$ (\autoref{fig:G3}) , such as the above simulated network in \autoref{fig:G2G3Metastasis}, generate cells controlled by second-order cancer networks $G2$, (\autoref{fig:G2}). In turn, $G2$-networks generate cells controlled by 1st-order geometric networks $G1$, (\autoref{fig:G1}).  $G1$-networks are linear cancer networks which generate terminal cells $G0$ that do not proliferate.  

\begin{figure}[H]
\subfloat[{\bf Active cells in 1st oder metastases from a primary 2nd order geometric cancer network.}  ]{
\includegraphics[scale=0.4]{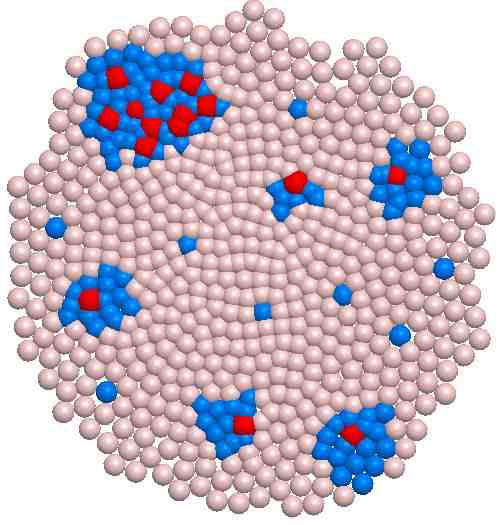}
\label{fig:G2Metsubfig1b}
}
\hspace{0.2cm}
\subfloat[{\bf Active cells in 2nd and 1st order metastases generated by a primary 3rd order geometric cancer network.}  ]{
\includegraphics[scale=0.4]{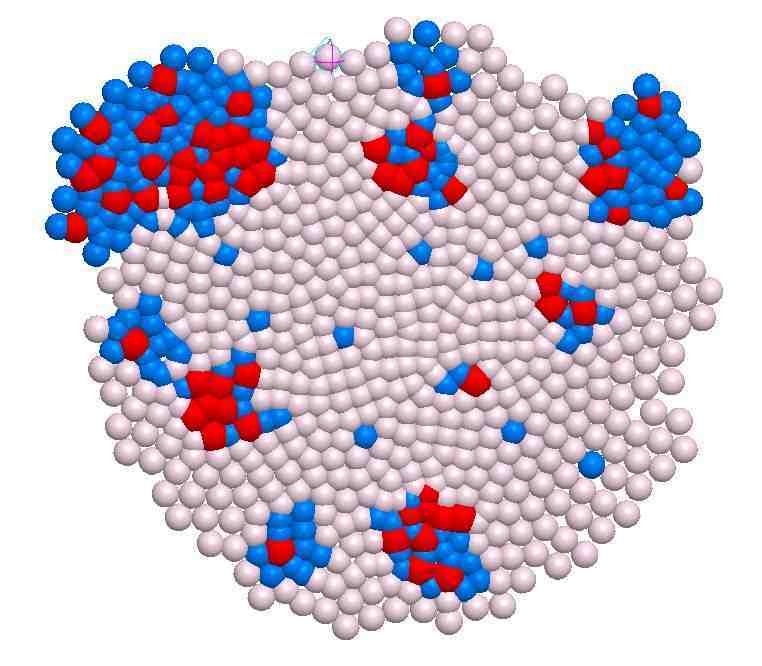}
\label{fig:G3metsubfig2}
}
\caption{
{\bf View of the active proliferating cells in G1, G2 and G3 geometric metastases in \autoref{fig:G2G3Metastasis}}. Cellular proliferation alone does not distinguish higher order from lower order geometric tumors including metastatic tumors. Only terminal cells of type G0 can be distinguished since they do not divide, or at most, are progenitor cells that ultimately lead to terminal cells that no longer divide.  Rather than cell proliferation alone, it is the control state and downstream proliferation potential that distinguishes geometric networks from each other. 
}
\label{fig:G3MetRedActiveCells}
\end{figure}

\subsection{Relating metastatic phenotype with geometric cancer networks}

Since any of the cells generated by a cancer network have the potential to metastasize by moving to other regions of the organism, each primary tumor controlled by a cancer network will have a distinct metastatic phenotype.  Therefore, we can use the phenotype of the metastases to draw inferences about the nature and architecture of the network generating that phenotype.  Moreover, if we find the metastases that together form related patterns as described for the geometric metastases above, it warrants looking for the primary tumor generating the secondary and tertiary metastases.  

\subsection{Treatment options for metastases generated by geometric cancer networks}

Treatment would focus on eliminating or transforming cells with the higher level control networks generating the less dangerous lower level networks.  Treatment can be by cell death or by network transformations that block or modify the cancer network.  Different transformations are necessary for different cancer networks.  Once metastases have been created by a higher level network, it is no longer sufficient to block or transform the higher level initiating subnetwork. Instead, each of the cells with lower level cancer subnetworks also have to be transformed or destroyed.  If they are 1st order linear networks, they may be relatively harmless since they only produce cells that no longer proliferate. However, this will depend on their location and effect on the other multicellular contexts in which they reside. Because of cell interactions such as cell signaling, stochastic reactivation of higher level cancer networks such cells, depending on their differentiation state and the cellular context, may still be dangerous. 

\subsection{Stochastic reactivation of terminal cells in geometric networks}

If terminal cells can reactivate a proliferative control state in a network, then the more numerous the number of terminal cells the more likely it is that one or more of them will become an active cancer cell.  Hence, given the possibility of stochastic activation of ancestral cancer networks,  formerly passive metastatic cells can become active tumors. 

\section{Stochastic differentiation in stem cell networks}
\label{sec:NSSC}
It has been argued that the traditional theory of linear stem cells is wrong \cite{Jones2007}.  On the traditional theory of stem cells, stem cells are immortal and constant in number.  However, experimental data appears to show that for some stem cells appear to generate further stem cells.  On a deterministic theory stem cells that produce stem cells are what we have termed meta-stem cells.  However, another model would have stem cells divide and differentiate stochastically either exponentially into two stem cells, linearly into a stem cell and a terminal cell, or into two terminal cells.  The evidence, suggests that epidermal cells divide stochastically according to the distribution 8\% double stem cells, 84\% one stem cell and one terminal cell, and 8\% two terminal cells.  The biological mechanism is not known.  

\subsection{Historical background: A network that generates Till's stochastic stem cell model}

One of the first models of stem cells was the stochastic model of hemopoietic cell proliferation  [Till et.al.~\cite{Till1964}].  Till distinguished what he called ``colony forming cells'' that have the capacity to form colonies from differentiated cells without that capacity.  When a colony develops from a single cell only a small number of these cells have ``colony-forming capacity''. The rest are differentiated cells without this capacity.   This he considered a "birth-and-death" process, the generation of colony cells being a birth process and the generation of differentiation a death process. According to the model, a given cell with colony-forming capacity can either divide into two colony forming cells (with probability $p_{2}$) or differentiate into a terminal cell (with probability $p_{0} = 1 - p_{2}$) having no colony forming capacity.  The case of generating mixtures of one colony-forming cells and one differentiated cell was considered a birth process followed by death and was handled by adjusting the probabilities.   
 
Production numbers of blood cells is relatively constant under normal conditions. Under stress or increased demand there is a rapid increase in production of blood cells.  The properties of hemopoiesis (blood cell creation) implies that the production of differentiated cells is precisely controlled.  Hence, Till argues, there must be control mechanisms.  Till then considers if cell proliferation (the numbers of cells produced) is also under precise control or lax control.  He argues for lax control and that the data suggests cell proliferation is stochastic. While a stem cell network was not given by Till, the behavior of the Till model of colony forming cells can be generated by following stochastic stem cell network:

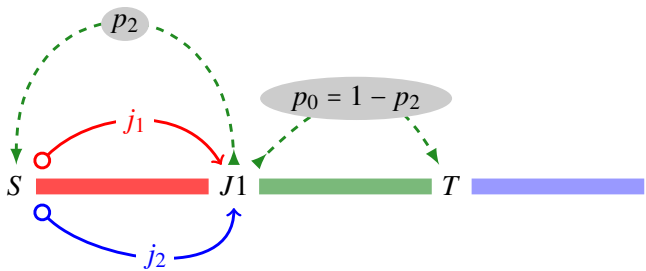
\begin{figure}[H]
\begin{tikzpicture}[style=mystyle]
\matrix (m) [matrix of math nodes, row sep=3em,
column sep=3em, text height=1.5ex, text depth=0.25ex]
{ \vphantom{a} && \vphantom{b} &&  \vphantom{c} \\ %&t_{1} & \vphantom{c}\\ %  & & &\\
 S  && J1 &&  T && \vphantom{b}\\%T1 && T2 &&  \\
 \vphantom{a} &&  \vphantom{b} && \vphantom{B} &&\\ };
 \path[\pot]
(m-2-1) edge [inPot1, red, cross line] node[nodedescr] {$j_1$} (m-2-3) %S to J1 
(m-2-1) edge [inPot2, in=-90, blue, cross line] node[nodedescr] {$j_{2}$} (m-2-3); %S to J1
\path[stochasticPathstyle]
(m-2-3) edge [inPot1, out=90, in=90,distance=5cm, green, cross line] node[stochasticNodestyle] {$p_{2}$} (m-2-1) %J1 to S 
(m-2-3) edge [in2Pot1,in=120,distance=3cm,green, cross line] node[stochasticNodestyle] {$p_{0} = 1 -p_{2}$} (m-2-5); %J1 to T
\path[solid,red!70, line width=6pt]
(m-2-1) edge (m-2-3);
\path[solid,green!60, line width=6pt]
(m-2-3) edge (m-2-5);
\path[solid,blue!40, line width=6pt]
(m-2-5) edge (m-2-7);
\end{tikzpicture}
\caption{
    {\bf Network TillSSC:  Classic stochastic stem cell network}. This network models one of the earliest stochastic stem cell theories (Till~\cite{Till1964}).  A stem cell (S) divides to produce two cell of type J1 that then stochastically change either into a terminal network cell T or ``self renew'' by dedifferentiating to the parent stem cell S type. The cell J1  stochastically either loops back to activate the stem cell S with probability $p_{2}$ or it activates the terminal network cell T with probability $p_{0}= 1- p_{2}$.  The network's behavior approaches a deterministic exponential stem cell network as the probability $p_{2}$  approach $1$.  Stem cells controlled by such a network can spontaneously stop because of the fact that for all points in all possible  paths in the network there is the possibility of reaching the terminal cell state T. The multicellular system controlled by this network consists of two main cell types:  Stem cells (S) that can proliferate and terminal cells (T) with no proliferative potential. 
  }
   \label{fig:TillSSC}
\end{figure}

Under this  network (\autoref{fig:TillSSC}) a stem cell divides into identical daughter cells J1. Each J1 cell  can with probability $p_{2}$  dedifferentiate into a stem cell S or differentiate with probability $p_{0} = 1-p_{2}$ into a terminal cell T.  Hence, unlike Till's original model, this network explicitly handles all the possible outcomes of stem cell division.  Upon stem cell division, the network allows the generation of two stem cells (S,S), as well as the mixed case (S,T) of one stem cell and one differentiated cell, and the case (T,T) of two differentiated cells.  

Till's model  has been criticized because it predicts that stem cells eventually differentiate into terminal cells and, thus, no longer self renew.  However, this depends on the probability distribution.  If $p$ is high then exponential proliferation outruns terminal cell differentiation. 

A central problem with this network is that the key to its behavior is dependent on the value of the probability $p_{2}$.  On the one hand, if the probability $p_{2}$ of stem cell self-renewal is high, it tends toward exponential growth of stem cells. This leads to too many stem cells versus progenitor and terminal, specialized cells. On the other hand, if the self-renewal probability $p_{2}$ is low, it eventually  leads to the elimination of all stem cells. In the latter case, this network does not represent a stem cell with unlimited proliferative potential.  In this network, no dedifferentiation is possible for terminal cells. Hence, the greater the probability $p_{0} = 1-p_{2}$ that the terminal network T is activated, the more likely is the permanent quiescence of the parent stem cell.     

If the probability $p_{2}=1$ we have exponential growth of stem cells.  If $p_{2}=0$ we have differentiation to the terminal cells T.  Hence, this network cannot model the developmental dynamics of deterministic linear networks of the type fig:G1~\autoref{fig:G1}.  However, with the right choice of the probability distribution, it can approximate a relatively constant production of stem cells and terminal cells. 

In their Monte Carlo simulation the probability used was $p_{2} = 0.6$ and $p_{0}=0.4$.  In our Monte Carlo simulation, using their probability distribution, the network proliferation dynamics tends toward a 21\% proportion of stem cells after 52 generations.  
This appears to be much higher than the empirical data for hemopoietic stem cells.  
However, these percentages depend on how many cells are generated by the terminal cell T.  If T is not a terminal cell and instead a progenitor cell controlled by a terminal network T*, and is instead a terminal cell state that does not divide further (as was the case in Till's original model), then the stem cell percentage is lower being 12\% after 20 generations of cell division. We add progenitor cell capacity to T in the network in (\autoref{fig:TillSSCpro}) below. 

Note, in this sort of modeling approach, precise differentiation networks (architectures) are coupled with stochastic activation networks.  The topology of the network architecture sets the boundaries for what stochastic paths are possible at all. 

\begin{figure}[H]
\subfloat[Probability $p_{2}$ = 0.52, gives 4\% stem cells]{
\includegraphics[scale=0.4]{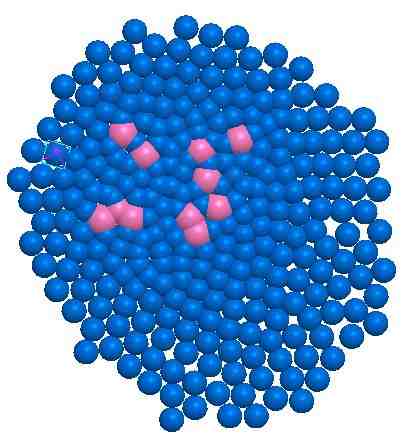}
\label{fig:StocStemSubfig1}
}
\hspace{0.3cm}
\subfloat[When $p_{2}$ = 0.7 gives 36\% stem cells]{
\includegraphics[scale=0.4]{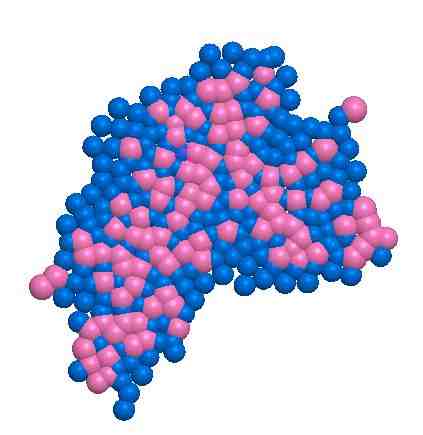}
\label{fig:StocStemSubfig2}
}
\hspace{0.3cm}
\subfloat[Probability $p_{2}$ = 0.9 gives 80\% stem cells]{
\includegraphics[scale=0.4]{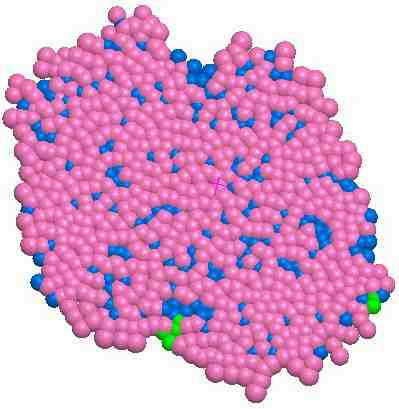}
\label{fig:StocStemSubfig3}
}
\caption{
{\bf Three views of stem cells under different probabilities.} All three tumors are  controlled by same Till network (\autoref{fig:TillSSC} ).  It shows the growth response to changes in the probability $p_{2}$.   As the probability $p_{2}$ is increased, more and more stem cells are generated due to increasing dedifferentiation of the daughter cells to their parent stem cell.  The stochastic exponential network begins to dominate as $p_{2}$ approaches 1. 
}
\label{fig:StocStemProb}
\end{figure}

\subsection{The Till stochastic network extended to generate progenitor cells}
It is very simple to extend Tills original model of hemopoietic (blood) stem cells (S)  [Till et.al.~\cite{Till1964}] to include progenitor networks that control the proliferative capacity of stem cell progeny.  We simply link the terminal cell state to a bounded network T* that in the case below is simply a network that results in one further cell division to produce two distinct daughter cells T1 and T2.  But, T* could be any bounded network.  Thus, in principle, given the right network T*, such progenitor cells could develop into arbitrarily complex structures.  

\begin{figure}[H]
\begin{tikzpicture}[style=mystyle]
\matrix (m) [matrix of math nodes, row sep=3em,
column sep=3em, text height=1.5ex, text depth=0.25ex]
{ \vphantom{a} && \vphantom{b} &&  \vphantom{c}  &t_{1} & \vphantom{c}\\ %  & & &\\
 S  && J1 &&  T*  && T1 && T2 &&  \\
 \vphantom{a} &&  \vphantom{b} && \vphantom{B} && t_{2}  && \vphantom{T}\\}; % & \vphantom{d} & \vphantom{D}\\ };
 \path[\pot]
(m-2-1) edge [inPot1, red, cross line] node[nodedescr] {$j_1$} (m-2-3) %S to J1 3
(m-2-1) edge [inPot2, in=-90, blue, cross line] node[nodedescr] {$j_{2}$} (m-2-3); %S to B 5
%\path[right hook->]
\path[stochasticPathstyle]%[>=latex,>->,dashed]
(m-2-3) edge [inPot1, out=90, in=90,distance=5cm, green, cross line] node[stochasticNodestyle] {$p_{2}$} (m-2-1) %J1 to S 
(m-2-3) edge [in2Pot1,in=120,distance=3cm,green, cross line] node[stochasticNodestyle] {$p_{0} = 1 -p_{2}$} (m-2-5); %J1 to T
 \path[\poti]
(m-2-5) edge [pot2,blue](m-3-7);%T to T2
 \path[\pote]
(m-3-7) edge [pot2,blue] (m-2-9); 
 \path[\poti]
(m-2-5) edge [pot2,red](m-1-6);%T to T1
 \path[\pote]
(m-1-6) edge [pot2,red] (m-2-7);

\path[solid,red!70, line width=6pt]
(m-2-1) edge (m-2-3);
\path[solid,green!60, line width=6pt]
(m-2-3) edge (m-2-5);
\path[solid,blue!40, line width=6pt]
(m-2-5) edge (m-2-7);
\path[solid,black!40, line width=6pt]
(m-2-7) edge (m-2-9);
\end{tikzpicture}
\caption{
    {\bf Network TillSSCpro:  Classic stochastic stem cell network (\autoref{fig:TillSSC}) extended to include progenitor networks}. This network extends one of the earliest stochastic stem cell theories (Till~\cite{Till1964}) to include progenitor cells.  A stem cell (S) divides to produce two cell of type J1 that then either divide into a terminal network cell T* or ``self renew'' by dedifferentiating to the parent stem cell S.  The cell J1 stochastically loops back to activate the stem cell S with probability $p_{2}$ or it activates the terminal network cell T* with probability $p_{0}= 1- p_{2}$.  In Till, T* is a terminal cell type T.  In this example, T* is a terminal network for a progenitor cell that generates two terminal cell types T1 and T2. As with the original model (\autoref{fig:TillSSC}), this network's behavior approaches a deterministic exponential stem cell network as the probability $p_{2}$  approach $1$.  Stem cells controlled by such a network can spontaneously stop because of the fact that for all points in all possible  paths in the network there is the possibility of reaching the terminal cell state T*. The multicellular system controlled by this network consists of three main cell types:  Stem cells (S), progenitor cells (T*) with limited proliferative potential, and terminal cells (T1, T2) with no proliferative potential. 
  }
   \label{fig:TillSSCpro}
\end{figure}
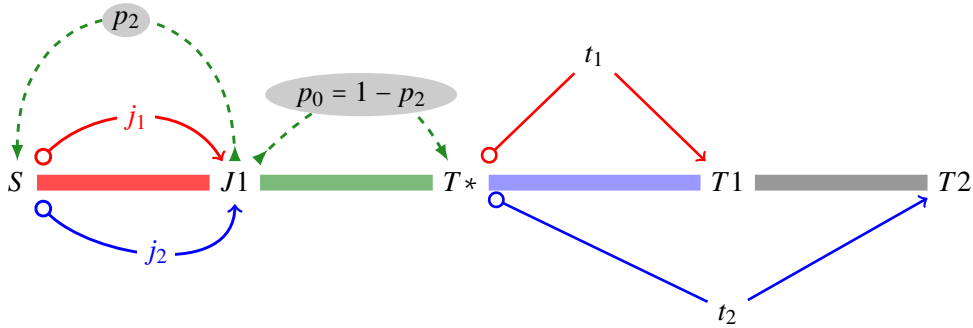

As with the previous network \autoref{fig:TillSSC}, the proliferative potential of stem cells under this network \autoref{fig:TillSSCpro} is dependent on the probability $p_{2}$.  Since no dedifferentiation is possible in this network for terminal cells T1 and T2.  Furthermore, it is not an inherent property of the network topology that stem cells self renew.  However, if the dedifferentiation probability $p_{2}$ of self renewal is high enough then stem cells do self renew and may even dominate the colonies they generate.  

If the probability $p_{2}=1$ we have exponential growth of stem cells, with no progenitor cells T*.  If $p_{2}=0$ we have differentiation to to the terminal progenitor network T* which in this case ends in just two terminal cells T1 and T2.  Hence, this network cannot model the developmental dynamics of deterministic linear networks of the type G1~\autoref{fig:G1}.  However, with the right choice of the probability distribution, it can approximate that a relatively constant ratio of stem cells, progenitor and terminal cells. 

It follows that under the Till model of stem cells we do not need dedifferentiation of terminal cells to maintain stem cell capacity.  All we need is for the 1st order stem cells to be physically loose, mixing in with the terminal cells in the tumor as it grows.  In this case, we would see proliferating stem cells within the tumor in the context of what appear to be only terminal cells.  It may appear as if terminal cells have spontaneously dedifferentiated into stem cells, but this need not be the case since it may be the result of stem cell mixing.  

What could also be happening is that we have a 2nd order geometric stem cell network that is generating 1st order stem cells which due to developmental control in conjunction with physics leads to a mixture of 1st order stem cells with terminal cells. Since the control state of a cell my be hidden by the overt differentiation state of the cell, the 1st order stem cells may be indistinguishable relative to a set of markers from terminal cells.  Stochastically or under particular conditions these 1st order stem cells may then begin to proliferate generating terminal or progenitor cells. 

\subsection{A flexible stochastic network architecture with exponential and linear potential}

\begin{figure}[H]
\begin{tikzpicture}[style=mystyle]
\matrix (m) [matrix of math nodes, 
row sep=2em,
column sep=3em, 
text height=1.5ex, text depth=0.25ex]
{ \vphantom{a} && \vphantom{b} &&  \vphantom{c}  && \vphantom{c}  & & &\\
 S  && J1 &&  J2  && T && \vphantom{c}  && \vphantom{c} \\ %TT\\
 \vphantom{a} &&  \vphantom{b} && \vphantom{B} &&  \vphantom{c}  && \vphantom{C} & \vphantom{d} & \vphantom{D}\\ };
 \path[\pot]
(m-2-1) edge [inPot1, red, cross line] node[nodedescr] {$j_1$} (m-2-3) %S to J1 5
(m-2-1) edge [inPot2, in=-90, blue, cross line] node[nodedescr] {$j_2$} (m-2-5); %S to 2
 % \path[->]
%\path[right hook->]
\path[stochasticPathstyle]
(m-2-3) edge [inPot1, out=90, in=90,distance=4cm, green, cross line] node[stochasticNodestyle] {$p$} (m-2-1) %J1 to S 1
(m-2-3) edge [in2Pot1,green, cross line] node[stochasticNodestyle] {$1-p$} (m-2-7) %J1 to T 7
(m-2-5) edge [inPot2, out=-125, in= -90, distance=6cm, green, cross line] node[stochasticNodestyle] {$q$} (m-2-1) %J2 to S 
(m-2-5) edge [inPot2, out=-60, in= -90, distance=3cm, green, cross line] node[stochasticNodestyle] {$1-q$} (m-2-7); %J2 to T 
\path[solid,red!70, line width=6pt]
(m-2-1) edge (m-2-3);
\path[solid,green!60, line width=6pt]
(m-2-3) edge (m-2-5);
\path[solid,blue!40, line width=6pt]
(m-2-5) edge (m-2-7);
\path[solid,black!40, line width=6pt]
(m-2-7) edge (m-2-9);
\end{tikzpicture}
\caption{
    {\bf Network XLSSC: Flexible exponential-linear stochastic stem cell network}.  One stem cell (S) divides to produce two cells (J1 and J2) that each stochastically activate either the stem cell itself or a terminal cell.     }
   \label{fig:XLSSC}
\end{figure}
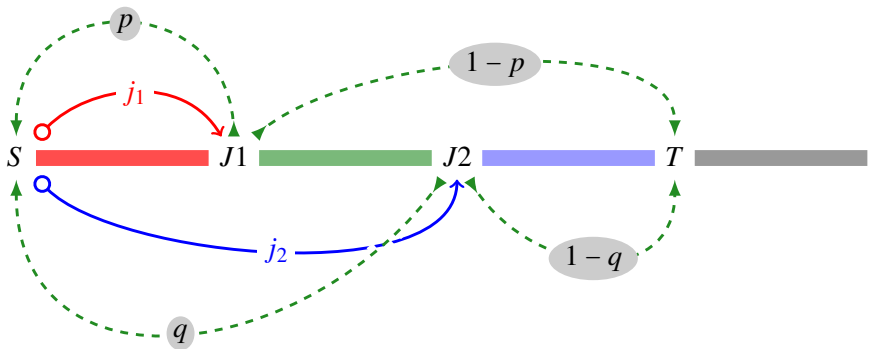

In the network in \autoref{fig:XLSSC} is very flexible. Depending on the probability distribution, the network can range between being exponential, linear, or terminal, as well as every mixture in between.  Furthermore, the network can be deterministic, stochastic or mixture of both.  This flexibility is partly the result of separating out the probability distributions for the behaviors of the two daughter cells.  It shows that the architecture of the network imposes constraints on what kinds of developmental dynamics are in principle possible.  The probability distribution presupposes a network architecture of possible developmental paths.  

The cell S is only a stem cell stochastically and not intrinsically when $p < 1$ and $q < 1$.   When $p = q = 1$ the network is deterministic  exponential. When $p = 1, q = 0$ or $p = 0, q = 1$ the network is deterministic linear, i.e., a deterministic 1st order geometric stem cell network. If $p = q = 0$ the network is terminal.  When $p = 1$  and $q < 1$, or when $q=1$ and $p<1$, then the network is mixed deterministic linear with stochastic exponential tendencies.  

If probabilities $p = q$ then as $p$ and $q$ approach $1$ the network  approaches the behavior of a deterministic exponential network.  However, if $p$ and $q$ are different then this network can simulate a linear stochastic network as well when, for example, $p$ approaches $1$ and $q$ approaches $0$, or vice versa. As the probabilities $p$ and $q$ decrease,  the more frequently the cancer stem cell results in a terminal tumor that does not develop further because it consists only of cells of terminal type T.  This shows that stochastic cancer stem cell networks can in some cases go into spontaneous remission. While this network can also exhibit exponential growth even in a stochastically linear probability distribution, because of the two backward loops, there is a diminishing probability that it remains exponential. Thus, whether this network results in linear or exponential proliferation depends on the probability distribution.  

For the network XLSSC, if the probabilities $p = 1- q$ then the higher the probability of $p$ the more the network approximates a deterministic linear developmental network.  Since, in this case the distribution is anti-symmetric,  the cell population partition of cell types consists of an equal number of exponential stem cells and terminal cells, with the majority of cells being linear stem cells. This corresponds to the observed distribution in epidermal basal stem cells.  If, on the other hand we have a symmetric distribution where $p = q$ then the higher the probability of $p$ the more the network approximates a deterministic exponential network.  Thus the type of cancer network we have depends on the probability distributions over the connecting stochastic links.  

The network XLSSC (\autoref{fig:XLSSC}) has some similarity to the Till model (TillSSC~\autoref{fig:TillSSC}), when $p=q$ (and value of probability $p=p_{2}$ as in the original model by Till~\cite{Till1964} model) and T.  The Till model forces that daughter cells of S are the same. In contrast, this model is more flexible in that, depending on the probability distribution ($p$, $q$), it can model both exponential and linear dynamics of multicellular development. 

\subsection{Linear stochastic stem cell network LSSC}

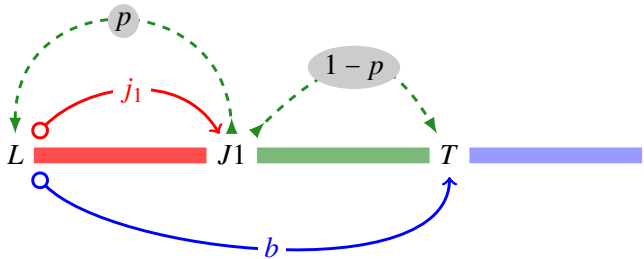
\begin{figure}[H]
\begin{tikzpicture}[style=mystyle]
\matrix (m) [matrix of math nodes, row sep=3em,
column sep=3em, text height=1.5ex, text depth=0.25ex]
{ \vphantom{a} && \vphantom{b} &&  \vphantom{c}  && \vphantom{c}\\ %  & & &\\
 L  && J1 &&  T  && %\vphantom{J2) &&  \vphantom{C) && \vphantom{c} \\ %TT\\
 \vphantom{a} &&  \vphantom{b} && \vphantom{B} &&  \vphantom{c}  && \vphantom{T}\\}; % & \vphantom{d} & \vphantom{D}\\ };
 \path[\pot]
(m-2-1) edge [inPot1, red, cross line] node[nodedescr] {$j_1$} (m-2-3) %S to J1 3
(m-2-1) edge [inPot2, in=-90, blue, cross line] node[nodedescr] {$b$} (m-2-5); %S to B 5
 % \path[->]
%\path[right hook->]
\path[stochasticPathstyle]
(m-2-3) edge [inPot1, out=90, in=90,distance=4cm, green, cross line] node[stochasticNodestyle] {$p$} (m-2-1) %J1 to S 
(m-2-3) edge [in2Pot1,in=120,distance=3cm,green, cross line] node[stochasticNodestyle] {$1-p$} (m-2-5); %J1 to B
\path[solid,red!70, line width=6pt]
(m-2-1) edge (m-2-3);
\path[solid,green!60, line width=6pt]
(m-2-3) edge (m-2-5);
\path[solid,blue!40, line width=6pt]
(m-2-5) edge (m-2-7);
\end{tikzpicture}
\caption{
    {\bf Network LSSC:  A linear stochastic stem cell  network}.  A linear stem cell (L) divides to produce a terminal cell T and cell J1. The cell J1  stochastically loops back to activate the stem cell T or it activates the terminal cell T.  The network's behavior approaches a deterministic linear stem cell network as the probability $p$  approach $1$.  Cancer stem cells controlled by such a network can go into spontaneous  remission because of the fact that for all points in all possible  paths in the network there is the possibility of reaching the terminal cell state T. 
  }
   \label{fig:LSSC}
\end{figure}

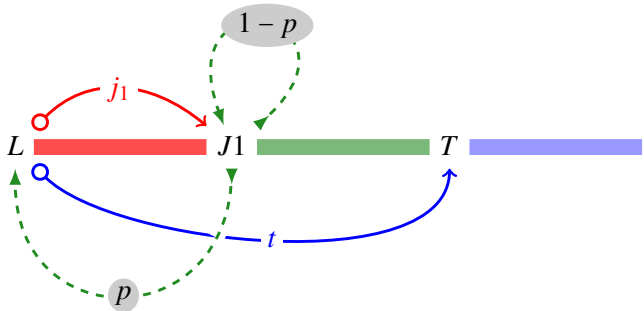
\begin{figure}[H]
\begin{tikzpicture}[style=mystyle]

\matrix (m) [matrix of math nodes, row sep=3em,
column sep=3em, text height=1.5ex, text depth=0.25ex]
{ \vphantom{a} && \vphantom{b} &&  \vphantom{c}  && \vphantom{c}\\ %  & & &\\
 L  && J1 &&  T  && %\vphantom{J2) &&  \vphantom{C) && \vphantom{c} \\ %TT\\
 \vphantom{a} &&  \vphantom{b} && \vphantom{B} &&  \vphantom{c}  && \vphantom{T}\\}; % & \vphantom{d} & \vphantom{D}\\ };
 \path[\pot]
(m-2-1) edge [inPot1, in=140, red, cross line] node[nodedescr] {$j_1$} (m-2-3) %S to J1 3
(m-2-1) edge [inPot2, in=-90, blue, cross line] node[nodedescr] {$t$} (m-2-5); %S to B 5

\path[stochasticPathstyle]
(m-2-3) edge [inPot2, out=-90, in=-90,distance=4.5cm, green, cross line] node[stochasticNodestyle] {$p$} (m-2-1) %J1 to S 
(m-2-3) edge [in2Pot1,in=110,out=40,distance=4.5cm,green, cross line] node[stochasticNodestyle] {$1-p$} (m-2-3); %J1 to B
\path[solid,red!70, line width=6pt]
(m-2-1) edge (m-2-3);
\path[solid,green!60, line width=6pt]
(m-2-3) edge (m-2-5);
\path[solid,blue!40, line width=6pt]
(m-2-5) edge (m-2-7);
\end{tikzpicture}
\caption{
    {\bf Network DLSSC: Deterministic linear stem cell network with stochastic delay}.  A linear stem cell (L) divides to produce a terminal cell T and cell J1. The cell J1  stochastically loops back to activate the stem cell T with probability $p$.  Unlike with a true linear stochastic network like LSS which stochastically self-differentiates to L with probability $p$ or differentiates to  T, this is a deterministic linear stem cell network that always self-differentiates back to L.  The stochasticity only effects the rate at which the cell divides.  The probability $p$ only effects the frequency of looping back to L and, thereby, changes the cycle time cell division.  Unlike LSSC, the linear cancer stem cells controlled by this network never go into spontaneous  remission since the cell endlessly self-renews.  The lower the probability $p$ the longer it takes for the cell to cell renew and divide. 
  }
  \label{fig:DLSSC}
\end{figure}

\subsection{1st-Order Geometric Cancer Networks with open stochastic dedifferentiation}

\label{sec:G1SD}
\begin{figure}[H]
\begin{tikzpicture}[style=mystyle]
\matrix (m) [matrix of math nodes, row sep=3em,
column sep=3em, text height=1.5ex, text depth=0.25ex]
{ \vphantom{a} & \vphantom{b}  &  \vphantom{c}  & \vphantom{c}  & \\
 S  &&  J  &&  T  &&  \\
  \vphantom{a} &   \vphantom{j} & \vphantom{b} &  \vphantom{c}  & \\ };
 \path[\pot]
(m-2-1) edge [selfloop1, cross line] node[nodedescr] {$ a $} (m-2-1)
(m-2-1) edge [inPot2, in=-110, blue, cross line] node[nodedescr] {$j$} (m-2-3); %S to J 
\path[stochasticPathstyle]
(m-2-3) edge [inPot2, out=-80, in=-90,distance=4cm, green, cross line] node[stochasticNodestyle] {$p$} (m-2-1) %J1 to S 
(m-2-3) edge [in2Pot1,in=90,out=80,distance=4cm,green, cross line] node[stochasticNodestyle] {$1-p$} (m-2-5); %J1 to B
\path[solid, line width=6pt]
(m-2-1) edge (m-2-3)
(m-2-3) edge (m-2-5);
%(m-2-5) edge (m-2-7);
\end{tikzpicture}
\caption{
    {\bf Network G1SD: A 1st-order geometric cancer with stochastic dedifferentiation:} In this linear network, a cell of type $S$ is a 1st-order stem cell. It produces progenitor cells of type J that have a stochastic dedifferentiation potential.  A cell J can either dedifferentiate to its parent state S or differentiate into the terminal cell T.  The effect of the stochasticity is seen in the population of progenitor cells J produced by the network.  The lower the probability $p$ of dedifferentiation, the greater the number of terminal cells in the cell population generated by the stem cell S.  
  }
    \label{fig:G1SD}
\end{figure}
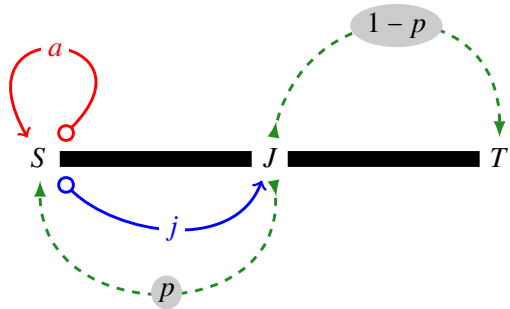 

\subsection{1st-Order Geometric Cancer Networks with closed stochastic dedifferentiation }
In the above network \autoref{fig:G1SD} dedifferentiation can occur once prior to permanent differentiation in a terminal cell state. The effect of the stochasticity can only be seen in the tumor as a whole that is by the cancer stem cell.  In the following network dedifferentiation is a constant possibility for the life of the cell. 

\label{sec:G1SDc}
\begin{figure}[H]
\begin{tikzpicture}[style=mystyle]
\matrix (m) [matrix of math nodes, row sep=3em,
column sep=3em, text height=1.5ex, text depth=0.25ex]
{ \vphantom{a} & \vphantom{b}  &  \vphantom{c}  & \vphantom{c}  & \\
 S  &&  J  &&  T  &&  \\
  \vphantom{a} &   \vphantom{j} & \vphantom{b} &  \vphantom{c}  & \\ };
 \path[\pot]
(m-2-1) edge [selfloop1, cross line] node[nodedescr] {$ a $} (m-2-1)
(m-2-1) edge [inPot2, in=-110, blue, cross line] node[nodedescr] {$j$} (m-2-3); %S to J 
\path[stochasticPathstyle]
(m-2-3) edge [inPot2, out=-80, in=-90,distance=4cm, green, cross line] node[stochasticNodestyle] {$p$} (m-2-1) %J1 to S 
(m-2-3) edge [in2Pot1,in=110,out=40,distance=4.5cm,green, cross line] node[stochasticNodestyle] {$1-p$} (m-2-3); %J1 to B
\path[solid, line width=6pt]
(m-2-1) edge (m-2-3)
(m-2-3) edge (m-2-5);
%(m-2-5) edge (m-2-7);
\end{tikzpicture}
\caption{
    {\bf Network G1SDc: A 1st-order geometric cancer with closed stochastic dedifferentiation}  }
    \label{fig:G1SDc}
\end{figure}
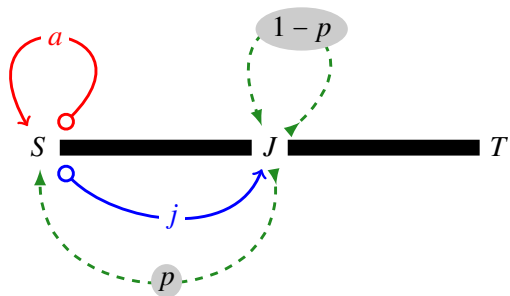 

In this linear network \autoref{fig:G1SDc}, a cell of type $S$ is a 1st-order stem cell. It produces progenitor cells of type J that have a continual stochastic dedifferentiation potential.  A cell J can either dedifferentiate to its parent state S or maintains its present state J.  In this closed stochastic network, a cell J continues to have a dedifferentiation potential for the life of that cell. There is no path to permanent terminal differentiation.  This contrasts with the network \autoref{fig:G1SD} where the cell only has a limited time frame to dedifferentiate prior to its permanent differentiation into a terminal cell T. Thus, all in a population of cells generated by S have the potential to dedifferentiate.  In contrast, in the network \autoref{fig:G1SD} if the probability $p$ is low, most cells in the population generated by S will be terminal cells with no dedifferentiation potential.

\subsection{Closed exponential stem cell network with stochastic delays}
Depending on the probability distribution, the next stochastic network proliferative properties can vary from deterministic terminal, linear, or exponential to stochastic terminal, linear or exponential. 

\begin{figure}[H]
\begin{tikzpicture}[style=mystyle]
\matrix (m) [matrix of math nodes, row sep=3em,
column sep=3em, text height=1.5ex, text depth=0.25ex]
{ \vphantom{a} && \vphantom{b} &&  \vphantom{c}  && \vphantom{c}  & & &\\
 S  && J1 &&  J2  && T && \vphantom{c}  && \vphantom{c} \\ %TT\\
 \vphantom{a} &&  \vphantom{b} && \vphantom{B} &&  \vphantom{c}  && \vphantom{C} & \vphantom{d} & \vphantom{D}\\ };
 \path[\pot]
(m-2-1) edge [inPot1, red, cross line] node[nodedescr] {$j_1$} (m-2-3) %S to J1 5
(m-2-1) edge [inPot2, in=-125, blue, cross line] node[nodedescr] {$j_2$} (m-2-5); %S to 2

 % \path[->]
%\path[right hook->]
\path[stochasticPathstyle]

(m-2-3) edge [inPot1, out=90, in=90,distance=4cm, green, cross line] node[stochasticNodestyle] {$p$} (m-2-1) %J1 to S 1
(m-2-3) edge [in2Pot1,in=130,out=60,distance=3cm,green, cross line] node[stochasticNodestyle] {$1-p$} (m-2-5) %J1 to T 7
(m-2-5) edge [inPot2, out=-90, in= -90, distance=6cm, red, cross line] node[stochasticNodestyle] {$q$} (m-2-1) %J2 to S 
(m-2-5) edge [inPot1, out=45, in= 110, distance=5cm, red, cross line] node[stochasticNodestyle] {$1-q$} (m-2-5); %J2 to T 

\path[solid,red!70, line width=6pt]
(m-2-1) edge (m-2-3);
\path[solid,green!60, line width=6pt]
(m-2-3) edge (m-2-5);
\path[solid,blue!40, line width=6pt]
(m-2-5) edge (m-2-7);
\path[solid,black!40, line width=6pt]
(m-2-7) edge (m-2-9);

\end{tikzpicture}
\caption{
    {\bf Network DXSSC modified LSSC: Closed exponential stochastic stem cell  network}.  If the probability $q > 0$ then this network is inherently nonterminal. All stochastic paths lead to potentially nonterminal nodes in the network.  $p = q = 0$ is terminal. $p = q = 1$ is exponential. $p = 1, q = 0$ is linear. $p = 0, q = 1$ is exponential. 
}
 \label{fig:DXSSC}
\end{figure}
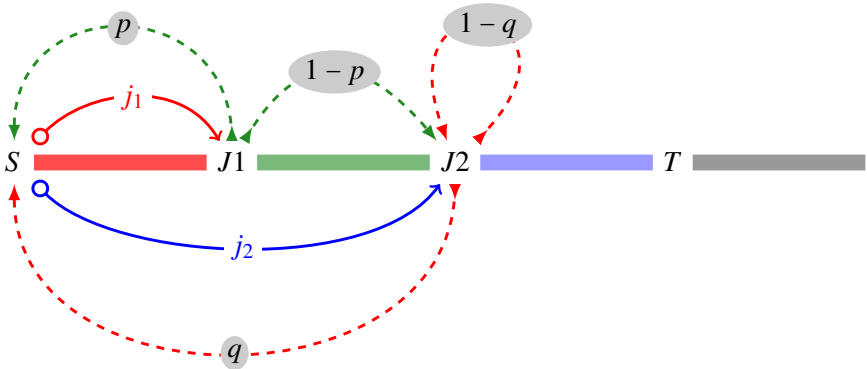

Unlike LSSC \autoref{fig:LSSC}, if the probability $q > 0$ in cell state J2 then there are no true, probability independent, terminal cells in DXSSC \autoref{fig:DXSSC}.  True terminal cells have no possible stochastically available paths that lead to a reversal of differentiation to a more dedifferentiated cell type.  Instead, all cell states are stochastic with possible developmental paths that lead to the original founder stem cell S. The cell in state J2 has a probability $q$ of dedifferentiating to S.  The higher the probabilities $q$ and $p$ the more this network mirrors the behavior of a deterministic exponential network.  If either $p$ or $q$ is low while the other high it is more similar to a linear developmental network.  Thus the probability distribution determines the similarity to linear or exponential networks. Even if the probability of  $q$ is very low, as the number of J2 cell increase it becomes more and more likely that one of them will dedifferentiate to S. Thus, the cancer pulls away from linearity with an increasing cell population size.  

If probabilities $q = p = 0$ then the network is deterministic terminal network leading to only one cell division. If $q = 0$ and $p = 1$ we have a deterministic linear stem cell network. However, if the probability $q = 0$ and $p > 0$, then the network is a stochastic linear network where the lower the probability $p$ the greater the chance that proliferation terminates.   If either $p = 1$ or $p = 0$, and if $q = 1$ then the network is a deterministic exponential stem cell network.  If $q > 0$ then no matter how small the nonzero probability $q$ is, proliferation increases with time as more and more cells are created. 

{\bf Remark:} There is another way to escape the maze of exponential proliferation. Cell proliferation can be influenced by the number of times the self-loop at J2 is allowed to repeat.  If it is it is only allowed to repeat only a limited number of times relative then become quiescent, then, depending on the probability $q$, it could result in a terminal differentiation state after some stochastic tries. In other words, if, in parallel with the repetitions of the self-entry loop at J2, there exists a separate simultaneous, parallel process of terminal differentiation by means of some counting mechanism (where the cell at J2 becomes quiescent after some finite number of counting steps), then this network could lead to the ultimate termination of cell proliferation or at most linear proliferation.  The counting mechanism could be dependent on the number of loops executed at J2 or it could be a function of some other temporal variable.  

\subsection{A 1st order geometric/exponential stochastic stem cell network with stochastic progenitor cell dedifferentiation}

\begin{figure}[H]
% \begin{narrow}{-1.5cm}{-1.5cm}%{-1cm}{-1cm}
\begin{tikzpicture}[style=mystyle]
\matrix (m) [matrix of math nodes, 
row sep=3em,
column sep=\ColSepTight, %\matrixsepless, %\matrixsep, %3em, 
text height=1.5ex, text depth=0.25ex]
{ \vphantom{a} &\vphantom{a}& \vphantom{b} &&  \vphantom{c}  && \vphantom{c}  & & &\\
 G  && J1 && J2 &&  B  && J3 &&  J4 && T^{*} &&\\ %TT\\
 \vphantom{a} &\vphantom{a}&  \vphantom{b} && \vphantom{B} &&  \vphantom{c}  && \vphantom{C} && \vphantom{d} & \vphantom{D}\\ };
 % \path[->]
%\path[right hook->]
\path[stochasticPathstyle]
(m-2-3) edge [inPot1, out=90, in=90,distance=4cm, green, cross line] node[stochasticNodestyle] {$p$} (m-2-1) %J1 to S 
(m-2-3) edge [in2Pot1,out=60,in=110,distance=3.5cm,green, cross line] node[stochasticNodestyle] {$1-p$} (m-2-7) %J1 to B
(m-2-5) edge [in2Pot1,out=-60,in=-110,distance=2.5cm,black, cross line] node[stochasticNodestyle] {$q$} (m-2-7) %J1a to B
(m-2-5) edge [in2Pot2,in=-90,out=-90, distance=3.5cm,black, cross line] node[stochasticNodestyle] {$1-q$} (m-2-1) %J1a to B

(m-2-9) edge [inPot1, out=60, in= 110, distance=3cm, green, cross line] node[stochasticNodestyle] {$s$} (m-2-11) %J3 to J4 
(m-2-9) edge [inPot1, out=90, in= 80, distance=4cm, green, cross line] node[stochasticNodestyle] {$1-s$} (m-2-13) %J2 to C
(m-2-11) edge [inPot2, out=-90, in= -120, distance=6cm, red] node[stochasticNodestyle] {$1-r$} (m-2-1) %J2 to B 
(m-2-11) edge [inPot2, out=-60, in= -70, distance=3cm, red, cross line] node[stochasticNodestyle] {$r$} (m-2-13); %J2 to T 

 \path[\pot]
(m-2-1) edge [inPot1, red, cross line] node[nodedescr] {$j_1$} (m-2-3) %S to J1 3
(m-2-1) edge [inPot2, out=-60, in=-120,distance=2.5cm, blue] node[nodedescr] {$j_{2}$} (m-2-5)%S to J2 5
(m-2-7) edge [inPot1, red, cross line] node[nodedescr] {$j_{3}$} (m-2-9) %S to J2 7
(m-2-7) edge [inPot2, in=-120, blue] node[nodedescr] {$j_{4}$} (m-2-11); %B to  C 9

\path[solid,red!80, line width=6pt]
(m-2-1) edge (m-2-3);
\path[solid,green!70, line width=6pt]
(m-2-3) edge (m-2-5);
\path[solid,black!40, line width=6pt]
(m-2-5) edge (m-2-7);
\path[solid,blue!40, line width=6pt]
(m-2-7) edge (m-2-9);
\path[solid,green!70, line width=6pt]
(m-2-9) edge (m-2-11);
\path[solid,red!40, line width=6pt]
(m-2-11) edge (m-2-13);
%\path[solid,red!80, line width=10pt]
%(m-2-6) edge (m-2-7);
\end{tikzpicture}

\caption{
    {\bf Network G1SSC:   A 1st-order stochastic geometric/exponential stem cell  network with stochastic dedifferentiation}. 
  }
  \label{fig:G1SSC}
%  \end{narrow}
\end{figure}
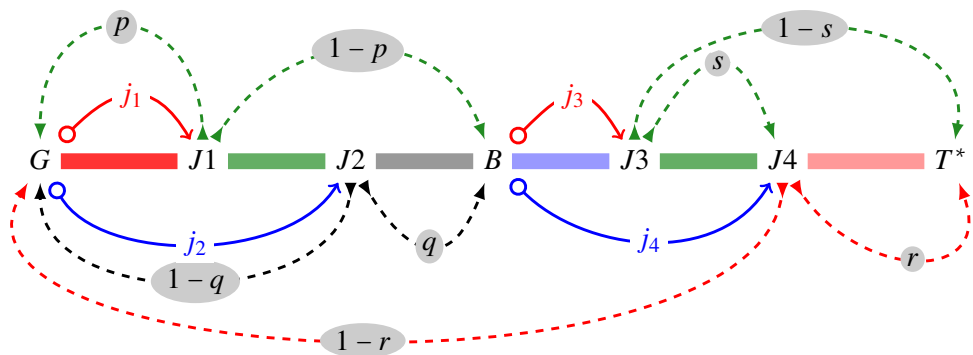

This stochastic stem cell network (\autoref{fig:G1SSC}) generates progenitor cells B that generate cells J2 that can dedifferentiate to stem cells G or differentiate to a terminal network $T^{*}$. Depending on the probability distribution, this network can exhibit terminal, linear-geometric, and exponential developmental dynamics. If the probability distribution is such that $p=q=r=s=1$ then this is equivalent to a deterministic 1st order geometric stem cell network. A 1st-order stem cell (G) stochastically divides to produce either two progenitor cell B, or or two stem cells G or one progenitor cell B and one stem cell G.   and cell J1. The cells in states J1 or J2  stochastically loop back to activate the stem cell G or to activate the progenitor cell $B$. The progenitor cell $B$ divides into a semi-terminal cell $T^{*}$ by way of J3 or in divides into two cells of type  J4.  J4 either terminates with $T^{*}$ or it dedifferentiates into a stem cell G.  

Its behavior approaches a deterministic 1st-order geometric stem cell network as the probabilities $p$ and $q$ approach $1$.  However, the subnetwork activated by this 1st-order geometric network can still stochastically dedifferentiate into one or two new stem cells G. Cancers controlled by such a network can go into spontaneous  remission because of the fact that for all points in all possible  paths in the network there is the possibility of reaching a terminal cell state.  However, because there is the possibility at J1, J2 and J4 to dedifferentiate to earlier, upstream stem cell states, the relative numbers of stem cells to terminal cells will be higher than in networks that have fewer dedifferentiation pathways. Note, too that the dedifferentiation pathways introduce several possibilities for exponential growth. Even if dedifferentiation probabilities are very small, any increase in these probabilities could have a significant destabilizing influence on the resulting tumor. 

\subsection{A broad spectrum Linear or 2nd order geometric or exponential stochastic stem cell network}

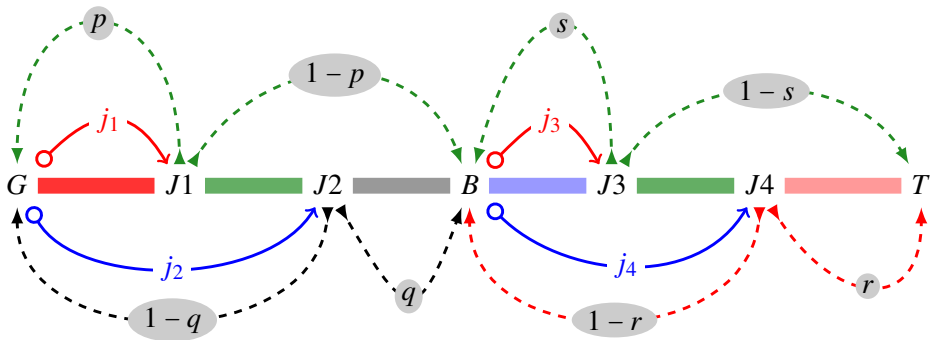
\begin{figure}[H]
% \begin{narrow}{-1.5cm}{-1.5cm}%{-1cm}{-1cm}
\begin{tikzpicture}[style=mystyle]
\matrix (m) [matrix of math nodes, 
row sep=3em,
column sep=\ColSepTight, %\matrixsepless, %\matrixsep, %3em, 
text height=1.5ex, text depth=0.25ex]
{ \vphantom{a} &\vphantom{a}& \vphantom{b} &&  \vphantom{c}  && \vphantom{c}  & & &\\
 G  && J1 && J2 &&  B  && J3 &&  J4 && T &&\\ %TT\\
 \vphantom{a} &\vphantom{a}&  \vphantom{b} && \vphantom{B} &&  \vphantom{c}  && \vphantom{C} && \vphantom{d} & \vphantom{D}\\ };
 % \path[->]
%\path[right hook->]
\path[stochasticPathstyle]
(m-2-3) edge [inPot1, out=90, in=90,distance=5cm, green, cross line] node[stochasticNodestyle] {$p$} (m-2-1) %J1 to S 
(m-2-3) edge [in2Pot1,out=60,in=110,distance=3.5cm,green, cross line] node[stochasticNodestyle] {$1-p$} (m-2-7) %J1 to B
(m-2-5) edge [in2Pot1,out=-60,in=-110,distance=3.5cm,black, cross line] node[stochasticNodestyle] {$q$} (m-2-7) %J1a to B
(m-2-5) edge [in2Pot2,in=-90,out=-90, distance=4cm,black, cross line] node[stochasticNodestyle] {$1-q$} (m-2-1) %J1a to B

(m-2-9) edge [inPot1, out=90, in= 75, distance=5cm, green, cross line] node[stochasticNodestyle] {$s$} (m-2-7) %J2 to S 
(m-2-9) edge [inPot1, out=60, in= 120, distance=3cm, green, cross line] node[stochasticNodestyle] {$1-s$} (m-2-13) %J2 to C
(m-2-11) edge [inPot2, out=-90, in= -90, distance=4cm, red] node[stochasticNodestyle] {$1-r$} (m-2-7) %J2 to B 
(m-2-11) edge [inPot2, out=-60, in= -90, distance=3cm, red, cross line] node[stochasticNodestyle] {$r$} (m-2-13); %J2 to T 

 \path[\pot]
(m-2-1) edge [inPot1, red, cross line] node[nodedescr] {$j_1$} (m-2-3) %S to J1 3
(m-2-1) edge [inPot2, out=-60, in=-120,distance=2.5cm, blue] node[nodedescr] {$j_{2}$} (m-2-5)%S to J2 5
(m-2-7) edge [inPot1, red, cross line] node[nodedescr] {$j_{3}$} (m-2-9) %S to J2 7
(m-2-7) edge [inPot2, in=-120, blue] node[nodedescr] {$j_{4}$} (m-2-11); %B to  C 9

\path[solid,red!80, line width=6pt]
(m-2-1) edge (m-2-3);
\path[solid,green!70, line width=6pt]
(m-2-3) edge (m-2-5);
\path[solid,black!40, line width=6pt]
(m-2-5) edge (m-2-7);
\path[solid,blue!40, line width=6pt]
(m-2-7) edge (m-2-9);
\path[solid,green!70, line width=6pt]
(m-2-9) edge (m-2-11);
\path[solid,red!40, line width=6pt]
(m-2-11) edge (m-2-13);
%\path[solid,red!80, line width=10pt]
%(m-2-6) edge (m-2-7);
\end{tikzpicture}

\caption{
    {\bf Network LG2XSSC:   A stochastic linear, 2nd order geometric or exponential stem cell  network}. If the probability distribution is such that $p=q=r=s=1$ then this is equivalent to a deterministic 2nd order geometric stem cell network. Thus, as these probabilities approach 1 the behavior approaches a 2nd order geometric stem cell network. 
  }
  \label{fig:LG2XSSC}
%  \end{narrow}
\end{figure}

In \autoref{fig:LG2XSSC}, a 2nd-order meta-stem cell (G) divides to produce a 1st-order stem cell $B$ and cell J1. The cell J1  stochastically loops back to activate the meta-stem cell G or it activates the first order linear stem cell $B$. The 1st-order stem cell $B$ divides into a semi-terminal cell C and a cell J2. The cell J2 stochastically activates either its parent stem cell $B$, the semi-terminal cell C or dedifferentiates to G.  The cell C also can stochastically dedifferentiate to any of the previous stem cell fates (G, B) or differentiate into the final terminal cell T. 

The network consists of two linked 1st-order linear stochastic networks. Its behavior approaches a deterministic 2nd-order geometric stem cell network as the probabilities $p$ and $q_1$ approach $1$.  Cancers controlled by such a network can go into spontaneous  remission because of the fact that for all points in all possible  paths in the network there is the possibility of reaching a terminal cell state.  

However, because there is the possibility at J2 and C to dedifferentiate to earlier, upstream network states, the relative numbers of stem cells to terminal cells will be higher than in networks that have fewer dedifferentiation pathways. Note, too that the dedifferentiation pathways introduce several possibilities for exponential growth. Even if their probabilities  ($q_3$, $r_2$, $r_3$) are very small, any increase in these probabilities could have a significant destabilizing influence on the resulting tumor.  

The deterministic possible behaviors include: $(p=0, q= 1, s=0, r=1)$ is terminal. $(p=1, q= 0)$ is exponential. $(p=1, q= 1, s=0, r=1)$ is linear.  $(p=1, q= 1, s=1, r=1)$ is 2nd order geometric.  $(p=1, q= 1, s=1, r=0)$ is a linear network linked to an exponential network. 

\subsection{A Master stochastic geometric and exponential network}

We now present a generalized, broad spectrum, master network where each probability distribution results in a particular geometric or exponential network.  One way to make the developmental network control theory more acceptable to traditional mathematicians and physicists is to formulate the networks in terms of differential equations.  Given that stochastic networks in the cancer paper can be made so general as to describe any network based on the probability distribution, perhaps one can use stochasticity in differential equations to attain one Master Equation that describes all the major types of networks.  Particular probability distributions would then give particular network behavior.  The Master Network and Master Equation can be made fully general by allowing arbitrarily many loops. 

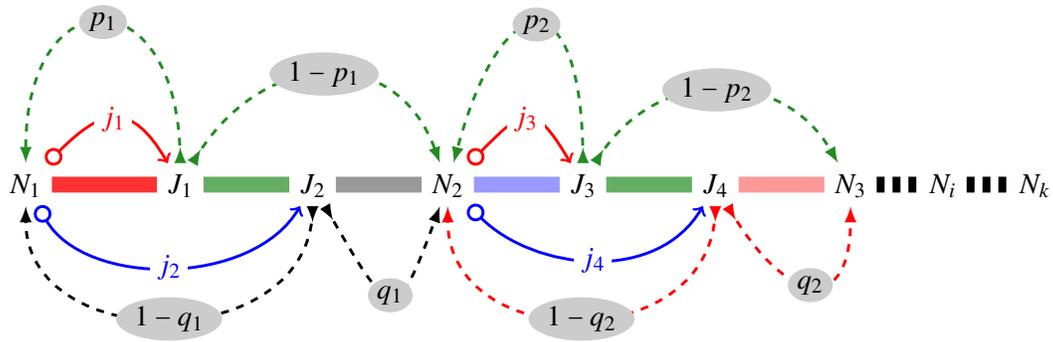
\begin{figure}[H]
% \begin{narrow}{-1.5cm}{-1.5cm}%{-1cm}{-1cm}
\begin{tikzpicture}[style=mystyle]
\matrix (m) [matrix of math nodes, 
row sep=3em,
column sep=\ColSepExtraTight, %\ColSepTight, %\matrixsepless, %\matrixsep, %3em, 
text height=1.5ex, text depth=0.25ex]
{ \vphantom{a} &\vphantom{a}& \vphantom{b} &&  \vphantom{c}  && \vphantom{c}  & &  \vphantom{c} &&  \vphantom{c} \\
 N_{1}  && J_{1} && J_{2} &&  N_{2}  && J_{3} &&  J_{4} && N_{3} & N_{i} & N_{k}\\ %TT\\
 \vphantom{a} &\vphantom{a}&  \vphantom{b} && \vphantom{B} &&  \vphantom{c}  && \vphantom{C} && \vphantom{d} && \vphantom{D}\\ };
 % \path[->]
%\path[right hook->]
\path[stochasticPathstyle]
(m-2-3) edge [inPot1, out=90, in=90,distance=5cm, green, cross line] node[stochasticNodestyle] {$p_{1}$} (m-2-1) %J1 to S 
(m-2-3) edge [in2Pot1,out=60,in=110,distance=3.5cm,green, cross line] node[stochasticNodestyle] {$1-p_{1}$} (m-2-7) %J1 to B
(m-2-5) edge [in2Pot1,out=-60,in=-110,distance=3.5cm,black, cross line] node[stochasticNodestyle] {$q_{1}$} (m-2-7) %J1a to B
(m-2-5) edge [in2Pot2,in=-90,out=-90, distance=4cm,black, cross line] node[stochasticNodestyle] {$1-q_{1}$} (m-2-1) %J1a to B

(m-2-9) edge [inPot1, out=90, in= 75, distance=5cm, green, cross line] node[stochasticNodestyle] {$p_{2}$} (m-2-7) %J2 to S 
(m-2-9) edge [inPot1, out=60, in= 120, distance=3cm, green, cross line] node[stochasticNodestyle] {$1-p_{2}$} (m-2-13) %J2 to C
(m-2-11) edge [inPot2, out=-90, in= -90, distance=4cm, red] node[stochasticNodestyle] {$1-q_{2}$} (m-2-7) %J2 to B 
(m-2-11) edge [inPot2, out=-60, in= -90, distance=3cm, red, cross line] node[stochasticNodestyle] {$q_{2}$} (m-2-13); %J2 to T 

 \path[\pot]
(m-2-1) edge [inPot1, red, cross line] node[nodedescr] {$j_1$} (m-2-3) %S to J1 3
(m-2-1) edge [inPot2, out=-60, in=-120,distance=2.5cm, blue] node[nodedescr] {$j_{2}$} (m-2-5)%S to J2 5
(m-2-7) edge [inPot1, red, cross line] node[nodedescr] {$j_{3}$} (m-2-9) %S to J2 7
(m-2-7) edge [inPot2, in=-120, blue] node[nodedescr] {$j_{4}$} (m-2-11); %B to  C 9

\path[solid,red!80, line width=6pt]
(m-2-1) edge (m-2-3);
\path[solid,green!70, line width=6pt]
(m-2-3) edge (m-2-5);
\path[solid,black!40, line width=6pt]
(m-2-5) edge (m-2-7);
\path[solid,blue!40, line width=6pt]
(m-2-7) edge (m-2-9);
\path[solid,green!70, line width=6pt]
(m-2-9) edge (m-2-11);
\path[solid,red!40, line width=6pt]
(m-2-11) edge (m-2-13);
\path[dashed,black, line width=6pt]
(m-2-13) edge (m-2-14);
\path[dashed,black, line width=6pt]
(m-2-14) edge (m-2-15);
%\path[solid,red!80, line width=10pt]
%(m-2-6) edge (m-2-7);
\end{tikzpicture}

\caption{
    {\bf Network MasterNet:   A stochastic linear, k-th order geometric or exponential stem cell network}. Each subnetwork $N_{i}$ for $i = 1 \ldots k$ has the same form as $N_{1}$.  If the probability distribution is such that $p_{i}=q_{i}=1$ for $i=1 \ldots i=k$ for each network $N_{i}$ then this is equivalent to a deterministic k-th order geometric stem cell network $N_{1} \ldots N_{i} \ldots N_{k}$ . Thus, as the probabilities $p_{i} = q_{i}$ approach 1 the behavior approaches a k-th order geometric stem cell network.  On the other hand, if any reachable network $N_{i}$ has probabilities $p_{i} = 1- q_{i} = 1$ the network has exponential potential.  Note, we assume $N_{k}$ links to a terminal network $N_{k+1} = T$. 
  }
  \label{fig:MasterNet}
%  \end{narrow}
\end{figure}

\subsection{2nd Order Geometric stochastic stem cell  network}

\begin{figure}[H]
\begin{tikzpicture}[style=mystyle]
\matrix (m) [matrix of math nodes, row sep=3em,
column sep=3em, text height=1.5ex, text depth=0.25ex]
{ \vphantom{a} && \vphantom{b} &&  \vphantom{c}  && \vphantom{c}  & & &\\
 G  && J1 &&  B  && J2 &&  C && \vphantom{c} \\ %TT\\
 \vphantom{a} &&  \vphantom{b} && \vphantom{B} &&  \vphantom{c}  && \vphantom{C} & \vphantom{d} & \vphantom{D}\\ };
 \path[\pot]
(m-2-1) edge [inPot1, red, cross line] node[nodedescr] {$j_1$} (m-2-3) %S to J1 3
(m-2-1) edge [inPot2, in=-90, blue, cross line] node[nodedescr] {$b$} (m-2-5) %S to B 5
(m-2-5) edge [inPot1, red, cross line] node[nodedescr] {$j_2$} (m-2-7) %B to J2 7
(m-2-5) edge [inPot2, in=-90, blue, cross line] node[nodedescr] {$c$} (m-2-9); %B to  C 9
 % \path[->]
%\path[right hook->]
\path[stochasticPathstyle]
(m-2-3) edge [inPot1, out=90, in=90,distance=5cm, green, cross line] node[stochasticNodestyle] {$p$} (m-2-1) %J1 to S 
(m-2-3) edge [in2Pot1,in=120,distance=3cm,green, cross line] node[stochasticNodestyle] {$1-p$} (m-2-5) %J1 to B
(m-2-7) edge [inPot1, out=90, in= 75, distance=5cm, green, cross line] node[stochasticNodestyle] {$q$} (m-2-5) %J2 to S 
(m-2-7) edge [inPot1, out=60, in= 90, distance=3cm, green, cross line] node[stochasticNodestyle] {$1-q$} (m-2-9); %J2 to T 
\path[solid,red!80, line width=6pt]
(m-2-1) edge (m-2-3);
\path[solid,green!70, line width=6pt]
(m-2-3) edge (m-2-5);
\path[solid,blue!40, line width=6pt]
(m-2-5) edge (m-2-7);
\path[solid,green!70, line width=6pt]
(m-2-7) edge (m-2-9);
\end{tikzpicture}
\caption{
    {\bf Network G2SSC:  A 2nd order geometric stochastic stem cell  network}.  A 2nd-order meta-stem cell (G) divides to produce a 1st-order stem cell $B$ and cell J1. 
  }
   \label{fig:G2SSC}
\end{figure}
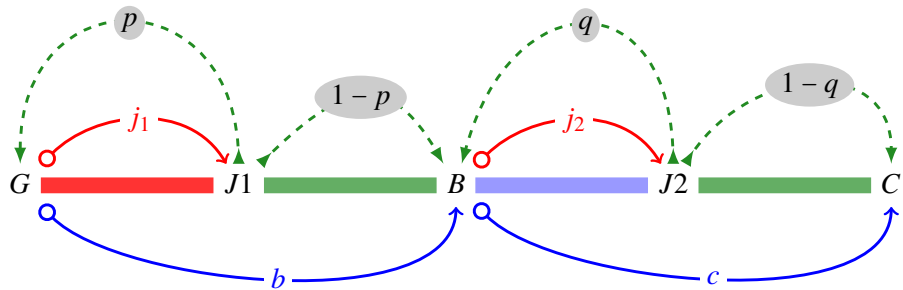

In \autoref{fig:G2SSC} The cell J1  stochastically loops back to activate the meta-stem cell G or it activates the first order linear stem cell $B$. The 1st-order stem cell $B$ divides into a terminal cell C and a cell J2. The cell J2 stochastically activates either the stem cell $B$ or the terminal cell $C$.  The network consists of two linked 1st-order linear stochastic networks. Its behavior approaches a deterministic 2nd-order geometric stem cell network as the probabilities $p$ and $q$ approach $1$.  Cancers controlled by such a network can go into spontaneous  remission because of the fact that for all points in all possible  paths in the network there is the possibility of reaching a terminal cell state. 

\subsection{Geometric stochastic stem cell  network with dedifferentiation}

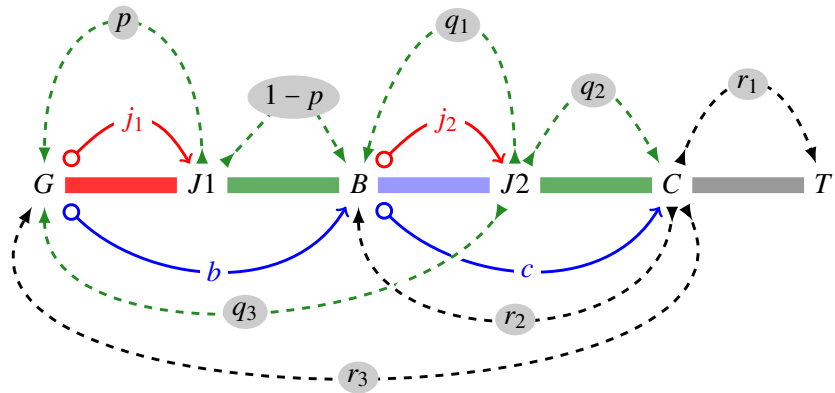
\begin{figure}[H]
\begin{tikzpicture}[style=mystyle]
\matrix (m) [matrix of math nodes, 
row sep=2.5em,
column sep=\matrixsep, %2.5em, 
text height=1.5ex, text depth=0.25ex]
{ \vphantom{a} && \vphantom{b} &&  \vphantom{c}  && \vphantom{c}  & & &\\
 G  && J1 &&  B  && J2 &&  C && T &&\\ %TT\\
 \vphantom{a} &&  \vphantom{b} && \vphantom{B} &&  \vphantom{c}  && \vphantom{C} && \vphantom{d} & \vphantom{D}\\ };
 \path[\pot]
(m-2-1) edge [inPot1, red, cross line] node[nodedescr] {$j_1$} (m-2-3) %S to J1 3
(m-2-1) edge [inPot2, in=-120, blue, cross line] node[nodedescr] {$b$} (m-2-5) %S to B 5
(m-2-5) edge [inPot1, red, cross line] node[nodedescr] {$j_2$} (m-2-7) %B to J2 7
(m-2-5) edge [inPot2, in=-120, blue, cross line] node[nodedescr] {$c$} (m-2-9); %B to  C 9
 % \path[->]
%\path[right hook->]
\path[stochasticPathstyle]
(m-2-3) edge [inPot1, out=90, in=90,distance=5cm, green, cross line] node[stochasticNodestyle] {$p$} (m-2-1) %J1 to S 
(m-2-3) edge [in2Pot1,in=120,distance=3cm,green, cross line] node[stochasticNodestyle] {$1-p$} (m-2-5) %J1 to B
(m-2-7) edge [inPot1, out=90, in= 75, distance=5cm, green, cross line] node[stochasticNodestyle] {$q_1$} (m-2-5) %J2 to S 
(m-2-7) edge [inPot1, out=60, in= 120, distance=3cm, green, cross line] node[stochasticNodestyle] {$q_2$} (m-2-9) %J2 to C
(m-2-7) edge [inPot2, out=-120, in= -90, distance=4cm, green, cross line] node[stochasticNodestyle] {$q_3$} (m-2-1) %J2 to G 
(m-2-9) edge [inPot1, out=70, in= 110, distance=3cm, black, cross line] node[stochasticNodestyle] {$r_1$} (m-2-11) %J2 to G 
(m-2-9) edge [inPot2, out=-90, in= -90, distance=4cm, black, cross line] node[stochasticNodestyle] {$r_2$} (m-2-5) %J2 to B 
(m-2-9) edge [inPot2, out=-60, in= -120, distance=7cm, black, cross line] node[stochasticNodestyle] {$r_3$} (m-2-1); %J2 to G 
\path[solid,red!80, line width=6pt]
(m-2-1) edge (m-2-3);
\path[solid,green!70, line width=6pt]
(m-2-3) edge (m-2-5);
\path[solid,blue!40, line width=6pt]
(m-2-5) edge (m-2-7);
\path[solid,green!70, line width=6pt]
(m-2-7) edge (m-2-9);\path[solid,black!40, line width=6pt]
(m-2-9) edge (m-2-11);
\end{tikzpicture}
%\end{center}
\caption{
    {\bf Network GSSCdediff:  A geometric stochastic stem cell  network with dedifferentiation}.   }
   \label{fig:GSSCdediff}
\end{figure}

In \autoref{fig:GSSCdediff} let probabilities $q_1 + q_2 +q_3 =  1$ and $r_1+r_2+r_3 = 1$. A 2nd-order meta-stem cell (G) divides to produce a 1st-order stem cell $B$ and cell J1. The cell J1  stochastically loops back to activate the meta-stem cell G or it activates the first order linear stem cell $B$. The 1st-order stem cell $B$ divides into a semi-terminal cell C and a cell J2. The cell J2 stochastically activates either its parent stem cell $B$, the semi-terminal cell C or dedifferentiates to G.  The cell C also can stochastically dedifferentiate to any of the previous stem cell fates (G, B) or differentiate into the final terminal cell T. The network consists of two linked 1st-order linear stochastic networks. Its behavior approaches a deterministic 2nd-order geometric stem cell network as the probabilities $p$ and $q_1$ approach $1$.  Cancers controlled by such a network can go into spontaneous  remission because of the fact that for all points in all possible  paths in the network there is the possibility of reaching a terminal cell state.  However, because there is the possibility at J2 and C to dedifferentiate to earlier, upstream network states, the relative numbers of stem cells to terminal cells will be higher than in networks that have fewer dedifferentiation pathways. Note, too that the dedifferentiation pathways introduce several possibilities for exponential growth. Even if their probabilities  ($q_3$, $r_2$, $r_3$) are very small, any increase in these probabilities could have a significant destabilizing influence on the resulting tumor.

\subsection{Modeling deterministic stem cell networks with stochastic networks}
For stochastic stem cell networks the probability distribution over a network topology determines their dynamic properties.  If the stochastic network allows all possible paths in a set of linear, 2nd-order geometric and exponential networks, then the dynamic behavior of each is approximated as the corresponding embedded links are assigned probabilities that match existence or nonexistence of links in the particular network type,  For example, if a link exists in a network embedded in the stochastic network, then it assigned a probability close to 1 or 1 itself if we want an exact match.  If the link does not exist in the embedded network, it is assigned a probability close to 0 or 0 itself for an exact match.  Hence, the probability distribution distinguishes different network types that are embeddable in the topology or architecture of a given stochastic network. 

\subsection{Transformations of probability distributions change stochastic stem cell behavior}
If we allow the probability distribution over a stochastic network to change with time because of external factors (such as ultra violet radiation) then the same stochastic stem cell network may exhibit various proliferative phenotypes, appearing alternatively as being in remission, linear, geometric or exponential.  Thus, meta-probability functions are at work here that conditionally change the probability distribution. 

\section{Communication in stem cell networks}
We distinguish one-way, reactive communication protocols in stem cell networks and two-way, interactive communication protocols in stem cell networks. Reactive communication stem cell networks only react to external signals but do not send signals themselves.  Interactive communicating stem cell networks interact with the networks of other cells via signaling protocols by both sending signals and receiving signals to and from other cells. 

\subsection{Reactive communication stem cell networks}

This class of stem cell networks reacts to external and internal cell signals and, thereby, activating the stem cell network if the cell receives the appropriate signal. Unlike more complex cases where there are intercellular signaling protocols involved, here the communication and the reaction are one-sided.   A cell sends a signal to which the stem cell reacts by activating its stem cell network. 

\subsection{A reactive signal based geometric network architecture with 1st and 2nd order geometric potential}

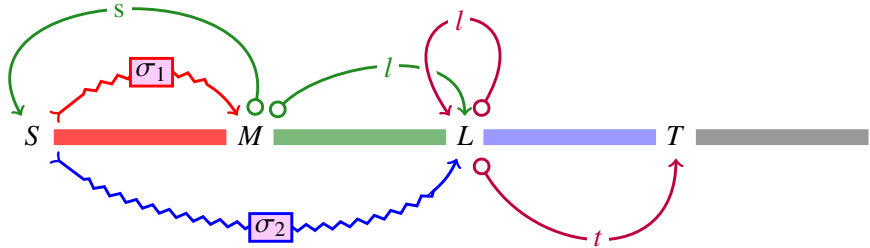
\begin{figure}[H]
\begin{tikzpicture}[style=mystyle]
\matrix (m) [matrix of math nodes, 
row sep=2em,
column sep=3em, 
text height=1.5ex, text depth=0.25ex]
{ \vphantom{a} && \vphantom{b} &&  \vphantom{c}  && \vphantom{c}  & & &\\
 S  && M &&  L  && T && \vphantom{c}  && \vphantom{c} \\ %TT\\
 \vphantom{a} &&  \vphantom{b} && \vphantom{B} &&  \vphantom{c}  && \vphantom{C} & \vphantom{d} & \vphantom{D}\\ };
  \path[\sigjump]
(m-2-1) edge [inPot1, red, receivestyle, cross line] node[receivesigstyle] {$\sigma_{1}$} (m-2-3) %S to J1 5
(m-2-1) edge [inPot2, in=-110, blue, receivestyle, cross line] node[receivesigstyle] {$\sigma_{2}$} (m-2-5); %S to 2

\path[\pot]
(m-2-3) edge [selfloop1, out=80, distance=4cm, green, cross line] node[nodedescr] {s} (m-2-1) %J1 to S 1
(m-2-3) edge [in2Pot1,green, cross line] node[nodedescr] {$l$} (m-2-5); %J1 to T 7
%\path[>=latex,>->,dotted]
\path[\pot]
(m-2-5) edge [selfloop1, out=60, in= 125, purple,cross line] node[nodedescr] {$l$} (m-2-5) %J2 to S 
(m-2-5) edge [inPot2, out=-60, in= -90, distance=3cm, purple,cross line] node[nodedescr] {$t$} (m-2-7); %J2 to T 
\path[solid,red!70, line width=6pt]
(m-2-1) edge (m-2-3);
\path[solid,green!60, line width=6pt]
(m-2-3) edge (m-2-5);
\path[solid,blue!40, line width=6pt]
(m-2-5) edge (m-2-7);
\path[solid,black!40, line width=6pt]
(m-2-7) edge (m-2-9);
\end{tikzpicture}
\caption{
    {\bf Network SigLSC: Reactive signal based geometric stem cell network with linear and meta-stem cell potential}.  On signal $\sigma_{1}$ one stem cell (M) divides to produce two cells (M and L).   M dedifferentiates back to the parent cell S.   On signal $\sigma_{2}$, the linear stem cell network L is activated. Once  L is activated it will continue to produce terminal cells T. Hence, this network is signal controlled. It will continue to produce M and hence S cells as long as it receives the signal $\sigma_{1}$.  S cells will differentiate to L cells on signal $\sigma_{2}$ stopping meta-stem cell production.  The signal $\sigma_{2}$ can thus eliminate all meta-stem cells.  Hence, it may be undesirable for systems that need to preserve their meta-stem cell line. Eliminating the signal  $\sigma_{2}$-path from S to L would insure that the supply of meta-stem cells can  always be increased as long as there are some meta-stem cells or S cells.  The terminal cell type T may be a single terminal cell or a terminal progenitor cell T* that generates a bounded number of further terminal cell types. 
  }
   \label{fig:SigLSC}
\end{figure}

\subsection{A pure reactive signal based geometric network architecture with 1st and 2nd order geometric potential}

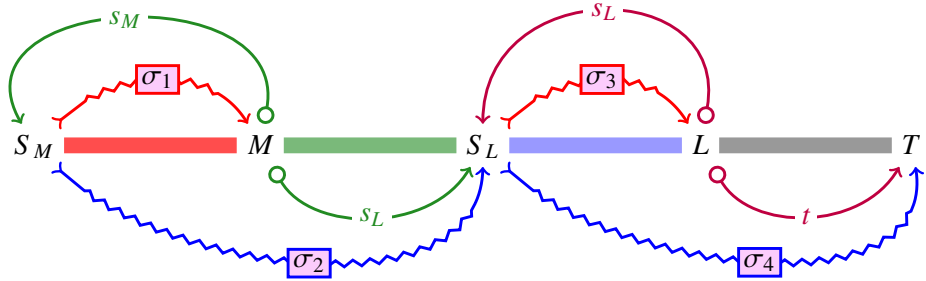
\begin{figure}[H]
\begin{tikzpicture}[style=mystyle]
\matrix (m) [matrix of math nodes, 
row sep=2em,
column sep=3em, 
text height=1.5ex, text depth=0.25ex]
{ \vphantom{a} && \vphantom{b} &&  \vphantom{c}  && \vphantom{c}  & & &\\
 S_{M}  && M &&  S_{L}  && L && T && \vphantom{c} \\ %TT\\
 \vphantom{a} &&  \vphantom{b} && \vphantom{B} &&  \vphantom{c}  && \vphantom{C} & \vphantom{d} & \vphantom{D}\\ };
  \path[\sigjump]
(m-2-1) edge [inPot1, red, receivestyle, cross line] node[receivesigstyle] {$\sigma_{1}$} (m-2-3) %S to J1 5
(m-2-1) edge [inPot2, in=-90, distance=4cm, blue, receivestyle, cross line] node[receivesigstyle] {$\sigma_{2}$} (m-2-5) %S to 2
(m-2-5) edge [inPot1, red, receivestyle, cross line] node[receivesigstyle] {$\sigma_{3}$} (m-2-7) %S to J1 5
(m-2-5) edge [inPot2, in=-80, distance=4cm, blue, receivestyle, cross line] node[receivesigstyle] {$\sigma_{4}$} (m-2-9); %S to 2

\path[\pot]
(m-2-3) edge [selfloop1, out=80, distance=4cm, green, cross line] node[nodedescr] {$s_{M}$} (m-2-1) %J1 to S 1
(m-2-3) edge [inPot2, out=-60, in=-120, green, cross line] node[nodedescr] {$s_{L}$} (m-2-5); %J1 to T 7
%\path[>=latex,>->,dotted]
\path[\pot]
(m-2-7) edge [selfloop1, out=80, in= 90, purple,cross line] node[nodedescr] {$s_{L}$} (m-2-5) %J2 to S 
(m-2-7) edge [inPot2, out=-60, in= -120, purple,cross line] node[nodedescr] {$t$} (m-2-9); %J2 to T 
\path[solid,red!70, line width=6pt]
(m-2-1) edge (m-2-3);
\path[solid,green!60, line width=6pt]
(m-2-3) edge (m-2-5);
\path[solid,blue!40, line width=6pt]
(m-2-5) edge (m-2-7);
\path[solid,black!40, line width=6pt]
(m-2-7) edge (m-2-9);
\end{tikzpicture}
\caption{
    {\bf Network SigG2SC: Pure reactive signal based geometric stem cell network with flexible geometric, linear and terminal potential}.  On signal $\sigma_{1}$, the signal dependent cell $S_{M}$ differentiates to the meta-stem cell (M) which then divides to produce two cells ($S_{M}$ and $S_{L}$).  Both $S_{M}$ and $S_{L}$ are controlled by equivalent signal dependent networks.  Their further behavior depends on the signals they receive. The signal combination $<\sigma_{1}, \sigma_{3}>$ produces a meta-stem cell network with 2nd order geometric potential. The signal combination $<\sigma_{1}, \sigma_{4}>$ and $<\sigma_{2}, \sigma_{3}>$ produce a 1st order stem cell network with linear potential.  The signal combination $<\sigma_{2}, \sigma_{3}>$ causes terminal differentiation stopping all cell production.  This example network shows that different signal combinations can change the architecture of signal based networks leading to distinct stem cell types with different generative potential.  The terminal cell type T may be a single terminal cell or a terminal progenitor cell T* that generates a bounded number of further terminal cell types. 
  }
   \label{fig:SigG2SC}
\end{figure}

\section{Stem cell networks with interactive communication}

% was \section{Communicating cancer networks in systems of cooperating social cells}
\label{sec:SigC}

A type of conditional cancer are {\em social cancers} that depend on cell signaling to be active.  This variety of cancer occurs when the genomic network interacts with receptors, signals and signal transduction pathways.  In that case we can have what might be called social cancers.  For example, a cell $A$ will signal $\alpha$ to cell $B$ and differentiates to A1, $B$ on receiving the signal divides into B1 and B1.  B1 sends a signal $\beta$ back to $A$ and then B1 dedifferentiates to $B$ (its network loops back).  When A1 receives $\beta$ its network loops back  into state $A$.  Now the process repeats with the $A$ cells sending signals $\alpha$ to $B$ cells.  One can see that this process is potentially exponential as long as we have sufficient cells of type $A$ with sufficient signal capacity to continue to activate all the developing $B$ cells.  

With conditional cancers that depend on social communication, the growth rate of the cancer will depend on the accessibility of the signal.  If the receiver of the signal requires direct contact with the sender then even if the cytogenic cell contains a conditional exponential network, since the signal may not be received by those daughter cells the exponential potential may not be realized. Furthermore, if it is a linear network then if the passive daughter cell is interposed between the sender and the receiving cytogenic cell then growth will stop after several divisions when the signal no loner can reach the cytogenic cells.  If later, because of physical pressure or other conditions, the cytogenic cell is again in close enough proximity to the sender then the cancer can start again. 

\subsection{Interactive signaling mono generative networks}
\label{sec:SigC1}
In the figure below we emphasize the communication links.   The network has one cell only signaling and the other cell signaling and conditionally dividing, however it  leaves open whether the cytogenic subnetwork is linear or exponential. The more detailed communication networks and their properties will described in the sections below. 

%was N13, was NSL
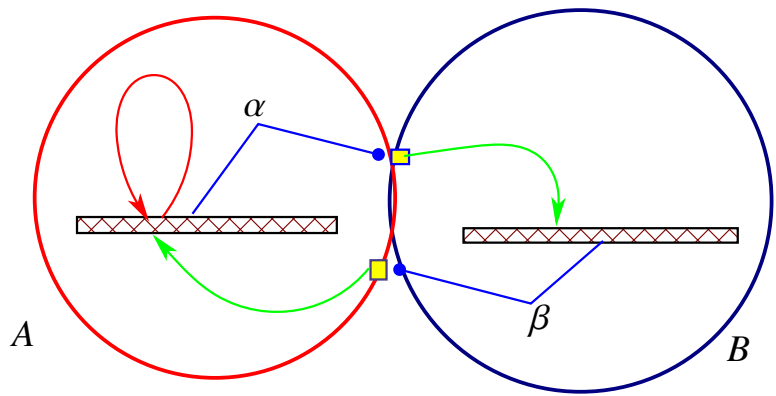
\begin{figure}[H]
\centering
\begin{tikzpicture}%scale= \PicSize, Crit. Scale does not work
[x=1.00mm, y=1.00mm, inner xsep=0pt, inner ysep=0pt, outer xsep=0pt, outer ysep=0pt]
\path[line width=0mm] (51.00,42.86) rectangle +(107.34,54.79);
\definecolor{L}{rgb}{0,0,0.502}
\path[line width= \CellWallThickness, draw=L] (128.68,70.26) circle (25.39mm);
\definecolor{L}{rgb}{1,0,0}
\path[line width= \CellWallThickness, draw=L] (80.06,70.68) circle (23.96mm);
\definecolor{L}{rgb}{0.502,0,0}
\path[line width=0.15mm, draw=L]  (64.12,65.99) -- (61.98,68.12) (66.94,65.99) -- (64.81,68.12) (69.77,65.99) -- (67.64,68.12) (72.60,65.99) -- (70.47,68.12) (75.43,65.99) -- (73.30,68.12) (78.26,65.99) -- (76.13,68.12) (81.09,65.99) -- (78.95,68.12) (83.92,65.99) -- (81.78,68.12) (86.74,65.99) -- (84.61,68.12) (89.57,65.99) -- (87.44,68.12) (92.40,65.99) -- (90.27,68.12) (95.23,65.99) -- (93.10,68.12) (96.27,67.78) -- (95.93,68.12);
\path[line width=0.15mm, draw=L]  (61.73,67.38) -- (62.47,68.12) (63.16,65.99) -- (65.30,68.12) (65.99,65.99) -- (68.12,68.12) (68.82,65.99) -- (70.95,68.12) (71.65,65.99) -- (73.78,68.12) (74.48,65.99) -- (76.61,68.12) (77.31,65.99) -- (79.44,68.12) (80.13,65.99) -- (82.27,68.12) (82.96,65.99) -- (85.09,68.12) (85.79,65.99) -- (87.92,68.12) (88.62,65.99) -- (90.75,68.12) (91.45,65.99) -- (93.58,68.12) (94.28,65.99) -- (96.27,67.98);
\definecolor{L}{rgb}{0,0,0}
\path[line width=0.30mm, draw=L] (61.73,65.99) rectangle +(34.54,2.13);
\definecolor{L}{rgb}{0.502,0,0}
\path[line width=0.15mm, draw=L]  (113.48,64.71) -- (113.11,65.08) (116.31,64.71) -- (114.39,66.63) (119.14,64.71) -- (117.22,66.63) (121.96,64.71) -- (120.05,66.63) (124.79,64.71) -- (122.87,66.63) (127.62,64.71) -- (125.70,66.63) (130.45,64.71) -- (128.53,66.63) (133.28,64.71) -- (131.36,66.63) (136.11,64.71) -- (134.19,66.63) (138.93,64.71) -- (137.02,66.63) (141.76,64.71) -- (139.84,66.63) (144.59,64.71) -- (142.67,66.63) (147.42,64.71) -- (145.50,66.63) (149.57,65.39) -- (148.33,66.63);
\path[line width=0.15mm, draw=L]  (113.11,65.03) -- (114.71,66.63) (115.62,64.71) -- (117.54,66.63) (118.45,64.71) -- (120.37,66.63) (121.28,64.71) -- (123.20,66.63) (124.11,64.71) -- (126.03,66.63) (126.94,64.71) -- (128.86,66.63) (129.77,64.71) -- (131.68,66.63) (132.59,64.71) -- (134.51,66.63) (135.42,64.71) -- (137.34,66.63) (138.25,64.71) -- (140.17,66.63) (141.08,64.71) -- (143.00,66.63) (143.91,64.71) -- (145.83,66.63) (146.74,64.71) -- (148.66,66.63) (149.57,64.71) -- (149.57,64.72);
\definecolor{L}{rgb}{0,0,0}
\path[line width=0.30mm, draw=L] (113.11,64.71) rectangle +(36.46,1.92);
\definecolor{L}{rgb}{0,0,1}
\definecolor{F}{rgb}{1,1,0}
\path[line width=0.30mm, draw=L, fill=F] (103.52,75.16) rectangle +(2.35,1.92);
\definecolor{L}{rgb}{0,1,0}
\path[line width=0.30mm, draw=L] (105.22,76.23) .. controls (108.27,76.71) and (111.33,77.14) .. (114.39,77.51) .. controls (117.44,77.87) and (120.71,78.10) .. (123.13,76.23) .. controls (124.60,75.10) and (125.46,73.37) .. (125.69,71.54) .. controls (125.80,70.69) and (125.78,69.83) .. (125.69,68.98) .. controls (125.64,68.48) and (125.57,67.98) .. (125.48,67.48);
\definecolor{F}{rgb}{0,1,0}
\path[line width=0.30mm, draw=L, fill=F] (125.48,67.48) -- (126.40,70.22) -- (125.65,69.58) -- (125.01,70.33) -- (125.48,67.48) -- cycle;
\definecolor{L}{rgb}{0.282,0.239,0.545}
\definecolor{F}{rgb}{1,1,0}
\path[line width=0.30mm, draw=L, fill=F] (100.75,59.81) rectangle +(2.13,2.56);
\definecolor{L}{rgb}{0,1,0}
\path[line width=0.30mm, draw=L] (100.53,61.30) .. controls (95.60,55.27) and (87.00,53.73) .. (80.28,57.68) .. controls (77.79,59.14) and (75.81,61.28) .. (73.88,63.43) .. controls (73.31,64.07) and (72.74,64.71) .. (72.17,65.35);
\definecolor{F}{rgb}{0,1,0}
\path[line width=0.30mm, draw=L, fill=F] (72.17,65.35) -- (73.52,62.80) -- (73.57,63.78) -- (74.56,63.73) -- (72.17,65.35) -- cycle;
\definecolor{L}{rgb}{0,0,1}
\path[line width=0.30mm, draw=L] (131.66,64.93) -- (122.07,56.61) -- (103.94,61.30);
\definecolor{F}{rgb}{0,0,1}
\path[line width=0.30mm, draw=L, fill=F] (104.62,61.13) circle (0.70mm);
\path[line width=0.30mm, draw=L] (77.08,68.55) -- (85.82,80.49) -- (101.81,76.44);
\path[line width=0.30mm, draw=L, fill=F] (101.81,76.44) circle (0.70mm);
\draw(83.90,81.56) node[anchor=base west]{\fontsize{14.23}{17.07}\selectfont $\alpha$};
\draw(122.07,53.20) node[anchor=base west]{\fontsize{14.23}{17.07}\selectfont $\beta$};
\draw(53.00,51.00) node[anchor=base west]{\fontsize{14.23}{17.07}\selectfont $A$};
\draw(148.00,49.00) node[anchor=base west]{\fontsize{14.23}{17.07}\selectfont $B$};
\definecolor{L}{rgb}{1,0,0}
\path[line width=0.30mm, draw=L] (73.00,68.00) .. controls (75.71,71.40) and (77.13,75.65) .. (77.00,80.00) .. controls (76.89,83.52) and (75.18,86.97) .. (72.00,87.00) .. controls (68.49,87.03) and (66.73,83.02) .. (67.00,79.00) .. controls (67.26,75.03) and (68.65,71.21) .. (71.00,68.00);
\definecolor{F}{rgb}{1,0,0}
\path[line width=0.30mm, draw=L, fill=F] (71.00,68.00) -- (70.15,70.76) -- (69.91,69.80) -- (68.95,70.03) -- (71.00,68.00) -- cycle;
\end{tikzpicture}%
%\fi
\caption{
{\bf A social signaling cancer loop where cell $A$ is cancerous} The cancer is social and conditional in that it depends on a signal from $B$ to divide.  Since the network shows only one loop, it leaves open the potential of the other daughter cell. Hence, it leaves open whether the cytogenic network has linear or exponential or some other potential.
}
\label{fig:SigC1}
\end{figure}

An {\em in vivo} instance of this network type is seen in bone cancer where a signal from the adjacent tissue is required for the other to become cancerous, for example, see (Logothetis~\cite{Logothetis2005}).  

%???This figure is still ambiguous.  

\subsection{Interactive signaling mono-linear network architecture}

The following network controls two cells types A and B that communicate by cell signaling.  The cell A divides only when it receives the signal $\beta$ from cell B.  The cell B only sends its signal after it has received the signal $\alpha$ from cell A. 

 \label{sec:SigL1}
 
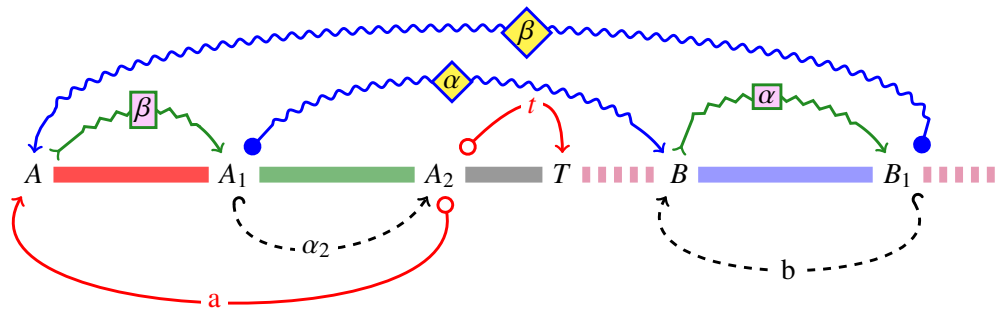
\begin{figure}[H]
\begin{tikzpicture}[style=mystyle]
\matrix (m) [matrix of math nodes, 
row sep=3em,
column sep=\ColSepNarrow, %\ColSepTight, 
text height=1.5ex, text depth=0.25ex]
{ \vphantom{a} && \vphantom{b} &&  \vphantom{c}  && \vphantom{c}  &&  \\
 A  && A_{1} &&  A_{2}  &T & B &\vphantom{c}&  B_{1}  & \vphantom{c} \\ %TT\\
 \vphantom{a} &&  \vphantom{b} && \vphantom{B} &&  \vphantom{c}  && \vphantom{C} & \vphantom{d} & \vphantom{D}\\ };
  \path[\sigjump]
(m-2-1) edge [inPot1, green, receivestyle, cross line] node[receivesigstyle] {$\beta$} (m-2-3); %S to J1 5
 \path[\sigjump]
(m-2-7) edge [inPot1, out=90, green, receivestyle, cross line] node[receivesigstyle] {$\alpha$} (m-2-9); 

\path[\pot]
(m-2-5) edge [selfloop2, out=-80, distance=4cm, red, cross line] node[nodedescr] {a} (m-2-1) %J1 to S 1
(m-2-5) edge [in2Pot1,red, cross line] node[nodedescr] {$t$} (m-2-6); %J1 to T 7
\path[\jump, dashed]%Differentiate after send signal
(m-2-3) edge [selfloop2, out=-90, in=-120, distance=2cm, black, cross line] node[nodedescr] {$\alpha_{2}$} (m-2-5)
(m-2-9) edge [selfloop2, out=-60, distance=3cm, black, cross line] node[nodedescr] {b} (m-2-7); 
\path[\sendsig]
%(m-2-5) edge [selfloop1, out=60, in= 125, purple,cross line] node[nodedescr] {$l$} (m-2-5) %J2 to S 
(m-2-3) edge [ in2Pot1, out= 60, in=125, blue,snakesendstyle,cross line] node[sendsigstyle] {$\alpha$} (m-2-7) %A1 sends sig to B
(m-2-9) edge [ in2Pot1, distance=5cm, blue,snakesendstyle,cross line] node[sendsigstyle] {$\beta$} (m-2-1);%B1 sends sig to A

\path[solid,red!70, line width=6pt]
(m-2-1) edge (m-2-3);
\path[solid,green!60, line width=6pt]
(m-2-3) edge (m-2-5);
\path[solid,black!40, line width=6pt]
(m-2-5) edge (m-2-6);
\path[dashed,purple!40, line width=6pt]
(m-2-6) edge (m-2-7);
\path[solid,blue!40, line width=6pt]
(m-2-7) edge (m-2-9);
\path[dashed,purple!40, line width=6pt]
(m-2-9) edge (m-2-10);
\end{tikzpicture}
\caption{
    {\bf Network SigL1: An interactive signal based stem cell network with linear potential.}  Cell A only divides if it receives a signal $\beta$ from cell B.  Cell B only sends a signal if it receives a signal from cell A.  On signal $\beta$ the cell A enters control state A$_{1}$.  Then A$_{1}$  sends a signal $\alpha$ to B. Subsequently A$_{1}$ switches to control state A$_{2}$ whereupon it divides to produce two cells (the dedifferentiated parent cell type A and a terminal cell T).  Hence, one of the daughter cells of  A$_{2}$ dedifferentiates back to the parent cell A.  The cell type B does not divide in response to the signal from A.  Instead, B responds to the signal $\alpha$ by switching to state B$_{1}$ and sending a signal $\beta$ to A.   After that B$_{1}$ dedifferentiates back to B.  The cell A will continue to divide as long the signal loop is maintained. {\bf Notation}:  The blue and green squiggly arrows between two cell states indicate cell signaling. The blue arrow $A_{1} \rightsquigarrow B$ labeled with the triangular $\alpha$ means that a cell in state $A_{1}$ sends a signal $\alpha$ that is received by a cell in state $B$.  The green arrow $B \rightsquigarrow B_{1}$ labeled with the rectangular $\alpha$ means that a cell in state $B$  on receiving the signal $\alpha$ differentiates or jumps to state $B_{1}$. 
  }
   \label{fig:SigL1}
\end{figure}

\subsection{Dual cancer signaling networks}
\label{sec:SigC2}

A the interacting between two cooperative cell types, called partners, is driven by a network where each partner has a separate role driven by a different subnetwork with signaling driving their actions.  In the figure below we emphasize the communication links.   The network has both cells signaling and conditionally dividing, however it  leaves open whether the network is linear or exponential. The more detailed communication networks and their properties will described in the following sections. 
 
\begin{figure}[H]
\centering
\begin{tikzpicture}%scale= \PicSize, 
[x=1.00mm, y=1.00mm, inner xsep=0pt, inner ysep=0pt, outer xsep=0pt, outer ysep=0pt]
\path[line width=0mm] (51.00,42.86) rectangle +(107.34,54.79);
\definecolor{L}{rgb}{0,0,0.502}
\path[line width= \CellWallThickness, draw=L] (128.68,70.26) circle (25.39mm);
\definecolor{L}{rgb}{1,0,0}
\path[line width= \CellWallThickness, draw=L] (80.06,70.68) circle (23.96mm);
\definecolor{L}{rgb}{0.502,0,0}
\path[line width=0.15mm, draw=L]  (64.12,65.99) -- (61.98,68.12) (66.94,65.99) -- (64.81,68.12) (69.77,65.99) -- (67.64,68.12) (72.60,65.99) -- (70.47,68.12) (75.43,65.99) -- (73.30,68.12) (78.26,65.99) -- (76.13,68.12) (81.09,65.99) -- (78.95,68.12) (83.92,65.99) -- (81.78,68.12) (86.74,65.99) -- (84.61,68.12) (89.57,65.99) -- (87.44,68.12) (92.40,65.99) -- (90.27,68.12) (95.23,65.99) -- (93.10,68.12) (96.27,67.78) -- (95.93,68.12);
\path[line width=0.15mm, draw=L]  (61.73,67.38) -- (62.47,68.12) (63.16,65.99) -- (65.30,68.12) (65.99,65.99) -- (68.12,68.12) (68.82,65.99) -- (70.95,68.12) (71.65,65.99) -- (73.78,68.12) (74.48,65.99) -- (76.61,68.12) (77.31,65.99) -- (79.44,68.12) (80.13,65.99) -- (82.27,68.12) (82.96,65.99) -- (85.09,68.12) (85.79,65.99) -- (87.92,68.12) (88.62,65.99) -- (90.75,68.12) (91.45,65.99) -- (93.58,68.12) (94.28,65.99) -- (96.27,67.98);
\definecolor{L}{rgb}{0,0,0}
\path[line width=0.30mm, draw=L] (61.73,65.99) rectangle +(34.54,2.13);
\definecolor{L}{rgb}{0.502,0,0}
\path[line width=0.15mm, draw=L]  (113.48,64.71) -- (113.11,65.08) (116.31,64.71) -- (114.39,66.63) (119.14,64.71) -- (117.22,66.63) (121.96,64.71) -- (120.05,66.63) (124.79,64.71) -- (122.87,66.63) (127.62,64.71) -- (125.70,66.63) (130.45,64.71) -- (128.53,66.63) (133.28,64.71) -- (131.36,66.63) (136.11,64.71) -- (134.19,66.63) (138.93,64.71) -- (137.02,66.63) (141.76,64.71) -- (139.84,66.63) (144.59,64.71) -- (142.67,66.63) (147.42,64.71) -- (145.50,66.63) (149.57,65.39) -- (148.33,66.63);
\path[line width=0.15mm, draw=L]  (113.11,65.03) -- (114.71,66.63) (115.62,64.71) -- (117.54,66.63) (118.45,64.71) -- (120.37,66.63) (121.28,64.71) -- (123.20,66.63) (124.11,64.71) -- (126.03,66.63) (126.94,64.71) -- (128.86,66.63) (129.77,64.71) -- (131.68,66.63) (132.59,64.71) -- (134.51,66.63) (135.42,64.71) -- (137.34,66.63) (138.25,64.71) -- (140.17,66.63) (141.08,64.71) -- (143.00,66.63) (143.91,64.71) -- (145.83,66.63) (146.74,64.71) -- (148.66,66.63) (149.57,64.71) -- (149.57,64.72);
\definecolor{L}{rgb}{0,0,0}
\path[line width=0.30mm, draw=L] (113.11,64.71) rectangle +(36.46,1.92);
\definecolor{L}{rgb}{0,0,1}
\definecolor{F}{rgb}{1,1,0}
\path[line width=0.30mm, draw=L, fill=F] (103.52,75.16) rectangle +(2.35,1.92);
\definecolor{L}{rgb}{0,1,0}
\path[line width=0.30mm, draw=L] (105.22,76.23) .. controls (108.27,76.71) and (111.33,77.14) .. (114.39,77.51) .. controls (117.44,77.87) and (120.71,78.10) .. (123.13,76.23) .. controls (124.60,75.10) and (125.46,73.37) .. (125.69,71.54) .. controls (125.80,70.69) and (125.78,69.83) .. (125.69,68.98) .. controls (125.64,68.48) and (125.57,67.98) .. (125.48,67.48);
\definecolor{F}{rgb}{0,1,0}
\path[line width=0.30mm, draw=L, fill=F] (125.48,67.48) -- (126.40,70.22) -- (125.65,69.58) -- (125.01,70.33) -- (125.48,67.48) -- cycle;
\definecolor{L}{rgb}{0.282,0.239,0.545}
\definecolor{F}{rgb}{1,1,0}
\path[line width=0.30mm, draw=L, fill=F] (100.75,59.81) rectangle +(2.13,2.56);
\definecolor{L}{rgb}{0,1,0}
\path[line width=0.30mm, draw=L] (100.53,61.30) .. controls (95.60,55.27) and (87.00,53.73) .. (80.28,57.68) .. controls (77.79,59.14) and (75.81,61.28) .. (73.88,63.43) .. controls (73.31,64.07) and (72.74,64.71) .. (72.17,65.35);
\definecolor{F}{rgb}{0,1,0}
\path[line width=0.30mm, draw=L, fill=F] (72.17,65.35) -- (73.52,62.80) -- (73.57,63.78) -- (74.56,63.73) -- (72.17,65.35) -- cycle;
\definecolor{L}{rgb}{0,0,1}
\path[line width=0.30mm, draw=L] (131.66,64.93) -- (122.07,56.61) -- (103.94,61.30);
\definecolor{F}{rgb}{0,0,1}
\path[line width=0.30mm, draw=L, fill=F] (104.62,61.13) circle (0.70mm);
\path[line width=0.30mm, draw=L] (77.08,68.55) -- (85.82,80.49) -- (101.81,76.44);
\path[line width=0.30mm, draw=L, fill=F] (101.81,76.44) circle (0.70mm);
\draw(83.90,81.56) node[anchor=base west]{\fontsize{14.23}{17.07}\selectfont $\alpha$};
\draw(122.07,53.20) node[anchor=base west]{\fontsize{14.23}{17.07}\selectfont $\beta$};
\draw(53.00,51.00) node[anchor=base west]{\fontsize{14.23}{17.07}\selectfont $A$};
\draw(148.00,49.00) node[anchor=base west]{\fontsize{14.23}{17.07}\selectfont $B$};
\definecolor{L}{rgb}{1,0,0}
\path[line width=0.30mm, draw=L] (128.00,67.00) .. controls (127.86,70.35) and (127.53,73.69) .. (127.00,77.00) .. controls (126.34,81.15) and (126.38,85.81) .. (130.00,87.00) .. controls (134.48,88.47) and (138.37,83.08) .. (137.00,77.00) .. controls (136.07,72.88) and (133.55,69.29) .. (130.00,67.00);
\definecolor{F}{rgb}{1,0,0}
\path[line width=0.30mm, draw=L, fill=F] (130.00,67.00) -- (132.61,68.23) -- (131.62,68.33) -- (131.72,69.32) -- (130.00,67.00) -- cycle;
\path[line width=0.30mm, draw=L] (73.00,68.00) .. controls (75.71,71.40) and (77.13,75.65) .. (77.00,80.00) .. controls (76.89,83.52) and (75.18,86.97) .. (72.00,87.00) .. controls (68.49,87.03) and (66.73,83.02) .. (67.00,79.00) .. controls (67.26,75.03) and (68.65,71.21) .. (71.00,68.00);
\path[line width=0.30mm, draw=L, fill=F] (71.00,68.00) -- (70.15,70.76) -- (69.91,69.80) -- (68.95,70.03) -- (71.00,68.00) -- cycle;
\end{tikzpicture}%
%\fi
\caption{
{\bf A dual cancer signaling network} where both communicating partner cells A and B are cancerous. The cancer is social and conditional in that each cell depends on a signal from the other to divide. The rate and extent of growth depends on whether the properties of the cancer subnetworks of the cell types A and B.
}
\label{fig:SigC2}
\end{figure}
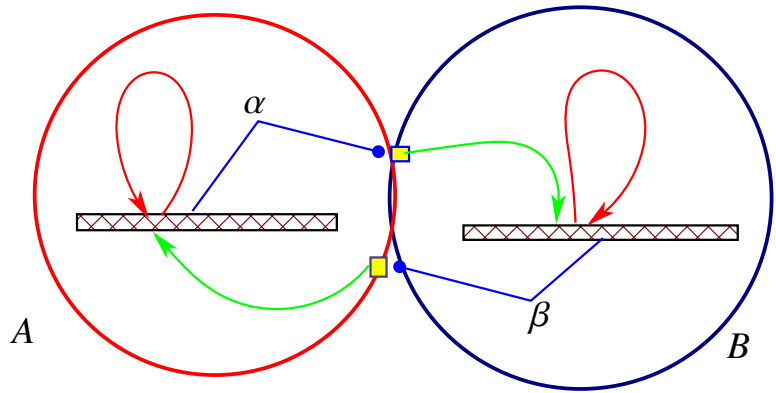

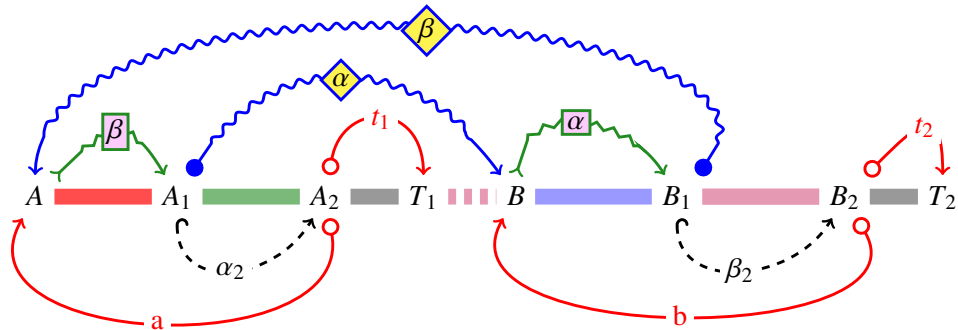
\begin{figure}[H]
\begin{tikzpicture}[style=mystyle]
\matrix (m) [matrix of math nodes, 
row sep=3em,
column sep=\ColSepTight, 
text height=1.5ex, text depth=0.25ex]
{ \vphantom{a} && \vphantom{b} &&  \vphantom{c}  && \vphantom{c}  && \vphantom{c}&&\vphantom{c}&\vphantom{c}&\vphantom{c}& \\
 A  && A_{1} &&  A_{2}  &T_{1}& B &\vphantom{c}& B_{1}  & \vphantom{c}& B_{2} &T_{2}\\ %TT\\
 \vphantom{a} &&  \vphantom{b} && \vphantom{B} &&  \vphantom{c}  && \vphantom{C} & \vphantom{d} & \vphantom{D}\\ };
  \path[\sigjump]
(m-2-1) edge [inPot1, green, receivestyle, cross line] node[receivesigstyle] {$\beta$} (m-2-3); %S to J1 5
 \path[\sigjump]
(m-2-7) edge [inPot1, out=90, green, receivestyle, cross line] node[receivesigstyle] {$\alpha$} (m-2-9); 

\path[\pot]
(m-2-5) edge [selfloop2, out=-80, distance=4cm, red, cross line] node[nodedescr] {a} (m-2-1) %J1 to S 1
(m-2-5) edge [in2Pot1,out=80, red, cross line] node[nodedescr] {$t_{1}$} (m-2-6); %J1 to T 7
\path[\pot]
(m-2-11) edge [in2Pot1, red, in=90, cross line] node[nodedescr] {$t_{2}$} (m-2-12)
(m-2-11) edge [selfloop2, out=-60, distance=4cm, red, cross line] node[nodedescr] {b} (m-2-7); 
\path[\jump, dashed]%Differentiate after send signal
(m-2-3) edge [selfloop2, out=-90, in=-120, distance=2cm, black, cross line] node[nodedescr] {$\alpha_{2}$} (m-2-5)
(m-2-9) edge [selfloop2, out=-90, in=-120,distance=2cm, black, cross line] node[nodedescr] {$\beta_{2}$} (m-2-11); 
\path[\sendsig] %fill=white,
%(m-2-5) edge [selfloop1, out=60, in= 125, purple,cross line] node[nodedescr] {$l$} (m-2-5) %J2 to S 
(m-2-3) edge [ in2Pot1, out= 60, in=125, distance=4cm, blue,  snakesendstyle,cross line] node[sendsigstyle] {$\alpha$} (m-2-7) %J2 to T 
(m-2-9) edge [ in2Pot1, distance=6cm, blue,snakesendstyle,cross line] node[sendsigstyle] {$\beta$} (m-2-1);
\path[solid,red!70, line width=6pt]
(m-2-1) edge (m-2-3);
\path[solid,green!60, line width=6pt]
(m-2-3) edge (m-2-5);
\path[solid,black!40, line width=6pt]
(m-2-5) edge (m-2-6);
\path[dashed,purple!40, line width=6pt]
(m-2-6) edge (m-2-7);
\path[solid,blue!40, line width=6pt]
(m-2-7) edge (m-2-9);
\path[solid,purple!40, line width=6pt]
(m-2-9) edge (m-2-11);
\path[solid,purple!40, line width=6pt]
(m-2-9) edge (m-2-11);
\path[solid,black!40, line width=6pt]
(m-2-11) edge (m-2-12);
\end{tikzpicture}
\caption{
    {\bf Network SigL2:  A signal based dual linear network with linear potential.} Two cells A and B have equivalent signaling protocols and behavior. Cell A only divides if it receives a signal $\beta$. Upon receiving the signal $\beta$, cell A changes to control state A$_{1}$.  Prior to dividing A$_{1}$ sends a signal $\alpha$ to B.  Then A$_{1}$ changes to control state A$_{2}$ whereupon it divides to produce two cells (the dedifferentiated parent cell type A and a terminal cell type T$_{1}$).  The communication protocol for B is similar to A.  Cell B only divides if it receives a signal $\alpha$.  Upon receiving the signal $\alpha$ cell B differentiates to control state B$_{1}$ which then sends a signal $\beta$ to A.  After that, cell B$_{1}$  switches to control state B$_{2}$.  Then cell B$_{2}$ divides to produce two different cell types (the dedifferentiated parent cell type B and the terminal cell type T$_{2}$).  The cell types A and B will continue to divide as long the signal loop interaction protocol is sustained. 
  }
   \label{fig:SigL2}
\end{figure}

What we have here is a signal loop that is essential for the whole cancer network.  The dual cancer signaling network in  \autoref{fig:SigC2} consists of two subnetworks, and A-subnetwork for cell type A and a B-subnetwork for cell type B. Once activated, the A-subnetwork (e.g., \autoref{fig:SigL2})  causes a signal $\alpha$ to be sent to cell $B$  followed by activation of subnetwork A$_{2}$ whereupon the cell divides into two daughter cells at least one of which enters the signal receptor state A which enables the cell to receive signals of type $\beta$.  The signaling partner subnetwork of cell type B is activated by a signal transduction cascade initiated by the receipt of  signal $\alpha$.  The B-subnetwork in state B$_{1}$ directs the sending of a signal $\beta$ to cell A.  After sending the signal, the B-subnetwork enters state $B_{2}$ that directs $B$ to divide into two daughter cells, one or both of which loop back, to enter the receptor state B.   

Clinically, we would observe depends on nature of the cytogenic subnetworks, as well as the developing morphology of the tumor in the tissue.  The cells A and B only proliferate if have a neighbor that is a signaling partner.  Thus, the growth of B cells depends on being adjacent to A cells and vice versa.  We would observe the stopping of cell proliferation if either the A or the B cells are removed or some other agent interferes with the signaling loop.  The latter can be effected either by inhibiting the receptor for $\alpha$ or $\beta$, the signal $\alpha$ or $\beta$, or by interfering with either the$\alpha$or $\beta$ signal transduction pathway.  Alternatively, one could interfere with the genome by inhibiting the areas responsible for initiation of signaling or the inhibiting the areas responsible for cellular division.  One can see there are many potential areas where one can break the cancer loop.  In actual in vivo systems one would choose that with minimal risk of side effects.  

We now investigate some of the particular types of dual cancer signaling networks. 

\subsection{A interactive signaling dual linear network architecture}
\label{sec:SigL2}

\section{Hybrid communicative and stochastic stem cell networks}

It is possible in principle for a stem cell network to become stochastic as a result of communication.  In other words, it is conditionally stochastic in response to cell signaling.  

So too a stochastic process can lead to cell signaling which can lead to the, possibly stochastic, activation of another developmental network.

\section{Discussion: Stem cell concepts}

\subsection{Formal and informal definitions of transpotency}
This section can be skipped or read later since the graphical formalism with which we will represent stem cell networks may be more intuitive. 

Given a cell type A, let $\theta_{1}\ldots\theta_{k}$ be the minimal number of external operations $\theta_{i}$ required to transform a cell type A into a cell type B. Then $B = \theta_{1}\ldots\theta_{k}A$. Then we say that cell type A is {\em k-distant} from cell type B and B is a {\em k-transform} of A. Note, if B is a k-transform of A that does not necessarily imply that A is a k-transform of B since the operations may be irreversible.  B is a {\em k-transform} of A if there exist a number of operations $\theta_{1}\ldots\theta_{k}$ such that $B = \theta_{1}\ldots\theta_{k}A$. B is a {\em transform} of A if there exists some $k$ operators such that B is a k-transform of A. The transpotency of a cell type A is then definable as the set of all ordered pairs $<B_{i}, k_{i}>$ where  $B_{i}$ is a $k_{i}$-transform of A. The {\em operational distance} from A to B is number of operators  $\theta_{i}, i = 1 \ldots k$ required to transform A to some type B.  

Another aspect of transpotency has to do with stochasticity. It can take many attempts to transform a cell type A into B. The operator sequence has to be applied to many cells of type A before one of them changes into type B.  Thus, the sequence of operators $\theta_{1}\ldots\theta_{k}$ applied to A only results in B with some probability $p$.  Thus the transformative distance of A from B is not just a one dimensional function of the operational distance, but also should include the probability that an application of the operators to A will result in B. The lower the transform probability $p$ the farther A is to B in transformational space.  And, if the probability $p = 0$ then B is {\em inaccessible} from A. If the probability $p=1$ then there is a deterministic relationship that insures $\theta_{1}\ldots\theta_{k}A \rightarrow B$, where $\rightarrow$ means ``yields'' or ``transforms into''.  If $0 \leq p \leq 1$ then $\theta_{1}\ldots\theta_{k}A \stackrel{p}{\rightarrow} B$, where $\stackrel{p}{\rightarrow}$ means ``The operations $\theta_{1}\ldots\theta_{k}$ on A {\em transform A into B with probability} $p$''.

\subsection{Problems with definitions of transpotency}
Clearly as experimental procedures evolve the operations required to transform A into B will change and the operational distance $k$ may decrease.  Still this formalization makes it clear that the larger operational distance the more resistant A is to becoming B.  It is also evident that artificial transpotency need not be correlated with natural transpotency. 

Moreover, the existence of operations via transcription factors that can dedifferentiate a cell to a transpotent totipotent or pluripotent cell and then induce that cell to differentiate become a specific cell type break down the difference between transpotent cells and normal cells making the definition of transpotent stem cell all inclusive and, thereby, useless.  

In normal development, the {\em natural transpotency} of a cell depends on the natural conditions and multicellular context of in which the cell finds itself, e.g., cell signaling in a multicellular context.  {\em Artificial transpotency} is experimentally induced transpotency resulting from viral, chemical or physical extraneous, experimental impingements. 

The degree of transpotency of a cell is not precisely defined because it is ultimately an operational definition that depends on what operations are performed on cells to make them differentiate into other cells. And since a few transcription factors can make some cells dedifferentiate into stem cells  the class of artificial transpotent cells has no clear boundaries. Hence, if sufficiently powerful experimental operations are allowed that can change any cell differentiation state to almost any other state, then this calls into question the whole concept of artificial transpotent, non self-renewing stem cells, since their transpotency can no longer operationally distinguished from normal cells . The only distinguishing feature would be the number and type of operators that are required to convert one cell type into another. The greater the number of steps and the more difficult the operations are the more resistant a cell is to transformation.
%****end of formal definitions of stem cell transpotency

\subsection{Conceptual ambiguities}

There is a conceptual ambiguity as to what stem cells are~\footnote{I am not referring to the controversy as to whether stem cells are deterministic or have a stochastic dedifferentiation potential (Dick~\cite{Dick2008}).}. One the one hand stem cells are viewed as having the capacity of endless self-renewal, and, other the other hand, stem cells are viewed as having the capacity to differentiate into multiple cell types. These two views of stem cells are distinct and refer to independent properties of cells. They are different conceptions of what stem cells are and what they can do.  The difference is in the capacity of a cell to generate multiple progeny while maintaining its original, unchanged cell type, versus the capacity of a cell to change its differentiation state by transforming into different possible cell types.  

The two attributes of self-renewal and transformative potential can coexist in the same cell.  However, if a stem cell has both capacities and differentiates into some other cell type it may loose its capacity to self-renew.  

Let us call stem cells that can endlessly generate other cells while maintaining their own cell type, {\em iteropotent stem cells}.  While stem cells that have the capacity to differentiate or transform into various cell types {\em transpotent stem cells}. 
% [*** see {\em Stem cell potency} below, dynamorphic (dyna=power), metamorphic, metapotent, epipotent, heteropoten, hyperpotent, metamorphic, transpotent,  iteropotent, aniteropotent (iter=repeat), isopotent (iso=same), anisopotent, suipotent, ansuitpotent (sui=self), egopotent, self-renewing, non-self-renewing***] 

Yet, a third conception of stem cells combines the self-renewal with a transformative capacity by distributing the capacities over parent and daughter cells.  It restricts the capacity of self-renewal to the parent cell and the transformative potential to daughter cells. In this version, stem cells have the capacity of endless self-renewal while they generate daughter cells that may, but need not, have the capacity to differentiate into multiple cell types. 

In summary, stem cells may be purely iteropotent having the capacity of self-renewal but not have transformative potency (totipotent, pluripotent, multipotent or oligopotent), or they may be both self-renewing and have transformative potency but they may loose their self-renewing capacity once they differentiate, or they may be strictly self-renewing but their daughter cells may have transformative potency.  A cell with transformative potency may generate a stem cell. A stem cell may generate another stem cell. As we will see, all these conceptions of stem cells are realizable by different developmental stem cell networks.    

\subsection{Iteropotent versus transpotent stem cells}
For example, a transpotent stem cell such as a totipotent or pluripotent stem cell may differentiate into a skin cell or heart cell or neuron. Or it may transform into a progenitor cell that generates a limited number of new cells of various types.  If, however, the stem cell is an iteropotent stem cell then it may, given the right conditions, continually generate skin cells, or heart cells or neurons. While the same stem cell may be both a transpotent as well as an iteropotent stem cell, it need not be. Some stem cells may be transpotent but not iteropotent while others may be iteropotent but not transpotent.  Some stem cells may generate transpotent stem cells that are not iteropotent.  And, as we will see iteropotent stem cells can generate other stem cells. In the latter case, we will call them {\em meta-stem cells}. 

It can be even more intertwined.  A daughter cell of a stem cell may stochastically or in response to an external signal, dedifferentiate into its parent stem cell control state. {\em Differentiation} includes a change in the cell's control state.  {\em Dedifferentiation} means that the cell control state jumps from its given control state to an earlier upstream control state in its global developmental network.  The {\em locality} of the stem cell network in the global network determines the particular phenotypic properties of the stem cell and that stem cell's progeny. 

Thus as we change the theoretical framework from a gene-centered view of development to a control network view of development a new conceptual framework becomes available that allows more precise  definitions of stem cells and their properties.  
 
The more immature a daughter cell is, the more cell types it can differentiate into.  The cell type it becomes may depend on the cell types of the cellular context into which it is born.  The more mature the parent stem cell is the more constrained is the cell type of its daughters\footnote{We need to distinguish stem cell network hierarchy from the stem cell maturity which has to do with network locality.} .  

The transpotent stem cells can be totipotent cells, pluripotent, multipotent or oligopotent.  A totipotent cell can generate an entire organism. However, it is not necessarily an iteropotent stem cell since it does not self-renew, only one embryo is formed from a given cell under normal development.  Stem cell that produced totipotent cells is a theoretical possibility. Then each daughter cell of such a stem cell could produce an embryo.  A pluripotent cell can differentiate into multiple cell types.  Multipotent cells are even more restricted in the cell types they can differentiate into.  These gradations are vague and difficult to define precisely but they are all based on the cell's capacity to differentiate into a set of cell types that then may or may not be able to generate entire organisms, organs, multicellular tissue or terminate in one or more cells of some type.  If we look at the control network guiding such cells then a totipotent cell is controlled by the global developmental control network or cenome of the organism. Transformative pluripotent cells are controlled by some subnetwork of the global network. Such transpotent networks need not have the architecture of iteropotent stem cells. Generative stem cells are defined by their developmental control network architecture or topology.  It is this architecture that is the basis of their capacity to self-renew.  

This hierarchy is based on a hierarchical network architecture.  This architecture is also the basis of metastatic hierarchies in cancer.  A group of totipotent stem cells interact with each other to prevent multiple embryos from forming, one potential embryo from each totipotent cell. 

The molecular implementation of stem cell networks involves transcription factors and possibly as yet undiscovered RNA based regulators.  The basis of all regulation is an addressing system that maps addresses to potential control areas of the genome and the cell.  

\subsection{Dedifferentiation of daughter cells into stem cells}
Dedifferentiation potential in the daughter cell introduces yet another layer of complexity. A stem cell may produce a daughter cell that has the capacity to dedifferentiate into its parent stem cell state. The dedifferentiation may be stochastic based on some probability or it may be induced by cell signaling or other cellular contextual information.  In that case the original stem cell is conditionally a meta-stem cell since it has parented a stem cell but it is also conditionally exponential since it has generated two cells of its own cell type.  However, even exponential networks that are communication dependent need not proliferate exponentially (Werner~\cite{Werner2011b}).  

\subsection{Stem cell locality} 
The global developmental control network \footnote{The global developmental network is called the cenome in (Werner~\cite{Werner2011a, Werner2011b}).} is the control network responsible for the embryogenesis and development of multicellular organisms. Stem cells can be of various types that depend on the position that their own local control network has in the global developmental network. The ability of daughter cells of stem cells to assume various cell types depends on the developmental context in which they find themselves.  The earlier the stem cell network is linked into and activated by the global developmental network the more likely that the stem cell daughter cells are immature and pluripotent, pluripotent meaning that they can differentiate into multiple cell types.  If the stem cell network is linked later into a developmental path within the global control network then the more likely the stem cell will produce daughter cells that are less pluripotent, being relatively fixed in their phenotype. 

\subsection{Distinct stem cell concepts} Stem cell properties are the result of the properties of the developmental cell network that controls their behavior. Thus, the attributes of stem cells are directly linked to the properties of their controlling developmental network.  Hence, using network properties we can distinguish different attributes of stem cells: 

Developmental networks by definition are proliferative, meaning that they lead to one or more cell divisions.  A network has potentially infinite proliferative capacity if it contains loops that lead to endless repetition of a developmental network. 
 
\begin{enumerate}%description}

\item {\bf Stem cell potency} is the capacity of a cell to differentiate into various cell types such as being totipotent, pluripotent and oligopotent. This capacity is the result of an underlying developmental subnetwork.  This developmental subnetwork may or may not be proliferative, but need not have self-renewing loops. 

\item {\bf Stem cell self-generating capacity} is determined by the stem cell network architecture which includes the  hierarchical order of the self-renewing stem cell network, e.g. 1st-order, 2nd-order, kth-order geometric network. 

\item {\bf Stem cell proliferative potential} is the capacity of a cell to generate a set of cells when the network it links to is run. 
\footnote{The ambiguity: A cell may divide into two daughter cells. These in turn can divide further. So we need to distinguish the whole {\bf downstream potential} of a cell from its immediate daughter generating potential.  e.g., a stem cell because it is self-renewing has infinite potential over time. The problem is giving labels to different properties of developmental networks and their dynamic behavior in a developing system of cells. }
\footnote{Distinguish network accessibility from the resulting execution of that accessible downstream network. Note, in the case of a first order stem cell that network can be very small, and yet has potentially infinite generative/proliferative potential. Because of the infinite loop, the downstream proliferative potential is of a 1st order stem cell is infinite.}
\item {\bf Downstream potential of a net state} Downstream accessible nodes of a given net state. A net state is a set of net nodes. The execution of a net state partitions the developmental space into a set of subspaces of multicellular systems. Each subspace is the result of different developmental paths due to various possible conditions. 
\item {\bf The execution potential} is the set of network paths that are accessible from a given cell network control state.  The downstream network from a cell state is the subnetwork that is accessible from that cell state. It includes every accessible cell history from the given cell state. 

A cell history is a path in the developmental network. Given a cell state $\sigma$ let $\Omega(\sigma) = \sigma^{*} = \{H \in \Omega \mid \exists t \in \Psi \wedge n \in \sigma, \text{ such that }   H_{t}= n \}$ \footnote{The $H$ in $\Omega$ can be infinite.  The problem is that a state is a set of nodes which may be in different parts of the network. Hence, the nodes need not be on the same history or path. While the cell state history is linear, the control state is nonlinear since it can be on several paths simultaneously. Hence, the history of a ``path'' of a cell control state is a set of histories or linear paths. It can branch out or shrink in the future. So at any point it is a set of nodes, and over time it is a set of  paths in the network.  It is like parallel programs. There can be interpretation conflicts.  Any control pointer can change as a result of cell signaling or a new one can be added. But the receptor and its interpretation has to be thrown first, unless it is part of the innate non-genomic or pre-genomic or epigenomic IES. }

\item {\bf Stem cell locality} describes the position, by way of the linkage relationships, of the local stem cell network within the global developmental network of the organism. Stem cell locality may influence stem cell differentiation potency. 
\end{enumerate}%description}

Note, that these distinct stem cell attributes are described in terms of the properties of the underlying stem cell network.  Through these network properties we can distinguish the overt dynamic stem cell phenotype: 
{\em Stem cell potency} to differentiate into various cells types (totipotent, pluripotent, oligopotent) versus {\em stem cell proliferative capacity} (the order of the meta-stem cell network) versus {\em stem cell network locality} (where the stem cell network is linked into the global developmental network).  

A so-called stem cell that has the capacity to differentiate into a subset of cell types may not even have stem cell proliferative capacity.  For example, a totipotent embryonic stem cell need not be self-renewing even though it can generate stem cells that are self-renewing. Hence, the term stem cell is being used equivocally for different network properties.  Differentiation potency is an independent dimension from stem cell proliferative capacity as governed by a stem cell network.  To clearly define stem cells we need to define them through their underlying developmental networks. 

\subsection{Stem cell classification and methods that induce cell differentiation}
Furthermore, when an experimentalist manipulates a cell by various chemical or physical means to differentiate into some specific cell type, it does not mean that that cell has a natural potential to differentiate into that cell type.  The {\em natural differentiation potential} is given by the cells local active developmental network. External experimental manipulations may force an {\em artificial differentiation potential} by forcing the activation of totally different developmental control network that then may give the cell a new proliferative dynamic and capacity to differentiate to a new, uncharacteristic phenotype.  

While forced cell differentiation may have bioengineering and medical applications, it should not be used to define the capacity of the unforced, natural stem cell.  That being said, placing the cell in a new multicellular context or the external introduction of cell signals such as hormones, while maintaining the cell's integrity,  can help decipher the conditional network activation potential of the cell.  Hence, the boundary between forced and unforced cell differentiation may be difficult to delineate.  

\nocite{Werner2003b, Werner2005, Werner2007a, Werner2009,Werner2010}
\nocite{Tyson1991, Tyson2003}
\nocite{Dick2010, Dick2009, Dick2008}
%\nocite{Simons2007a, Simons2007b}
\nocite{Shackleton2010, Shackleton2010a, Shackleton2010b, Shackleton2010c}
\nocite{Gardner1999, Gardner2001, Gardner2006, Gardner2007, Tesar2007}
\nocite{Sonnenschein1999}
\nocite{Bizzarri2008}
%\nocite{*}

\addcontentsline{toc}{section}{References}% for showing references in table of contents 
\begin{multicols}{2}

\footnotesize %\small %\footnotesize %\tiny
\bibliographystyle{abbrv}%nature}%plain}%for number e.g., [1] citations
\bibliography{StemCells} %BibCancerPaperArXiv} %Refs2,CancerModelingByrne}
\end{multicols}
\end{document}